\documentclass{aa}

\usepackage{graphicx}
\usepackage{txfonts}
\usepackage{colortbl}
\usepackage{multirow}
\usepackage[squaren, Gray, cdot]{SIunits}
\usepackage{epstopdf}
\begin{document}

\title{A near-infrared interferometric survey of debris-disk stars}

\subtitle{VII. The hot/warm dust connection\thanks{Based on observations made with ESO Telescopes at the La Silla Paranal Observatory under program IDs 093.C-0712 and 094.C-0325.}}

\titlerunning{A near-infrared interferometric survey of debris-disk stars. VII.}

\author{O.~Absil\inst{1}\fnmsep\thanks{F.R.S.-FNRS Research Associate} \and L.~Marion\inst{1} \and S.~Ertel\inst{2,3} \and D.~Defr\`ere\inst{4} \and G.~M.~Kennedy\inst{5} \and A.~Romagnolo\inst{6} \and J.-B.~Le Bouquin\inst{7} \and V.~Christiaens\inst{8} \and J.~Milli\inst{7} \and A.~Bonsor\inst{9} \and J.~Olofsson\inst{10,11} \and K.~Y.~L.~Su\inst{3} \and J.-C.~Augereau\inst{7}}

\authorrunning{Absil et al.}

\institute{STAR Institute, Universit\'e de Li\`ege, 19c All\'ee du Six Ao\^ut, 4000 Li\`ege, Belgium
\and Large Binocular Telescope Observatory, 933 North Cherry Avenue, Tucson, AZ 85721, USA
\and Steward Observatory, Department of Astronomy, University of Arizona, 993 N. Cherry Ave, Tucson, AZ, 85721, USA
\and Institute of Astronomy, KU Leuven, Celestijnlaan 200D, 3001 Leuven, Belgium
\and Department of Physics, University of Warwick, Gibbet Hill Road, Coventry, CV4 7AL, UK
\and Nicolaus Copernicus Astronomical Center, Polish Academy of Sciences, Bartycka 18, 00-716, Warsaw, Poland
\and Univ.\ Grenoble Alpes, CNRS, IPAG, 38000 Grenoble, France
\and School of Physics and Astronomy, Monash University, Clayton, Vic 3800, Australia
\and Institute of Astronomy, University of Cambridge, Madingley Road, Cambridge CB3 0HA, UK
\and Instituto de F\'isica y Astronom\'ia, Facultad de Ciencias, Universidad de Valpara\'iso, Av. Gran Breta\~na 1111, Playa Ancha, Valpara\'iso, Chile
\and N\'ucleo Milenio Formaci\'on Planetaria - NPF, Universidad de Valpara\'iso, Av. Gran Breta\~na 1111, Valpara\'iso, Chile
}

   \date{Received 6 June 2017; accepted 21 April 2021}

 
  \abstract
   {Hot exozodiacal dust has been shown to be present in the innermost regions of an increasing number of main sequence stars over the past fifteen years. The origin of hot exozodiacal dust and its connection with outer dust reservoirs remains however unclear.}
   {We aim to explore the possible connection between hot exozodiacal dust and warm dust reservoirs ($\geq 100$~K) in asteroid belts.}
   {We use precision near-infrared interferometry with VLTI/PIONIER to search for resolved emission at H band around a selected sample of 62 nearby stars showing possible signposts of warm dust populations.}
   {Our observations reveal the presence of resolved near-infrared emission around 17 out of 52 stars with sufficient data quality, four of which are shown to be due to a previously unknown stellar companion. The 13 other H-band excesses are thought to originate from the thermal emission of hot dust grains, close to their sublimation temperature. Taking into account earlier PIONIER observations, where some stars with warm dust were also observed, and after re-evaluating the warm dust content of all our PIONIER targets through spectral energy distribution modeling, we find a detection rate of $17.1^{+8.1}_{-4.6} \%$ for H-band excess around main sequence stars hosting warm dust belts, which is statistically compatible with the occurrence rate of $14.6^{+4.3}_{-2.8} \%$ found around stars showing no signs of warm dust. After correcting for the sensitivity loss due to partly unresolved hot disks, under the assumption that they are arranged in a thin ring around their sublimation radius, we however find tentative evidence at the $3\sigma$ level that H-band excesses around stars with outer dust reservoirs (warm or cold) could be statistically larger than H-band excesses around stars with no detectable outer dust.}
   {Our observations do not suggest a direct connection between warm and hot dust populations, at the sensitivity level of the considered instruments, although they bring to light a possible correlation between the level of H-band excesses and the presence of outer dust reservoirs in general.}


   \maketitle
%

\section{Introduction}

Studying the formation and evolution of potentially habitable Earth-like planets requires a good knowledge of the environment close to the habitable zone, and thus of the \emph{exozodiacal} dust residing in this region (similar to our zodiacal dust). The presence of exozodiacal dust around other stars may represent a major obstacle for future terrestrial planet-finding missions \citep{defrere10, defrere12, roberge12, stark14b}. Indeed, exozodiacal dust disks (``exozodis'') not only add a significant amount of photon noise to the observations, but may also result in confusion, where the structures of the exozodis mimic the expected signal of an Earth-like planet as seen by future coronagraphic or interferometric space-based observatories. Usually, when referring to exozodiacal dust, one considers primarily the dust in the habitable zone \citep[e.g.][]{defrere10, roberge12, stark08}. However, in our Solar system, zodiacal dust is much more extended than the habitable zone, and actually shows an increasing density down to the F-corona, with a possible dust-free zone within 0.1--0.2~au from the Sun \citep[e.g.][]{Dikarev2015,Howard2019}. Likewise, it is expected that exozodiacal dust can extend over a broad range of separations from its host star, much larger than just the habitable zone.

The capability of near-infrared interferometry to probe the presence of hot dust in the innermost regions around nearby stars was first demonstrated by \citet{ciardi01} and by \citet{absil06}. The study of \citet{absil06} was then followed by a series of papers, which have extended the search to about 150 nearby stars, mostly using the CHARA/FLUOR and VLTI/PIONIER instruments \citep{absil06,difolco07,Absil08,Absil13,ertel14,nunez17}. These studies have shown that near-infrared excesses can be resolved around about 10\% to 30\% of nearby main sequence stars depending on the observing wavelength. \citet{ertel16} have also demonstrated the repeatability of the detections, showing that near-infrared excesses are not spurious and caused by poorly understood instrumental or astrophysical errors. Our current understanding is that near-infrared excesses around main sequence stars are related to the thermal emission from hot dust grains close to their sublimation temperature ($\sim$1500~K for silicate dust grains). The contribution of scattered light cannot be excluded in some cases \citep{difolco07,mennesson11,defrere12,ertel14}, although recent polarimetric, interferometric, and theoretical studies argue against scattered light as a prominent contributor to the detected excesses \citep{kennedy15,kennedy2015,marshall16,kirchschlager17,kirchschlager20}. These previous studies have highlighted a tentative correlation between spectral type and near-infrared excess detection rate, but could not formally identify any correlation between the presence of hot dust and of cold, distant dust reservoirs detected by far-infrared and submillimeter photometry. The factors influencing the presence of hot exozodiacal dust around nearby main sequence stars are therefore still unclear, which calls for more observational constraints. 

\begin{figure}[t]
\centering
\includegraphics[width=0.48\textwidth]{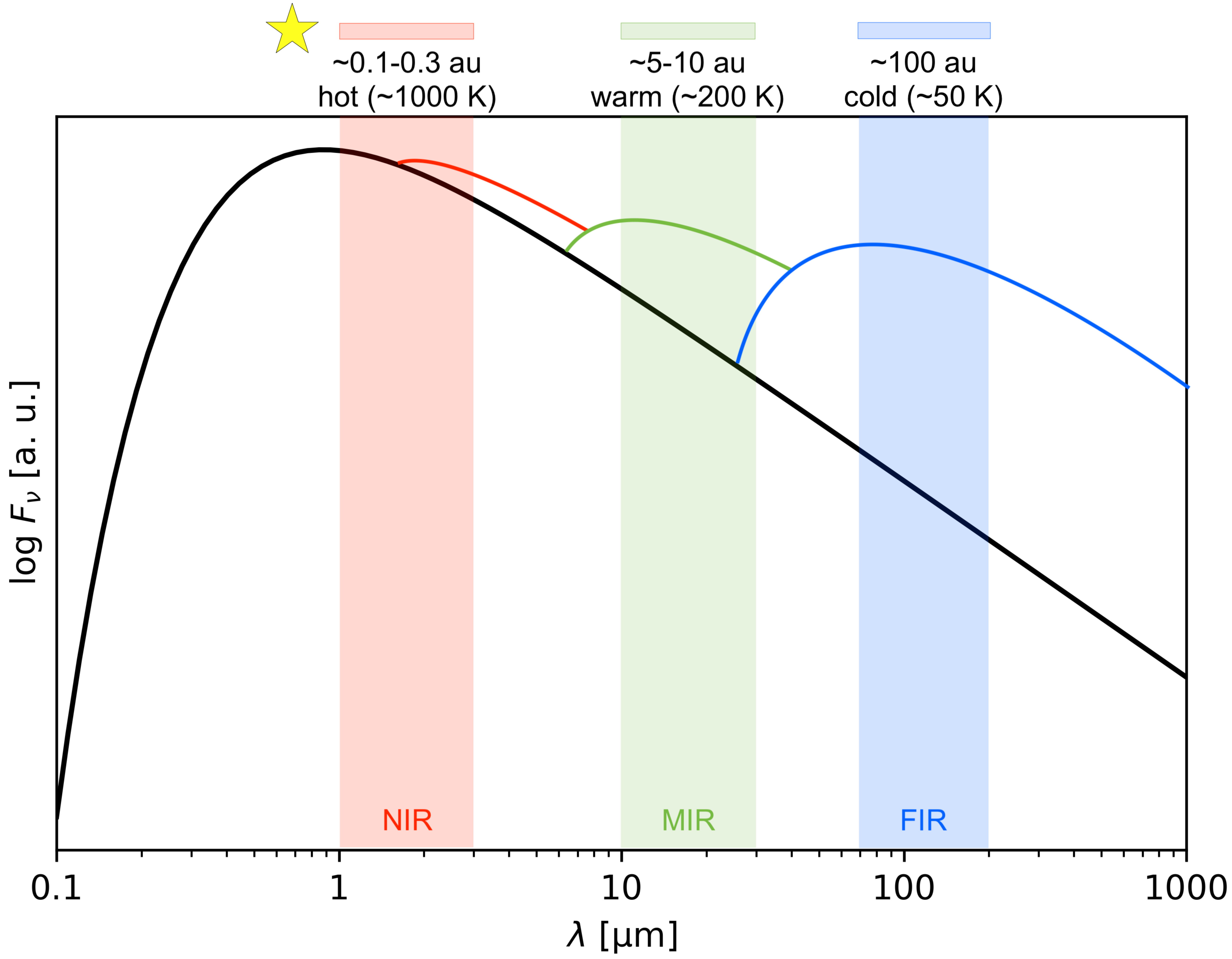}
\caption{Illustration of the typical orbital distances and temperatures for the hot, warm, cold dust belts considered in this work, and of the corresponding wavelength ranges in terms of spectral energy distribution. Adapted from \citet{kirchschlager17}.}
\label{fig:hotwarmcold}
\end{figure}

Here, we study the possible correlation of the hot dust phenomenon with the presence of warm asteroid belts around nearby main sequence stars. We define warm dust as dust producing a detectable excess in the mid-infrared but not in near-infrared (typical temperatures in the range 100--500~K), while hot dust is defined as dust producing an excess in the near-infrared (see Fig.~\ref{fig:hotwarmcold}). Our main goal is to determine whether the presence of hot exozodiacal dust could be directly related to the presence of a large reservoir of planetesimals in an asteroid belt, in an attempt to improve our understanding of the origin, architecture, and evolution of bright exozodiacal dust disks, as well as of the factors influencing their detection rate. To this aim, we build a sample of stars known to have a mid-infrared excess attributed to debris disks based on infrared space missions such as Spitzer and WISE (Sect.~\ref{sec:stelsamp}). After detailing the PIONIER observations and data reduction in Sect.~\ref{sec:obsandred}, we present in Sect.~\ref{sec:comp} the search for unknown companions in this sample -- a necessary step to remove possible contamination in our sample. In Sect.~\ref{sec:exozodi}, we present the search for hot exozodis in this sample, detailing the search method and the results. Finally, in Sect.~\ref{sec:discussion}, we discuss the connection between hot and warm dust. We also challenge the standard hypothesis of fully resolved exozodis in interferometric observations, and explore the consequences of partly resolved disks on the measured detection rates.


\section{Stellar sample}
\label{sec:stelsamp}

\begin{figure*}[!t]
\centering
\includegraphics[width=1\textwidth]{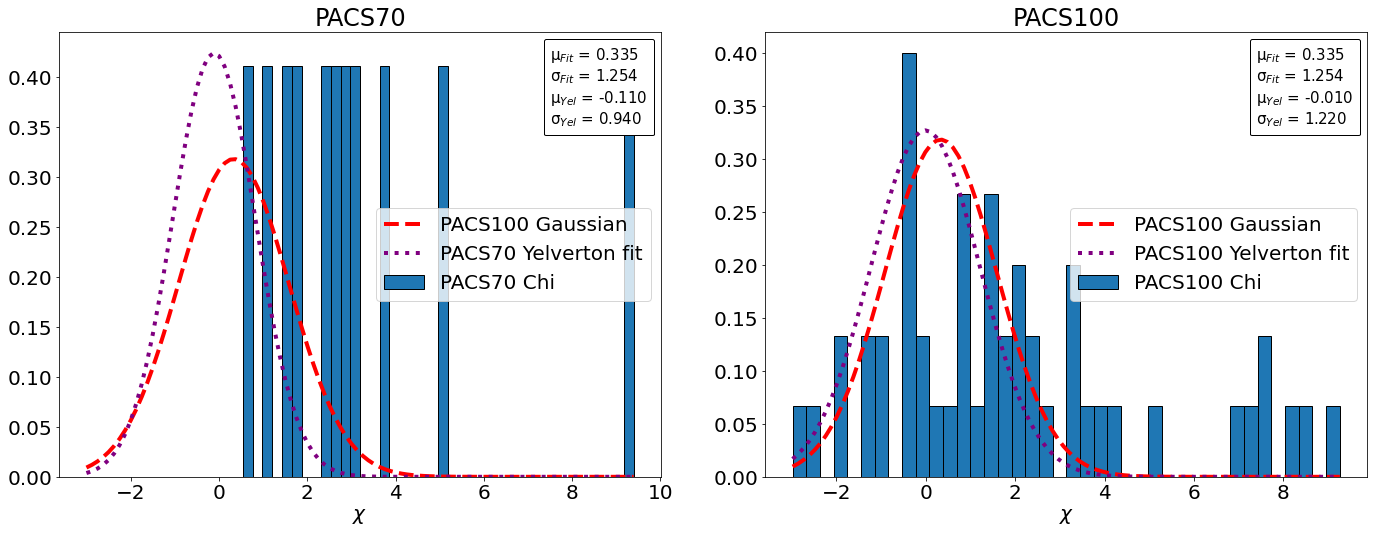}
\caption{Histograms of excess significance for PACS70 (left) and PACS100 (right) measurements. The purple dotted curve shows the noise distribution derived by \citet{yelverton19} for their PACS70 (left) and PACS100 (right) data sets, while the red dashed curve shows the same noise distribution derived from our PACS100 data set---our PACS70 data set was not large enough to robustly fit the noise distribution. The values noted as $\mu_{\rm Fit}$ and $\sigma_{\rm Fit}$ are respectively the mean and standard deviation of the significance in our PACS100 distribution, while $\mu_{\rm Yel}$ and $\sigma_{\rm Yel}$ are the ones from \citet{yelverton19}.} 
\label{fig:PacsYelComparison}
\end{figure*}

Searching for correlations between hot and warm dust populations first requires to build a large enough sample of nearby stars hosting warm dust. Three main space-based missions have been used to search for warm dust around nearby stars: Spitzer, AKARI, and WISE. We searched the literature for warm excesses around nearby stars, focusing mostly on these three missions \citep{carpenter09, chen06, hillenbrand08, ishihara10, morales12, ballering13, fujiwara13, chen14, vican14, patel14}. To identify warm dust, these missions rely on spectrophotometric observations at wavelengths shorter than 25~$\mu$m. Showing a mid-infrared excess is however not a sufficient condition to infer the presence of warm dust, as excesses in this wavelength range can sometimes correspond to the short-wavelength end of a bright but cold circumstellar emission. We originally built our sample based on the dust temperature estimated in the literature, using a threshold of 130~K\footnote{This temperature threshold of 130~K was only used to select our target stars, based on the available literature in 2013. We will discuss later that a temperature of 100~K was finally chosen to classify between warm and cold dust populations. A significant fraction of the selected targets actually turned out not to show the presence of any warm dust after re-evaluation of their mid- to far-infrared excess, as described in Sect.~\ref{sub:warmcold}.} as a criterion to distinguish warm from cold populations, following \citet{ballering13}.
In several cases, the warm excesses could only be detected at a single wavelength, making an accurate temperature determination impossible. In these cases, the authors generally quote the highest possible temperature compatible with their data set. Lacking more precise information, we decided to use these upper limits as a criterion to select stars with possible warm dust populations, where applicable. While our previous near-infrared interferometric surveys were targeting stars brighter than $H=5$, here to build a sufficiently large sample we include stars up to $H=7$, which remains comfortably within the magnitude limit of VLTI/PIONIER. Stars with visual companions within the interferometric field of view of PIONIER on the VLTI Auxiliary Telescopes ($\sim$400~mas full width at half maximum in H band) are not appropriate for detecting weak, extended circumstellar emission. Even light from companions outside the field of view may enter the optical path in case of bad seeing. Thus, as in \citet{ertel14}, all known binary systems with angular separation $<5\arcsec$ are removed from our sample. We identified a total of 62 stars meeting our criteria, which had not been observed yet with precision near-infrared interferometry. The main properties of these 62 targets to be observed with PIONIER are summarized in Table~\ref{tab:allsurv}. We collected PIONIER data of sufficient quality for only 52 of them, as described in Sect.~\ref{sec:obsandred}. Furthermore, four of these 52 stars turned out to be binaries, based on our PIONIER observations (see Sect.~\ref{sec:comp}). These binary stars are not amenable to a search for exozodiacal dust, and are therefore removed from our sample, so that we are left with 48 new stars to study the correlation between warm and hot dust.

\longtab{
\begin{longtable}{cccccccc}
\caption{Main properties of the 62 newly observed stars.} \label{tab:allsurv}\\
\hline\hline
Star & Type & Dist. & $V$ & $H$  & $\theta_{LD}$ & Age & References\\
 & & (pc) & (mag) & (mag)  & (mas) & (Gyr) & \\
\hline
\endfirsthead
\caption{continued.}\\
\hline\hline
Star & Type & Dist. & $V$ & $H$ & $\theta_{LD}$ & Age & References\\
 & & (pc) & (mag) & (mag)  & (mas) & (Gyr) & \\
\hline
\endhead
\hline
\endfoot
\object{HD 203} & F2IV & 39.0 & $6.181^{0.003}$ & $5.32^{0.05}$ & $0.347^{0.005}$& 0.021 & 1, 2, \textbf{3}, 31, 32\\
\object{HD 2834} & A0V & 53.0 & $4.751^{0.008}$ & $4.76^{0.07}$ & $0.381^{0.006}$ & 0.22 & \textbf{1}, 2, 3, 4\\
\object{HD 3126} & F5V & 41.0 & $6.907^{0.009}$ & $5.85^{0.05}$ & $0.284^{0.004}$& 1.59 & \textbf{1}, 2, 3, 4\\
\object{HD 4113} & G5V & 44.0 & $7.889^{0.009}$ & $6.34^{0.02}$ & $0.240^{0.003}$ & 5.8 & 5, \textbf{7}  \\
\object{HD 4247} & F0V & 27.0 & $5.218^{0.003}$ & $4.46^{0.01}$ & $0.513^{0.006}$ & 1.7 & 6  \\
\object{HD 9672} & A1V & 59.0 & $5.611^{0.004}$ & $5.53^{0.02}$ & $0.273^{0.004}$ & 0.1 & \textbf{1}, 3, 4, 8 \\
\object{HD 10008} & K0/1V & 24.0 & $7.66^{0.01}$ & $5.90^{0.04}$ & $0.324^{0.005}$ & 4.2 & 1, \textbf{3}, 9\\
\object{HD 10269} & F5V & 48.0 & $7.078^{0.004}$ & $5.90^{0.04}$ & $0.252^{0.003}$ & 1.6 & \textbf{1}, 4\\
\object{HD 10939} & A1V & 62.0 & $5.033^{0.003}$ & $5.03^{0.02}$ & $0.339^{0.006}$ & 0.2 & \textbf{1}, 2, 3\\
\object{HD 15427} & A2/3V & 47.0 & $5.124^{0.002}$ & $5.03^{0.02}$ & $0.349^{0.005}$ & 0.24 & \textbf{1}, 2\\
\object{HD 17848} & A2V & 50.5 & $5.252^{0.004}$ & $5.16^{0.08}$ & $0.350^{0.005}$ & 0.28 & 1, 2, \textbf{3}, 4, 8\\
\object{HD 23484} & K1V & 16.0 & $6.982^{0.004}$ & $5.09^{0.02}$ & $0.484^{0.006}$ & 6.9 & 2, 3, 8, 10, 11\\
\object{HD 24649} & F6V & 41.0 & $7.217^{0.007}$ & $6.09^{0.03}$ & $0.261^{0.004}$ & 4.8 & \textbf{1}, 6 \\
\object{HD 28287} & K0V & 38.0 & $8.77^{0.01}$ & $6.87^{0.04}$ & $0.210^{0.003}$ & 0.1 & \textbf{1}, 3, 7  \\
\object{HD 29137} & G5V & 52.0 & $7.663^{0.008}$ & $6.16^{0.03}$ & $0.258^{0.004}$ & 6.29 & \textbf{1}, 12 \\
\object{HD 31203} & F0IV & 37.1 & $5.606^{0.004}$ & $4.88^{0.02}$ & $0.414^{0.006}$ & 0.70 & 4, 13 \\
\object{HD 31392} & G9V & 26.0 & $7.600^{0.008}$ & $5.89^{0.04}$ & $0.317^{0.005}$ & 3.70 & 3, \textbf{8} \\
\object{HD 36187} & A0V & 87.8 & $5.557^{0.003}$ & $5.51^{0.02}$ & $0.264^{0.004}$ & 0.25 & 4, \textbf{14}\\
\object{HD 37306} & A2V & 63.0 & $6.087^{0.004}$ & $5.992^{0.02}$ & $0.215^{0.003}$ & 0.16 & 1, \textbf{2}, 3, 4, 8 \\
\object{HD 37484} & F3V & 57.0 & $7.249^{0.008}$ & $6.29^{0.02}$ & $0.217^{0.003}$ & 0.7 & \textbf{1}, 2, 3, 6, 8 \\
\object{HD 38949} & G1V & 43.3 & $7.808^{0.008}$ & $6.48^{0.04}$ & $0.215^{0.003}$ & 0.9 & 2, 3, \textbf{8} \\
\object{HD 41278} & F5V & 56.0 & $7.394^{0.008}$ & $6.36^{0.03}$ & $0.220^{0.003}$ & 2.3 & \textbf{1}, 6 \\
\object{HD 43879} & F5V & 64.1 & $7.494^{0.008}$ & $6.46^{0.04}$ & $0.216^{0.003}$ & 2.0 & 4, 15\\
\object{HD 44524} & F3V & 102.3 & $7.012^{0.01}$ & $6.46^{0.03}$ & $0.193^{0.003}$ & 1.6 & 6, \textbf{14}, 15 \\
\object{HD 59967} & G3V & 21.8 & $6.657^{0.004}$ & $5.25^{0.02}$ & $0.412^{0.006}$ & 0.63 & \textbf{1}, 2, 3, 16, 17 \\
\object{HD 60491} & K2V & 25.0 & $8.15^{0.01}$ & $6.14^{0.02}$ & $0.298^{0.005}$ & 0.08 & \textbf{1}, 3, 18 \\
\object{HD 61005} & G3/5V & 35.3 & $8.215^{0.008}$ & $6.58^{0.04}$ & $0.228^{0.004}$ & 0.1 & \textbf{1}, 2, 3, 19 \\
\object{HD 71722} & A0V & 71.7 & $6.05^{0.004}$ & $5.91^{0.02}$ & $0.225^{0.003}$ & 0.4 & \textbf{1}, 2, 3, 4, 8, 20 \\
\object{HD 76143} & F5IV & 52.0 & $5.328^{0.003}$ & $4.42^{0.02}$ & $0.536^{0.009}$ & 2.2 & \textbf{1}, 3, 4 \\
\object{HD 80133} & K1/2V & 68.5 & $7.76^{0.01}$ & $5.90^{0.03}$ & $0.337^{0.005}$ & 0.4 & \textbf{7}, 11 \\
\object{HD 80883} & K0V & 36.2 & $8.59^{0.01}$ & $6.63^{0.05}$ & $0.242^{0.004}$ & 6.3 & \textbf{ 7}, 21 \\
\object{HD 89886} & F7V & 167.0 & $7.44^{0.01}$ & $6.09^{0.05}$ & $0.273^{0.004}$ & 1.6 & 6, \textbf{7} \\
\object{HD 90781} & F3V & 77.0 & $7.448^{0.008}$ & $6.51^{0.03}$ & $0.198^{0.003}$ & 1.2 & 4, \textbf{14}, 15 \\
\object{HD 90874} & A2V & 68.0 & $5.991^{0.004}$ & $5.86^{0.04}$ & $0.237^{0.003}$ & 0.25 & \textbf{1}, 2, 3, 4, 8 \\
\object{HD 92945} & K1V & 21.4 & $7.708^{0.007}$ & $5.77^{0.05}$ & $0.347^{0.005}$ & 0.21 & 7, 8 \\
\object{HD 93453} & A4IV & 72.0 & $6.288^{0.004}$ & $5.91^{0.03}$ & $0.244^{0.003}$ & 0.4 & \textbf{1}, 4 \\
\object{HD 105850} & A1V & 56.1 & $5.447^{0.003}$ & $5.35^{0.04}$ & $0.290^{0.004}$ & 0.2 & \textbf{1}, 2, 3, 4, 8, 22 \\
\object{HD 105912} & F2/3V & 50.0 & $6.940^{0.007}$ & $5.96^{0.06}$ & $0.258^{0.004}$ & 2.7 & 1, 2, 3, \textbf{8}, 23\\
\object{HD 106906} & F5V & 59.0 & $7.798^{0.008}$ & $6.76^{0.04}$ & $0.184^{0.003}$ & 0.015 & 1, 2, 3\\
\object{HD 109573} & A0V & 67.1 & $5.777^{0.004}$ & $5.79^{0.04}$ & $0.231^{0.003}$ & 0.01 & 2, 3, 4, \textbf{8}, 22\\
\object{HD 109704} & A3V & 68.8 & $5.869^{0.003}$ & $5.77^{0.05}$ & $0.245^{0.003}$ & 0.4 & 1, 2, 3, 4, \textbf{8}, 22 \\
\object{HD 112603} & F2V & 61.0 & $6.952^{0.004}$ & $6.14^{0.05}$ & $0.232^{0.003}$ & 1.5 & \textbf{1}, 6 \\
\object{HD 117716} & A0/1V & 72.0 & $5.690^{0.004}$ & $5.67^{0.03}$ & $0.255^{0.003}$ & 0.3 & 1, 2, 3, \textbf{8}\\
\object{HD 118972} & K1V & 15.6 & $6.918^{0.004}$ & $5.14^{0.05}$ & $0.480^{0.006}$ & 0.3 &  \textbf{1}, 2, 3, 7, 8, 11, 24 \\
\object{HD 136544} & F6V & 74.0 & $7.43^{0.01}$ & $6.35^{0.03}$ & $0.221^{0.004}$ & 2.0 & \textbf{1}, 9 \\
\object{HD 141378} & A5IV & 54.0 & $5.522^{0.003}$ & $5.27^{0.03}$ & $0.306^{0.004}$ & 0.3 & 1, 2, \textbf{3}, 8, 25 \\
\object{HD 141943} & G0/2V  & 67.0 &  $7.85^{0.01}$ & $6.41^{0.03}$ & $0.231^{0.004}$ & 0.03 & 3, \textbf{8}, 33 \\
\object{HD 142139} & A3V & 66.0 & $5.747^{0.003}$ & $5.66^{0.05}$ & $0.261^{0.004}$ & 0.2 & \textbf{1}, 2, 3, 4, 8 \\
\object{HD 161612} & G6/8V & 26.9 & $7.18^{0.01}$ & $5.6^{0.1}$ & $0.344^{0.005}$ & 0.8 & \textbf{7}, 11, 26 \\
\object{HD 174474} & A2V & 82.0 & $6.169^{0.004}$ & $5.89^{0.04}$ & $0.236^{0.004}$ & 0.6 & \textbf{3}, 4, 14, 15\\
\object{HD 175073} & K1V & 24.0 & $7.96^{0.01}$ & $5.95^{0.03}$ & $0.324^{0.005}$ & 4.1 & \textbf{1}, 27 \\
\object{HD 178606} & F5V & 53.0 & $6.520^{0.007}$ & $5.49^{0.02}$ & $0.323^{0.004}$ & 1.7 & \textbf{1}, 23 \\
\object{HD 179520} & F3V & 62.0 & $7.092^{0.007}$ & $6.24^{0.02}$ & $0.232^{0.003}$ & 0.6 & \textbf{1}, 3, 23 \\
\object{HD 181327} & F5/6V & 52.0 & $7.035^{0.008}$ & $5.98^{0.04}$ & $0.263^{0.004}$ & 0.021 & 1, 2, 3, 28, 31, 32 \\
\object{HD 184932} & F2V & 65.0 & $8.03^{0.01}$ & $6.95^{0.02}$ & $0.166^{0.003}$ & 2.1 & \textbf{1}, 4 \\
\object{HD 185615} & G6IV & 43.5 & $8.11^{0.01}$ & $6.54^{0.03}$ & $0.286^{0.004}$ & 9.2 &\textbf{7}, 9, 15 \\
\object{HD 191089} & F5V & 52.0 & $7.178^{0.007}$ & $6.09^{0.03}$ & $0.243^{0.004}$ & 0.021 & 1, 3, 8, 22, 31, 32 \\
\object{HD 192758} & F0V & 62.0 & $7.013^{0.008}$ & $6.30^{0.04}$ & $0.217^{0.004}$ & 0.04 & 3, \textbf{8}, 29\\
\object{HD 196141} & G3V & 37.0 & $8.09^{0.01}$ & $6.58^{0.03}$ & $0.213^{0.004}$ & 0.4 & \textbf{7} \\
\object{HD 205674} & F3/5IV & 52.0 & $7.178^{0.007}$ & $6.25^{0.03}$ & $0.228^{0.003}$ & 2.2 & \textbf{1}, 3, 4, 8 \\
\object{HD 220476} & G5V & 30.0 & $7.611^{0.009}$ & $6.11^{0.04}$ & $0.276^{0.004}$ & 0.4 & \textbf{7} \\
\object{HD 224228} & K3V & 22.0 & $8.237^{0.009}$ & $6.01^{0.03}$ & $0.325^{0.005}$ & 0.1-0.2 & \textbf{1}, 30 \\
\end{longtable}
\tablefoot{$1\sigma$ error bars are given as superscripts. V and H magnitudes are from \citet{kharchenko09}. Limb-darkened stellar diameters ($\theta_{\rm LD}$) are computed from surface-brightness relationships based on the V and K magnitudes, following \citet{Kervella04}. References include previous searches for warm and cold dust around the target stars, with the reference in bold highlighting the origin of the warm dust classification that led to their addition to our sample, where applicable.}
\tablebib{ (1) \citet{patel14}; (2) \citet{ballering13}; (3) \citet{cotten16} ;
(4) \citet{david15}; (5) \citet{bonfanti16}; (6) \citet{holmberg09}; (7) \citet{vican14}; (8) \citet{chen14}; (9) \citet{pace13}; (10) \citet{eiroa13}; (11) \citet{valenti05}; 
(12) \citet{delgado14}; (13) \citet{huensch98}; (14) \citet{wu13}; (15) \citet{mcdonald12}; (16) \citet{durkan16}; (17) \citet{tuccimaia16}; (18) \citet{maldonado12}; (19) \citet{desidera15}; 
(20) \citet{pawellek14}; (21) \citet{delgado15}; (22) \citet{mittal15}; (23) \citet{feltzing01}; (24) \citet{mamajek08}; (25) \citet{DeRosa14}; (26) \citet{tsantaki13}; (27) \citet{casagrande11}; 
(28) \citet{stark14}; (29) \citet{wahhaj13}; (30) \citet{maire14}; (31) \citet{zuckerman04}; (32) \citet{binks14}; (33) \citet{chauvin10}
}}

To assess a possible correlation between the presence of hot and warm dust, we also need to build a control sample. Our control sample is based on the VLTI/PIONIER survey for hot exozodiacal dust carried out by \citet{ertel14}. The reader is referred to that paper for a detailed description of the stellar parameters of this sample. Among the 85 single and non-evolved stars included in that sample, we expected from the literature that a large majority does not host any warm dust population, based on the absence of mid-infrared excesses. We however noted that the warm vs.\ cold dust classification derived from the dust temperatures described in the literature was inconsistent, because of the various assumptions made in the publications of these mid-infrared surveys. We therefore decided to re-assess the presence of warm dust around the 133 stars included in both the \citet{ertel14} sample (85 stars) and our new sample (48 stars).

	\subsection{Reassessing the presence of warm and cold dust}
	\label{sub:warmcold}

\begin{figure*}[t]
\centering
\includegraphics[width=0.8\textwidth]{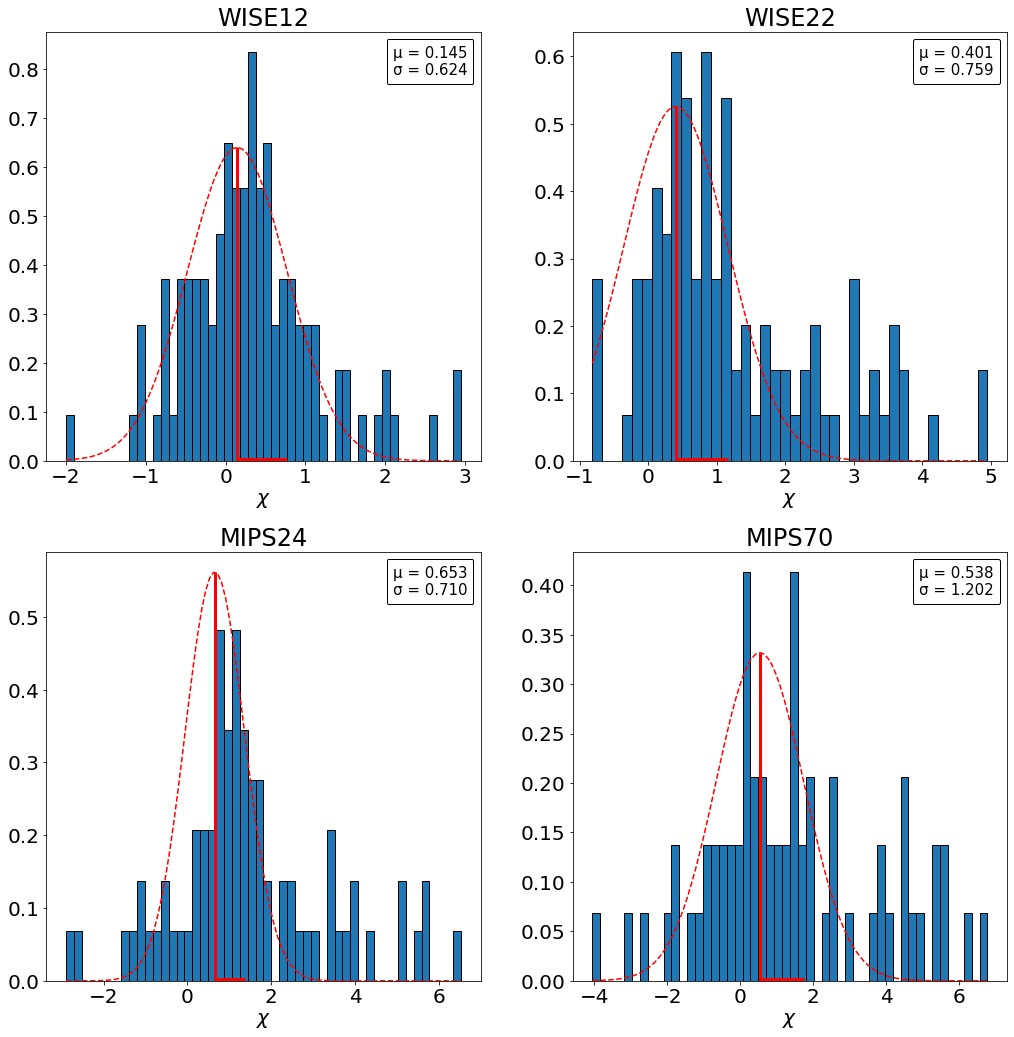}
\caption{Histograms of excess significance for WISE12 (top left), WISE22 (top right), MIPS24 (bottom left), and MIPS70 (bottom right), together with their best-fit Gaussian noise distribution (red dashed curve). The mean ($\mu$) and standard deviation ($\sigma$) of the noise distributions are also plotted respectively as red vertical and horizontal lines. Significance values lower than $-4$ and higher than $8$ are not displayed for the sake of clarity.}
\label{fig:ChiHist}
\end{figure*}

In order to reassess the presence of warm and/or cold dust around our combined sample of 133 stars, we collected photometry at all available optical, mid-, and far-infrared wavelengths for all targets. In most cases these are available in literature catalogues, for example for optical $UVB$ and $ubvy$ data \citep{1987A&AS...71..413M,2015A&A...580A..23P}, and 2MASS and WISE IR data \citep{2003tmc..book.....C,2010AJ....140.1868W}. However, far-IR photometry for our targets is either unpublished or spread across many different analyses. To maximize consistency, Spitzer/MIPS photometry at 24 and 70~$\mu$m uses our own updated PSF fitting for all targets, as described in \citet{2014ApJ...785...33S} and \citet{yelverton19}, except four bright targets, which use 70~$\mu$m photometry from \citet{chen14}. Herschel/PACS photometry at 70, 100, and 160~$\mu$m, and SPIRE photometry at 250, 350, and 500~$\mu$m, also uses our PSF fitting, as described in \citet{2018MNRAS.475.3046S}. We also use Spitzer/IRS spectra from the CASSIS archive \citep{2011ApJS..196....8L}, where available.

The data for each star are initially fit with a star + disk model using the \texttt{sdf} code, as described by \citet{yelverton19}. The star is a BT-Settl photosphere model \citep{2012RSPTA.370.2765A} and the disk is a modified blackbody (a Planck function that is multiplied by $\lambda_0^{-\beta}$ at wavelengths longer than $\lambda_0$), and the fitting is done using the \texttt{multinest} code \citep{2009MNRAS.398.1601F}. The fitting serves two purposes, firstly to provide an estimate of the stellar flux at all wavelengths to allow the presence of any IR excess to be quantified, and secondly to estimate the temperature and luminosity of any disk if the excess is deemed significant as described below (if no excess is present the disk component has negligible flux by definition, and is not used). In some cases the single-component disk provides a poor fit to the IR excess, in which case a second disk component is added. Whether a second component is needed is somewhat subjective, since the true dust spectrum is unknown and might mimic a two-component disk \citep{kennedy14}. Our assessment primarily considers whether two disk components are needed to fit all photometry and the IRS spectrum, but also considers whether the dust temperatures of a two-component fit are sufficiently different \citep{kennedy14}.

To assess whether IR excesses are significant, we use the empirical method used by many previous studies \citep[e.g.][]{su06,ertel14,yelverton20}, and which we also use below for the PIONIER observations. Essentially, we assume that most stars do not have a significant excess, and therefore that the distribution of excess significance in a given band $i$
$$ \chi_i = \frac{F_{obs}- F_\star}{\sqrt{\sigma_{obs}^2+\sigma_\star^2}} $$
should be approximately Gaussian with zero mean and unity standard deviation. Any positive outliers with $\chi_i>3$ would therefore be considered to have a significant excess. If the standard deviation is greater or smaller than unity, as is commonly the case, the uncertainties for that band are considered to be under- or overestimated respectively, and the threshold for an excess adjusted accordingly. For example, \citet{yelverton20} found the standard deviation for PACS 100~$\mu$m to be 1.68, and therefore set the threshold for an excess to be $3 \times 1.68$. In general the IRS spectra were used as important input for deriving disk temperatures, but in a few cases (HD~40307, HD~90874, HD~91324) were also used to confirm an excess that was not significant photometrically (due to low signal-to-noise ratio and/or poor wavelength coverage).


We decided to classify as dusty any stellar system whose excess significance was beyond four standard deviations from the average ($\chi>\mu+4\sigma$) in at least one band. In order to keep an average excess significance around zero in our noise distributions, the Gaussian fitting of the noise distribution was done without any star with $\chi>3$. In the case of PACS70, there were not enough data points to fit a decent Gaussian distribution. We therefore explored the possibility to use the noise distributions from \citet{yelverton19} for that filter. In Fig.~\ref{fig:PacsYelComparison}, we compare our Gaussian fit to the PACS100 significance with the ones obtained by \citet{yelverton19} for their PACS70 and PACS100 data sets. The difference, in terms of outliers, was found to be marginal: only the excesses of two additional stars (namely HD~7570 for PACS100, and HD~174474 for PACS70) were significant with the use of \citet{yelverton19} parameters but not with ours. We therefore decided to adopt the Gaussian noise distributions fitted by \citet{yelverton19} for our PACS70 and PACS100 data sets. The $\chi$ histograms for the remaining four WISE and MIPS filters are plotted in Fig.~\ref{fig:ChiHist} with their respective noise distributions.



\begin{figure*}[t]
\centering
\includegraphics[width=0.45\textwidth]{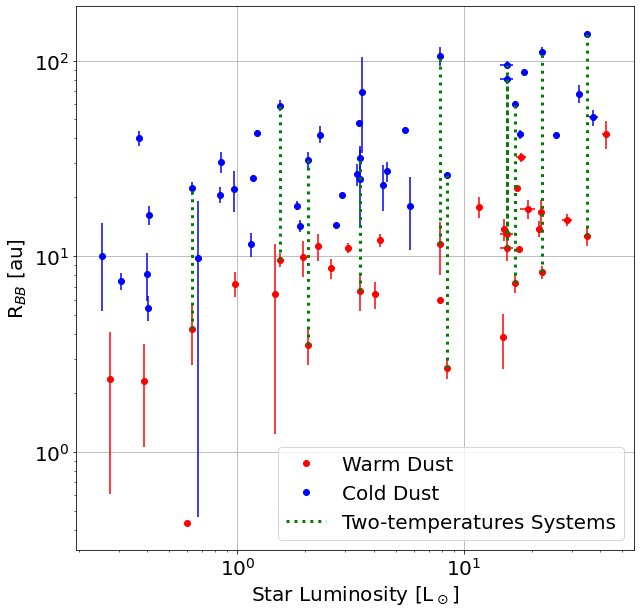} \hspace*{5mm} \includegraphics[width=0.45\textwidth]{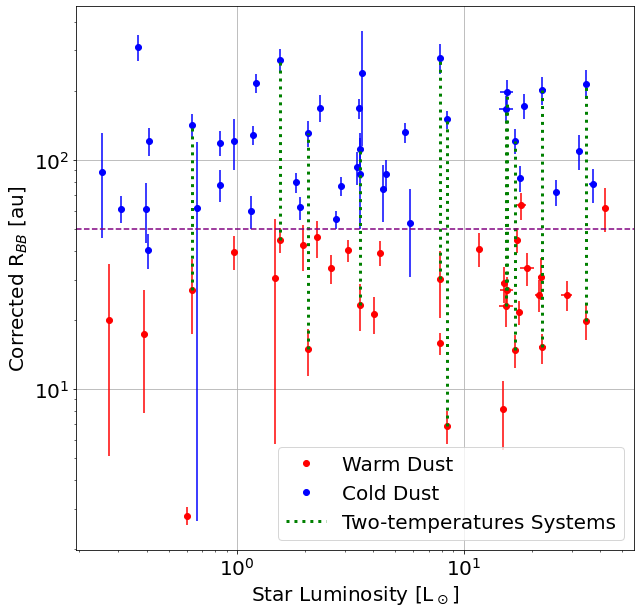}
\caption{(Left). Equivalent black body radius of the dust disks around the 66 stars in our combined sample showing a significant mid- to far-infrared excess, for which a dust temperature could be derived. Disks classified as warm ($>100$~K) and as cold ($<100$~K) are respectively colored red and blue. (Right). Same as left, for the corrected black body radius, using the 50\% astrosilicate + 50\% ice composition of \citet{pawellek15}. A purple dashed line at 50~au shows a rough separation between warm and cold dust populations. The vertical green dotted lines link the warm and the cold dusts that were found in the same stellar system.}
 \label{fig:diskradius}
\end{figure*}

Among the 133 stars in our combined sample, 68 show the presence of circumstellar dust in their mid- to far-infrared SED based on our analysis (see Appendix~\ref{appendix} for an illustration of all 133 SEDs). In order to separate these 68 stars into two categories (warm and cold dust), we set an arbitrary limit of 100~K to distinguish between the two populations. The new warm/cold dust classification for these 68 stars is given in Table~\ref{tab:class}. We note that two stars (HD~36187 and HD~89886) do not have sufficient mid- to far-infrared data to conclude on the temperature of the detected excess. These two stars are therefore removed from our warm/cold dust samples, leaving us with a total of 131 stars, among which 66 show the presence of circumstellar dust.

\longtab{
\begin{longtable}{cccccccccccc}
\caption{Warm vs.\ cold dust classification for the 68 stars showing the presence of a debris disk in our SED modeling. Symbols X and O represent respectively a significant excess and the absence of an excess in the considered filter, while a dash denotes a filter where no data (or data of insufficient quality) was available for the considered target. For the IRS data, symbol X is used when the $\pm 1 \sigma$ error interval around the measured spectrum is located above the photospheric model for a significant, contiguous part of the wavelength range spanned by the IRS spectrum. The last four columns give the inferred temperature and luminosity of the detected disks (``nc'' stands for non-constrained). Asterisks denote the newly observed stars from this study.} \label{tab:class}\\
\hline\hline
name & W12 & W22 & M24 & IRS & M70 & P70 & P100 & $T_{\rm warm}$ & $T_{\rm cold}$ & $L_{\rm warm}$/$L_{\ast}$ & $L_{\rm cold}$/$L_{\ast}$  \\
 & & & & & & & & (K) & (K) & ($\times 10^{-4}$)  & ($\times 10^{-4}$)  \\
\hline
\endfirsthead
\caption{continued.}\\
\hline\hline
name & W12 & W22 & M24 & IRS & M70 & P70 & P100 & $T_{\rm warm}$ & $T_{\rm cold}$ & $L_{\rm warm}$/$L_{\ast}$ & $L_{\rm cold}$/$L_{\ast}$  \\
 & & & & & & & & (K) & (K) & ($\times 10^{-4}$)  & ($\times 10^{-4}$)  \\
\hline
\endhead
\hline
\endfoot
$^{\ast}$HD 203	    & O	& X	& X	& X	& X	& X	& X	& 115 & $-$ & 1.3 & $-$\\
HD 2262 	&   & O	& X	& X	& X	&$-$& X	& 122 & $-$ & 0.095 & $-$ \\
$^{\ast}$HD 2834 	& O	& O	& X	& X	& X	&$-$&$-$& $-$ & 96 & $-$ & 0.12 \\
$^{\ast}$HD 3126 	& O	& O	& O	& O	& X	&$-$&$-$& $-$ & 53 & $-$ & 1.3 \\
HD 7570 	& O	& O	& X	& X	& O	&$-$& X	& 100 & $-$ & 0.066 & $-$\\
$^{\ast}$HD 9672	& X	& X	&$-$& X	& X	& X	& X	& 153 & 57 & 1.6 & 7.0\\
$^{\ast}$HD 10269 	& O	& X	&$-$&$-$&$-$& X	&$-$& 120 & $-$ & 1.3 & $-$\\
HD 10647 	& O	& O	& X	& X	& X	& X	& X	& 100 & 40 & 0.62 & 2.5 \\
$^{\ast}$HD 10939 	& O	& X	& X	& X	& X	& X	& X	& 190 & 58 & 0.15 & 0.73 \\
HD 11171 	& O	& O	& O	&$-$& X	& X	& X	& $-$ & 100 & $-$ & 0.11\\
$^{\ast}$HD 15427 	& O	& X	& X	& X	& X	&$-$&$-$& 101 & $-$ & 0.33 & $-$\\
$^{\ast}$HD 17848 	& O	& O	& X	& X	& X	& X	& X	& $-$ & 61 & $-$ & 0.42\\
HD 17925 	& 	& O	& 	& X	& X	& X	& X	& $-$ & 80 & $-$ & 0.33\\
HD 20794$^{a}$ 	&   &$-$& O	&$-$& O	& O	& O	& $-$ & 80 & $-$ & 0.022\\
$^{\ast}$HD 23484 	& O	& O	& O	& O	& X	& X	& X	& $-$ & 55 & $-$ & 0.93\\
$^{\ast}$HD 24649 	& O	& X	& X	&$-$&$-$& X	&$-$& $-$ & 76 & $-$ & 1.5 \\
HD 25457 	& 	& X	& X	& X	& X	&$-$&$-$& 180 & 60 & 0.40 & 0.97\\
$^{\ast}$HD 28287 	& O	& O	&$-$&$-$&$-$& X	&$-$& 140 & $-$ & 1.0 & $-$\\
HD 28355 	& O	& X	& X	& X	& X	&$-$& X	& $-$ & 88 & $-$ & 0.41\\
HD 30495 	& O	& O	& O	& O	& X	& X	& X	& $-$ & 59 & $-$ & 0.32\\
HD 31295 	& O	& X	& X	& X	& X	&$-$& X	& 165 & 61 & 0.19 & 0.54\\
HD 33262 	& 	& O	& O	&$-$& X	&$-$& X	& 120 & $-$ & 0.16 & $-$\\
$^{\ast}$HD 36187$^{b}$	& O	& X	&$-$&$-$&$-$&$-$&$-$& nc & nc & nc & nc \\
$^{\ast}$HD 37306 	& O	& X	& X	& X	& X	& X	& X	& 120 & $-$ & 0.70 & $-$\\
$^{\ast}$HD 37484 	& O	& X	& O	& X	& X	&$-$&$-$& 150 & 67 & 1.4 & 2.2\\
HD 38858 	& O	& O	& O	& X	& X	& X	&$-$& $-$ & 52 & $-$ & 22 \\
HD 39060 	& X	& X	&$-$&$-$& X	& X	& X	& 290 & 86 & 8.1 & 0.76\\
HD 40307 	& O & O	& O	& X	& O	& O	& O	& $-$ & 60 & $-$ & 0.071 \\
HD 45184 	& O	& O	& O	& X	& X	& X	&$-$& $-$ & 58 & $-$ & 0.77\\
$^{\ast}$HD 60491 	& O	& O	& X	&$-$& X	&$-$&$-$& $-$ & 76 & $-$ & 2.1 \\
$^{\ast}$HD 61005 	& O	& X	& X	& X	& X	& X	& X	& 120 & 53 & 2.1 & 24 \\
HD 69830 	& X	& X	& X	& X	& O	&$-$& X	& 373 & $-$ & 2.0 & $-$\\
HD 71155 	& X	& X	& X	&$-$& X	& X	& X	& 109 & $-$ & 0.28 & $-$\\
$^{\ast}$HD 71722 	& O	& X	& X	& X	& X	&$-$& X	& 210 & 73 & 0.28 & 0.76 \\
$^{\ast}$HD 76143$^{c}$	& O	& X	&$-$&$-$&$-$&$-$&$-$& >100 & $-$ & $\sim$2 & $-$\\
HD 76151 	& O	& O	& O	& X	& X	&$-$& X	& 103 & $-$ & 0.14 & $-$\\
$^{\ast}$HD 89886$^{b}$	& O	& X	&$-$&$-$&$-$&$-$& & nc & nc & nc & nc\\
$^{\ast}$HD 90874 	& O	& O	& O	& X	& O	&$-$&$-$& 148 & $-$ & 0.12 & $-$\\
HD 91324 	& 	& O	& O	& X	& O	&$-$&$-$& $-$ & 80 & $-$ & 0.053\\
$^{\ast}$HD 92945 	& O	& O	& O	& O	& X	& X	&$-$& $-$ & 34 & $-$ & 6.8 \\
$^{\ast}$HD 105850 	& X	& X	& X	& X	& O	&$-$&$-$& 161 & $-$ & 0.28 & $-$\\
$^{\ast}$HD 105912 	& O	& X	& X	& X	& X	&$-$&$-$& 111 & $-$ & 0.84 & $-$\\
$^{\ast}$HD 109573 	& X	& X	&$-$& X	& O	& X	& X	& $-$ & 97 & $-$ & 44 \\
HD 109704 	& O	& X	& X	& X	& O	&$-$&$-$& 140 & $-$ & 0.40 & $-$\\
HD 115617 	& O	& X	& O	& X	& X	& X	& X	& $-$ & 59 & $-$ & 0.23 \\
$^{\ast}$HD 117716 	& X	& O	& X	& X	& O	&$-$&$-$& 150 & $-$ & 0.085 & $-$\\
$^{\ast}$HD 118972 	& O & O	& O	& X	& X	&$-$& X	& $-$ & 95 & $-$ & 0.41\\
HD 135379 	& 	& X	& X	& X	& O	&$-$&$-$& 173 & $-$ & 0.44 & $-$\\
HD 139664 	& O	& O	&$-$&$-$& X	& X	&$-$& $-$ & 74 & $-$ & 1.26\\
$^{\ast}$HD 141943 	& O	& X	& X	& X	& O	&$-$&$-$& 102 & $-$ & 0.83 & $-$ \\
HD 160032 	& 	& O	& O	& X	& O	&$-$& X	& $-$ & 78 & $-$ & 0.053\\
HD 172555 	& X	& X	& X	& X	& X	& X	& X	& 190 & $-$ & 5.14 & $-$\\
$^{\ast}$HD 174474 	& O	& X	&$-$&$-$&$-$& X	&$-$& 280 & $-$ & 0.92 & $-$ \\
HD 178253 	& O	& X	& X	& X	& X	&$-$&$-$& 164 & $-$ & 0.17 & $-$ \\
$^{\ast}$HD 179520 	& O	& X	&$-$&$-$&$-$& X	&$-$& 160 & $-$ & 1.48 & $-$ \\
$^{\ast}$HD 181327 	& O	& X	& X	& X	& X	&$-$& X	& $-$ & 80 & $-$ & 27 \\
HD 188228 	& O	& O	& O	& X	& X	& X	& X	& $-$ & 80 & $-$ & 0.041 \\
$^{\ast}$HD 191089 	& O	& X	& X	&$-$& X	&$-$& X	& $-$ & 94 & $-$ & 15 \\
HD 192425 	& O	& X	& X	& X	& X	&$-$& X	& 210 & 57 & 0.26 & 0.36 \\
$^{\ast}$HD 192758 	& O	& X	& X	& X	& X	&$-$& X	& $-$ & 64 & $-$ & 5.0 \\
HD 195627 	& O	& O	& X	& X	& X	& X & X	& 140 & 45 & 0.15 & 0.90\\
$^{\ast}$HD 205674 	& O	& O	& X	& X	& X	&$-$& X	& $-$ & 55 & $-$ & 3.4 \\
HD 206860 	& O	& O	& O	& X	& O	&$-$& X	& $-$ & 85 & $-$ & 0.082 \\
HD 207129 	& O	& O	& O	& X	& X	& X	& X	& $-$ & 45 & $-$ & 0.91\\
HD 213845$^{a}$ 	& O	& O	& O	& O & O &$-$& O & $-$ & 80 & $-$ & 0.032\\
HD 216435 	& O	& O	&$-$& O	& X	& X	&$-$& $-$- & 50 & $-$ & 0.18 \\
HD 219482 	& O	& O	& X	& X	& X	& X	& X	& $-$ & 86 & $-$ & 0.27 \\
$^{\ast}$HD 224228 	& O	& O	& X	&$-$& O	&$-$&$-$& 130 & $-$ & 0.61 & $-$ \\
\end{longtable}
\tablefoot{($^{a}$) no formal excess detected, but the combination of marginal (close to significant) excesses at 70 and 100~$\mu$m is clear evidence for a cold dust population (see also SEDs in Fig.~\ref{seds0} to \ref{seds-1}); ($^{b}$) no data available beyond 22~$\mu$m, preventing us from determining the temperature and luminosity of the disk; ($^{c}$) temperature constrained to be higher than 100~K by a combination of WISE12, WISE22, and archival IRAS60 \& IRAS100 photometry, also giving a rough estimate of disk luminosity (with a large error bar $\sim 10^{-4}$ due to the temperature vs.\ luminosity degeneracy).}
}

	\subsection{Statistics of our combined sample}

Based on our SED modeling, our final sample of 131 stars contains a total of 35 stars that show the presence of warm dust populations ($>100$~K), among which 11 also show the presence of a cold dust reservoir ($<100$~K). Another 31 stars show the presence of cold dust only, while 65 stars show no sign of circumstellar dust, at a sensitivity level that depends on the stellar brightness and spectral type, and on the quality of the mid- to far-infrared observations available for each star. For the 66 stars that show the presence of circumstellar dust, we plot in Fig.~\ref{fig:diskradius} the equivalent black-body radius of the dust disk, as well as the corrected black-body radius following \citet{pawellek15}, using a 50\% astrosilicate + 50\% ice composition. The correction takes into account the blow-out of the smallest dust grains, and is meant to provide a more representative estimate of the true radius of a spatially unresolved dust disk. The right-hand side plot in Fig.~\ref{fig:diskradius} shows that our 100~K criterion corresponds more or less to a 50~au limit in terms of corrected black body radius between the warm and cold dust disks, although some A-type stars show warm dust up to about 60~au, while late-type stars show cold dust down to about 40~au. This relatively good agreement between our temperature classification and a classification based on the corrected black body distance to the host star provides an independent justification to our classification strategy. While a 50~au limit between warm and cold dust populations may seem large, it ensures that the two populations are balanced in size, and we argue in Sect.~\ref{sub:temp_corr} that choosing a higher temperature threshold would not change the conclusions of our study.


While the dusty and non-dusty samples are relatively equally spread between A-type, F-type and G/K-type stars, with $33\% \pm 8\%$ of stars in each of the three spectral type categories, it must be noted that the warm dust sample is largely biased towards A-type stars (with 18 A-type stars out 35), while the cold dust sample is biased towards G/K-type stars (14 G/K-type stars out of 31). This imbalance likely arises because disks tend to be warmer around earlier type stars \citep[e.g.][]{kennedy14}. This could also be partly due to the fact that warm dust appears up to larger orbital distances around A-type stars, even after taking into account the black body correction of \citet{pawellek15}.

\begin{table*}[t]
\caption{Summary of the new VLTI/PIONIER observations.}
\label{tab:log}
\centering
\begin{tabular}{cccccc}
\hline \hline
Run & Night  & \# stars & Seeing (\arcsec) & $\tau_0$ (ms) & Notes\\
\hline
093.C-0712(A) & 07-04-2014 & 7 & 1.6 (0.8 - 2.5) & 1.4 (1.1 - 1.7) & Strong seeing at the end of the night\\
093.C-0712(A) & 08-04-2014 & 9 & 1.7 (0.6 - 2.9) & 2.2 (0.8 - 3.7) & Strong seeing in the middle of the night\\
093.C-0712(A) & 09-04-2014 & 9 & 1.3 (0.6 - 2.0) & 2.5 (1.3 - 3.8) & Some clouds\\
093.C-0712(B) & 30-08-2014 & 8 & 1.2 (0.6 - 1.9) & 1.7 (1.1 - 2.3) & Good conditions\\
093.C-0712(B) & 31-08-2014 & 9 & 1.2 (0.6 - 1.9) & 2.1 (1.0 - 3.2) & Some clouds\\
093.C-0712(B) & 01-09-2014 & 4 & 1.1 (0.7 - 1.6) & 3.2 (1.9 - 4.5) & Thin clouds, dome closed for a part of the night\\
094.C-0325(A) & 22-12-2014 & 7 & 1.0 (0.5 - 1.5) & 2.3 (1.4 - 3.2) & Good conditions\\
094.C-0325(A) & 23-12-2014 & 11& 1.4 (0.5 - 2.4) & 2.3 (1.1 - 3.6) & Strong seeing at the beginning of the night\\
094.C-0325(A) & 24-12-2014 & 8 & 1.5 (0.4 - 2.5) & 3.1 (0.9 - 5.3) & Strong seeing at the beginning of the night\\
\hline
\end{tabular}
\end{table*}

Our target stars mostly consist of old, main sequence field stars, generally not younger than a few hundreds of million years. A handful of stars in our sample are somewhat younger: HD~141943 \citep[field star, 30~Myr,][]{chen14}; HD~203, HD~39060, HD~172555, HD~181327 and HD~191089 \citep[part of the $\beta$~Pic moving group, 21~Myr,][]{binks14}; HD~192758 \citep[part of IC~2391, 40~Myr,][]{wahhaj13}; HD~109573 \citep[part of the TW Hya association, 10~Myr,][]{mittal15}; HD~106906 \citep[part of the Lower Centaurus Crux association, 15~Myr,][]{Pecaut16}; and HD~188228 \citep[part of the Argus association, 38~Myr,][]{booth13}. We note that the average stellar age of the warm dust sample (0.83 Gyr) is significantly lower than for the control sample (2.7 Gyr), which is not unexpected as the presence of warm dust is known to be correlated with the system age \citep[e.g.,][]{su06,vican14}. Any correlation between hot and warm dust will therefore also be tested for a possible age bias, although previous studies \citep[e.g.][]{Absil13,ertel14} do not suggest a significant age dependence in the hot exozodi phenomenon.






\section{Observations and data reduction}
\label{sec:obsandred}

Observations were carried out with VLTI/PIONIER \citep{LeBouquin11} at H band in April, August and December 2014, each run consisting of three consecutive observing nights. An observing log of all nights can be found in Table~\ref{tab:log}. We used the four 1.8-m ATs to obtain six visibility measurements simultaneously. We used for all observing runs an array configuration (D0-H0-G1-I1) with baselines ranging between 41~m and 82~m. This configuration is larger than the one used for the \citet{ertel14} survey, because the stars in the present sample are more distant in average and therefore require a higher angular resolution to resolve their sublimation radius (see Sect.~\ref{subsec:sublimation} for more details). After the August run, the detector of PIONIER was changed, which implied a change in the read-out mode. The read-out mode was set to FOWLER with SMALL dispersion (three spectral channels) for the observations of April and August 2014, and to HIGH SENS with a GRISM dispersion (six spectral channels) for the observations of December 2014. Four calibrator stars were selected from \citet{merand2005} for each science target, typically within 10\degree \ on sky to minimize the effects of pupil rotation or instrumental polarization \citep[see][]{LeBouquin12}. Additional selection criteria were an H-band magnitude similar to the science target, and a small angular diameter. Most of the targets were observed in a CAL1-SCI-CAL2-SCI-CAL3-SCI-CAL4 sequence, where two non-consecutive calibrators can be the same. 

Out of the 62 stars in our observing list, a total of ten stars had to be removed from our final sample for various reasons. Four could not be appropriately observed: HD~141378 and HD~43879 due to incomplete observing sequences (not enough data), HD~59967 because of inappropriate calibrators, and HD~93453 because two out of the three SCI observations were obtained during a burst of bad seeing ($>2\arcsec$). The other six stars were removed after the data reduction and calibration procedure, as detailed below. The data reduction consists of the conversion of raw observations into calibrated interferometric observables. We use the exact same method as in \citet{ertel14}. The first step of the calibration is to calibrate the instrumental visibility within the CAL-SCI-$\dots$-CAL sequence. To do so, we calibrate each SCI individually by pairing it with either the preceding or the following CAL. During this process, we also make sure to discard all calibrators with low S/N or with a clear closure phase signal \citep[see][for details]{ertel14}. After calibration, six stars had to be rejected from our sample (HD~4247, HD~10008, HD~31392, HD~142139, HD~178606, and HD~184932), because of large discontinuities in the interferometric transfer function\footnote{The interferometric transfer function monitors the instrumental visibility, or instrumental closure phase, as a function of time.} due to poor seeing conditions, low coherence time, or clouds. The actual number of new stars added through this observing program therefore amounts to 52.

The last step in the data analysis procedure is to assess the systematic polarization effects of PIONIER. This part is automatically corrected by a dedicated option in the standard PIONIER pipeline \citep[\texttt{pndrs},][]{LeBouquin11} used for the reduction. A detailed explanation of the polarization effects can be found in \citet{ertel14}.


\section{Searching for companions}
\label{sec:comp}

Before searching for hot exozodiacal disks based on our interferometric observations, we first need to identify possible unknown (sub-)stellar companions that could also produce an infrared excess. This will ensure that all our target stars are single.

\subsection{Principle of the search}
\label{subsec:princip}

Following the same lines as in \citet{absil11} and \citet{marion14}, we use the full information delivered by PIONIER (squared visibilities, $V^2$, and closure phases, CP) to systematically search for companions around all the stars. Doing so, we are able to discriminate whether the small near-infrared excess detected around some of the target stars is due to an extended, mostly symmetric source, or to a point-like companion. The method used to detect companions is fully described in \citet{marion14}. We provide here a brief summary. 

First, we define the search region, taking into account three main factors. The first one is the Gaussian profile of the single-mode fiber used in PIONIER (FWHM $\approx$ 400 mas). The second one is the finite scan length of the optical path delay. With the medium configuration used here and for a typical scan length of 100~$\mu$m, the maximum separation that can be probed is $\Delta\theta_{\rm max} \simeq 200$~mas, although we recognize that companions with separations larger than 70~mas may not be simultaneously visible on all baselines.
The third one is the sufficient sampling of the closure phase signal as a function of wavelength, which depends on the baseline and on the spectral resolution. Following \citet{absil11}, the well-sampled field-of-view is about 50~mas in radius. Taking all of this into account, we consider a search region 50~mas in radius in this study. We note however that our search is actually sensitive to companions out to about 200~mas, with point-like sources between 50 and 200~mas creating aliasing within our 50~mas search region. As explained in Sect.~\ref{sec:stelsamp}, companions beyond 200~mas may bias our observations, but we expect that such companions (with contrasts of 1\% or more) would have already been identified in the literature.

\begin{table*}[t]
\caption{Summary of the stars showing a significance level higher than $5\sigma$ based on the analysis of the combined $\chi^2$ (CP+$V^2$). The significance of the detection based on the separate analysis of the CP and the $V^2$ is also given. The nature of the detection is either a disk or a point-like source, in which case its main properties are given in the last three columns.}
\label{tab:detcomp}
\centering
\begin{tabular}{cccccccc}
\hline \hline
 &  & & & & \multicolumn{3}{c}{Point-like source} \\ \cline{6-8}
Name & \multicolumn{3}{c}{Significance} & Nature & Separation & P.A. & Contrast \\
 &(CP+$V^2$) & (CP) & ($V^2$)& & (mas) & (deg) & (\%)\\
\hline
HD~203 & 5.4 & 3.7 & 7.3 & disk &  -- & -- & --\\
HD~31203 &14.5 & 20.2 & 25.6 &point-like&  $64.6 \pm 1.3$ & $-50.2\pm 0.4$ & $4.3 \pm 0.6$\\
HD~36187 & 9.3 & 2.2 & 23.4 & disk & -- & -- & --\\
HD~61005 & 6.7 & 2.5 & 7.7 & disk & -- & -- & --\\
HD~76143 & 6.4 & 5.3 & 4.9 & disk&  -- & -- & -- \\
HD~80133  &4642.7 & 318.5 & 14616.4 & point-like&$6.0\pm 0.2$ & $13.2 \pm 0.7$ & $85.2 \pm 2.6$\\
HD~106906 &145.0 & 12.0 & 509.1 & point-like& $1.4 \pm 0.1$ & $95.2 \pm 5.1$ & $95.0 \pm 7.4$ \\
HD~175073 &64.7 & 59.6 & 391.3 & point-like& $31.2 \pm 0.6$ & $-84.7 \pm 0.4$ & $13.0 \pm 0.9$\\
\hline
\end{tabular}
\end{table*}

To detect the presence of a companion, we use the closure phases and the squared visibilities in a combined way. As in \citet{marion14}, we compute a binary model considering the primary star at the center of the search region with an off-axis companion of various contrast $c$ at each point $(x,y)$ of the search region. In the present case, we can safely assume that both the primary and the secondary stars are unresolved, since all of our targets have an angular diameter $\lesssim 0.5$~mas. Then, we compute the $V^2$ and CP for each model and derive a combined goodness of fit that we normalize and collapse along the contrast axis to keep only the best-fitting companion contrast (i.e., minimum $\chi^2$ value) at each position in the search region. The resulting $\chi^2$ map can then be used to derive the probability for the single-star model to adequately represent the data, based on the $\chi^2$ distribution under a Gaussian noise assumption. If this probability is below a predefined threshold, the single-star model can be rejected and the best-fit binary solution is then considered as statistically significant. The detection criterion is defined as a threshold on the significance of the detection, which can be translated into a confidence level if the underlying probability distribution function is known. 

\begin{figure}[t]
\begin{center}
\includegraphics[scale=0.45]{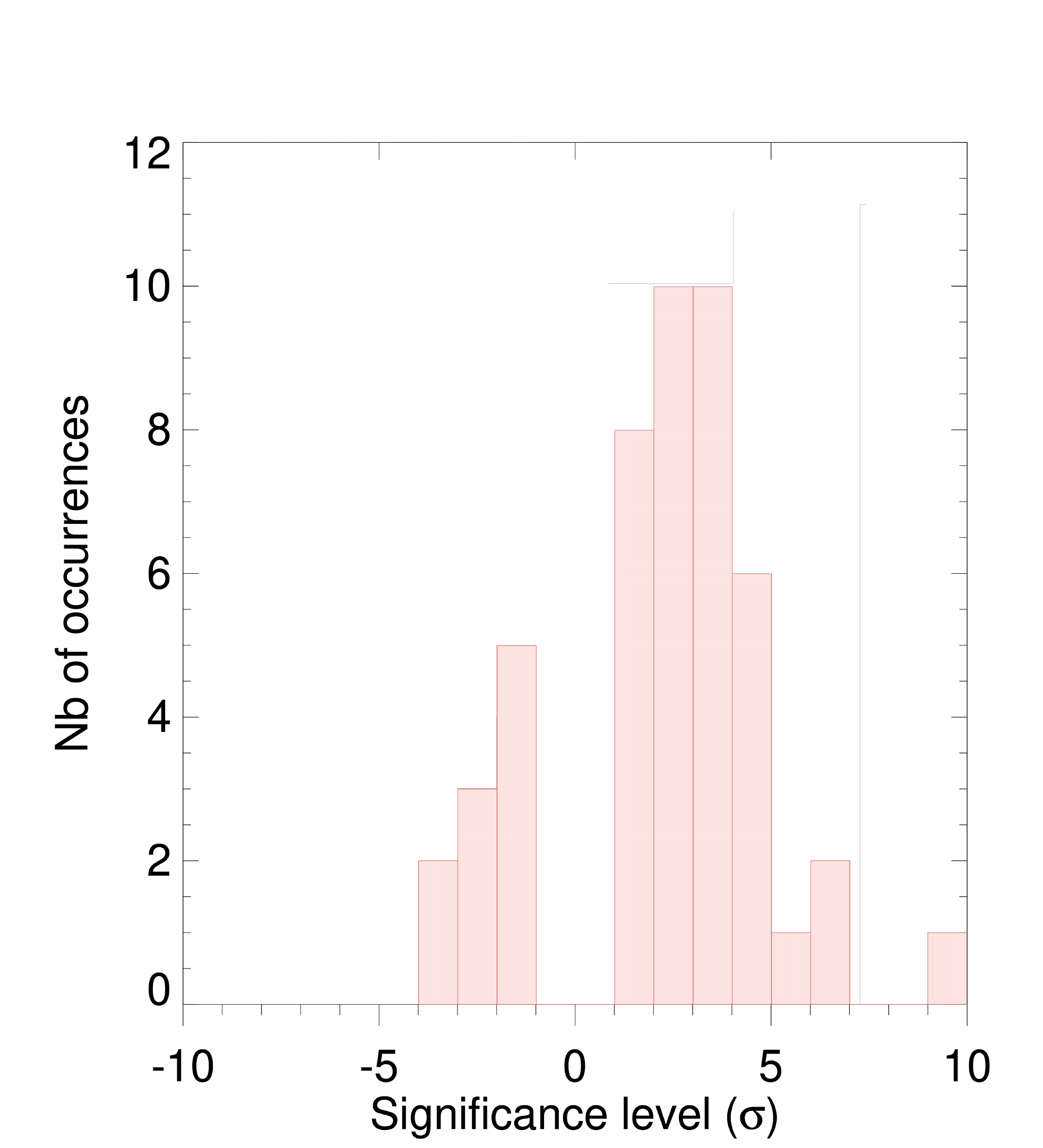}
\caption{Statistics of the (signed) significance level for the 52 stars based on the combined $\chi^2$, taking into account the CP and $V^2$. Four stars with a significance level higher than $10\sigma$ are not represented here for the sake of clarity.} 
\label{fig:histosignif}
\end{center}
\end{figure}

To determine the significance level to use as a detection threshold, we study the noise properties of the data set by including negative contrasts in our model for the off-axis companions. While non-physical, negative companions can be used to represent positive fluctuations in the $V^2$ (i.e., situations where the measured $V^2$ is higher than the expected $V^2$ from the photosphere). Negative companions can also be attributed to noise fluctuations in the CP, which can take both positive and negative values. In the following, we associate negative significance levels to negative companions. The histogram of the significance levels for our complete sample is illustrated in Fig.~\ref{fig:histosignif}, where the range of the plot has been limited to $[-10,10]$ for the sake of clarity. The negative significance levels in the histogram are purely due to noise fluctuations, and can therefore be used as a reference to study the noise properties of our sample. The absence of significance levels close to 0 in the histogram can be explained by the fact that, in presence of noise and due to the limited number of observations, it is generally possible to obtain a better fit to our data sets by inserting a companion somewhere in the field-of-view than by using a single-star model. Out of the 52 stars in our observed sample, ten show a negative significance level, but none are below $-5\sigma$. We therefore decide to use $5 \sigma$ as our empirical companion detection threshold based on the combined analysis of $V^2$ and CP. We note that a $4\sigma$ threshold would also have been appropriate for this analysis, owing to the distribution of the negative excesses. However, after a more detailed inspection of the closure phases, all the stars with a significance level between $4\sigma$ and $5\sigma$ actually turned out to be surrounded by extended emission, and not by point-like sources (see below for details on how this inspection was performed).

\subsection{Results of the search}
\label{subsec:rescomp}

\begin{figure*}[t]
\centering
\includegraphics[scale=0.5]{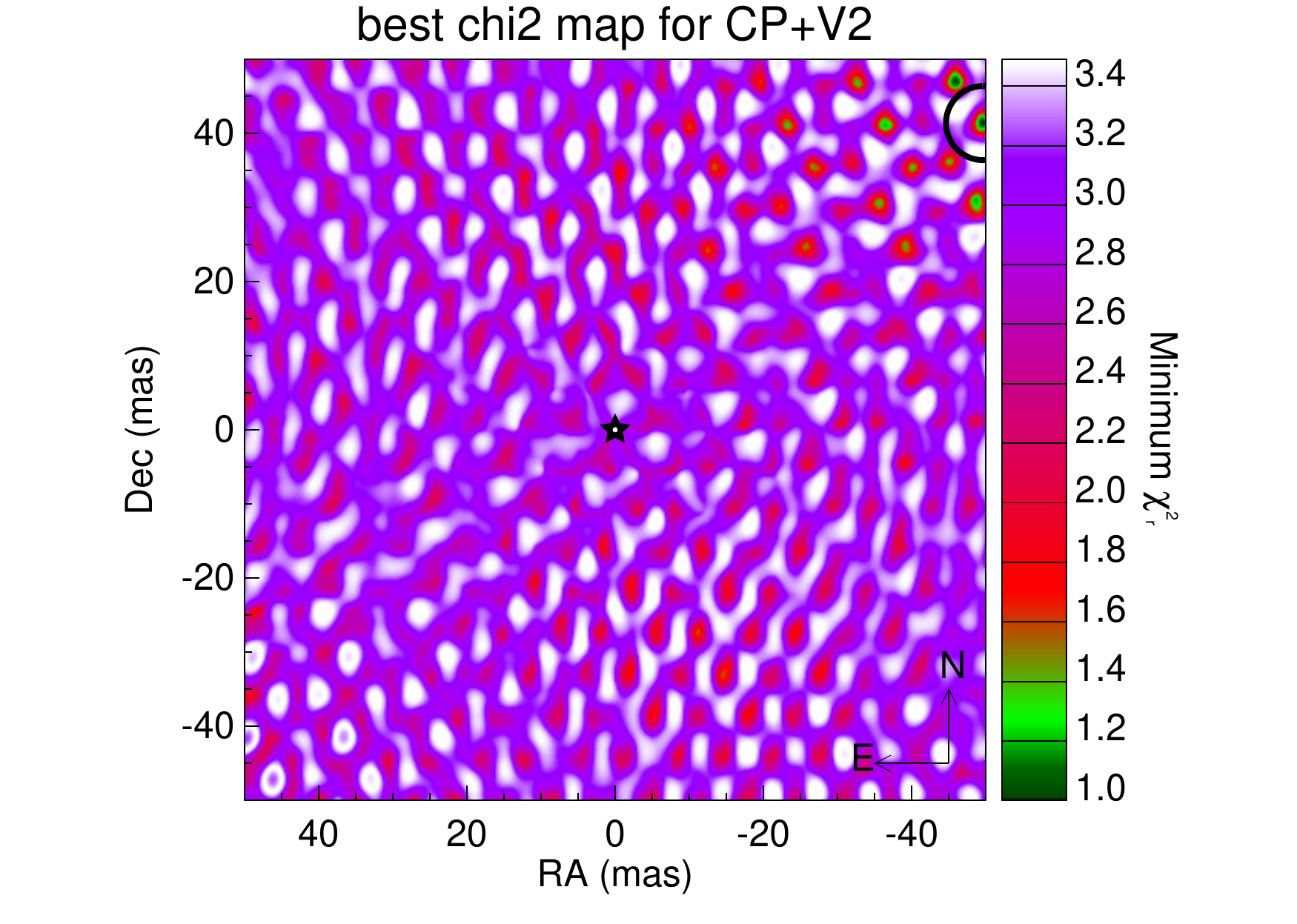}
\includegraphics[scale=0.5]{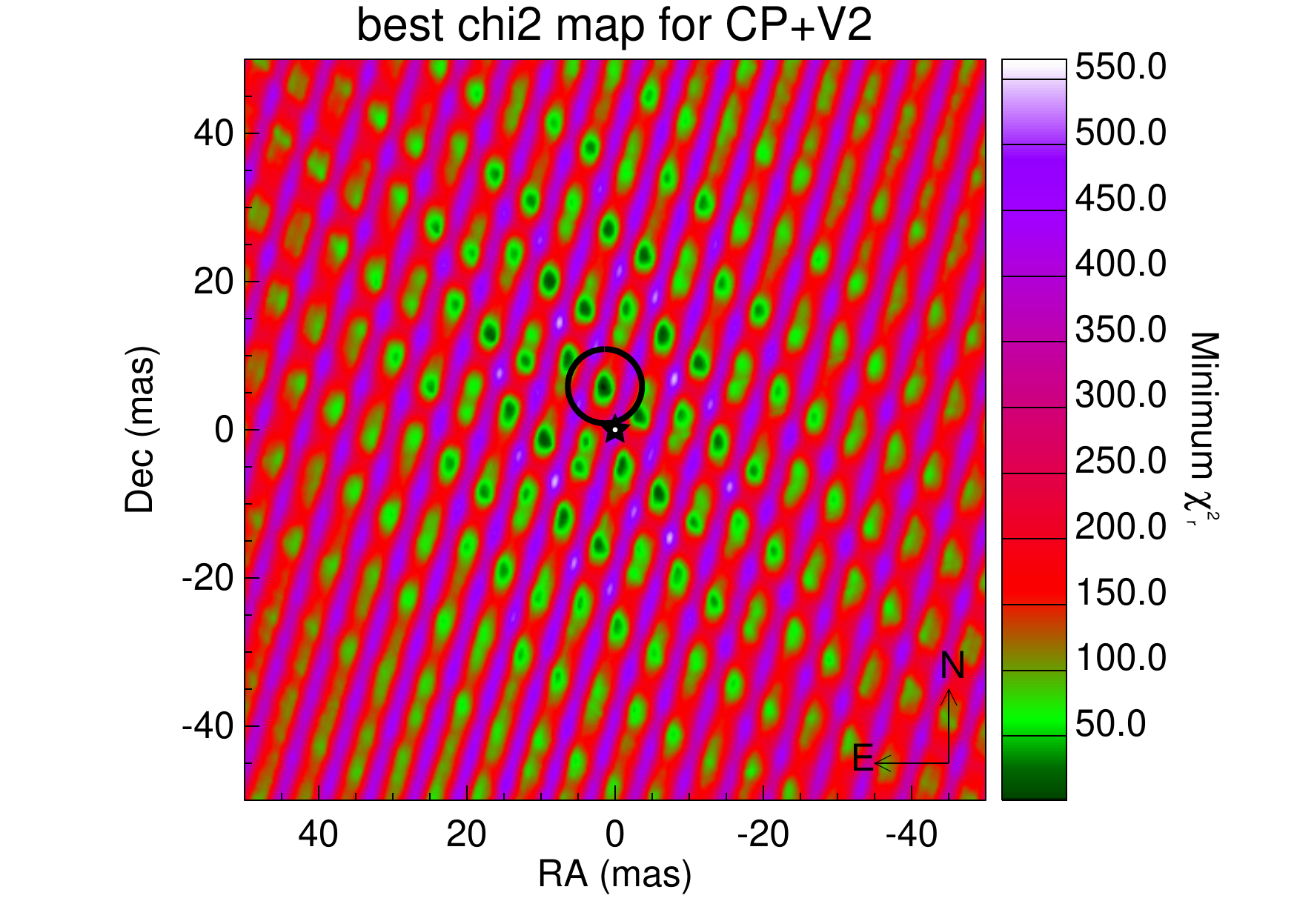}\\
\includegraphics[scale=0.5]{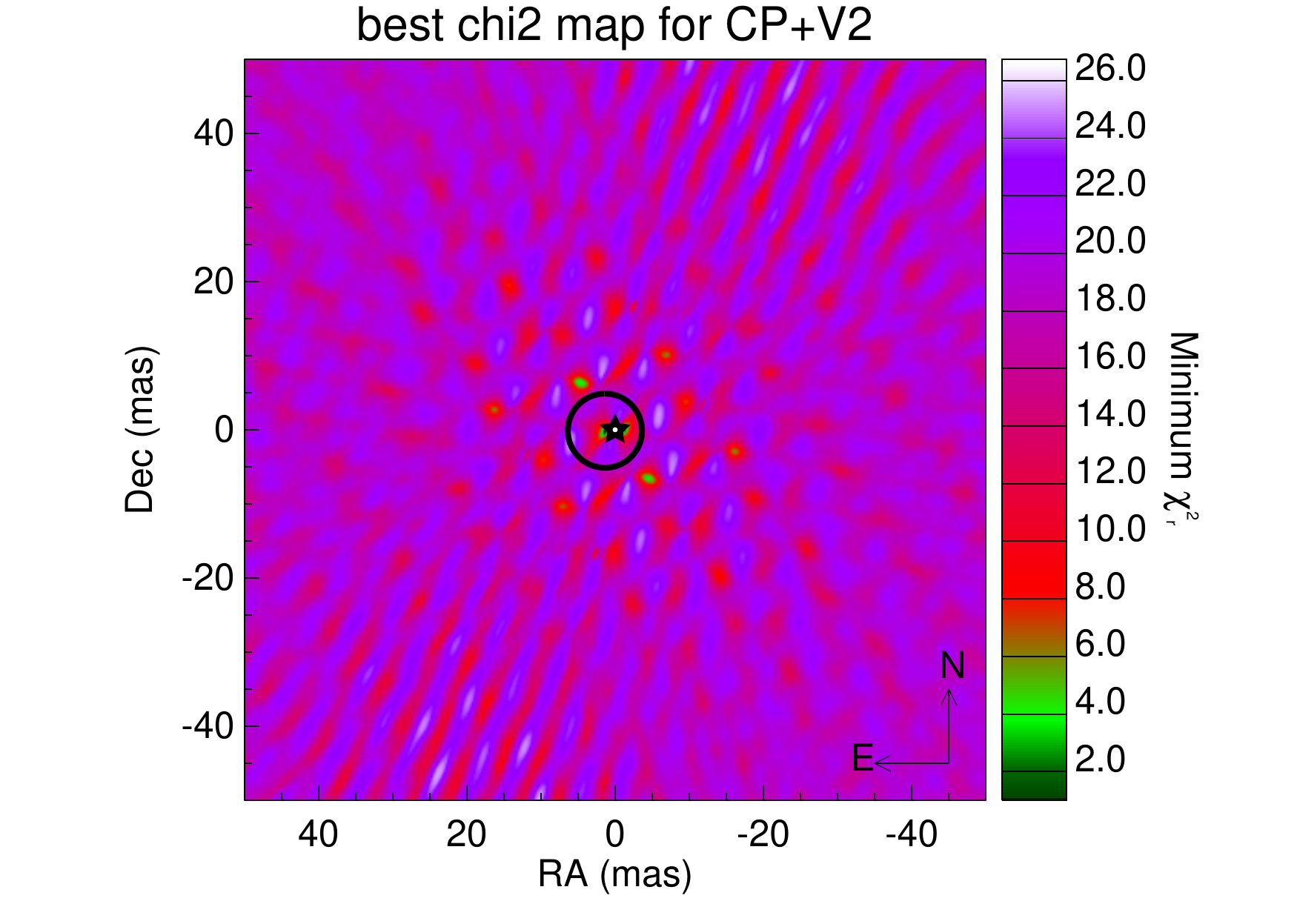}
\includegraphics[scale=0.5]{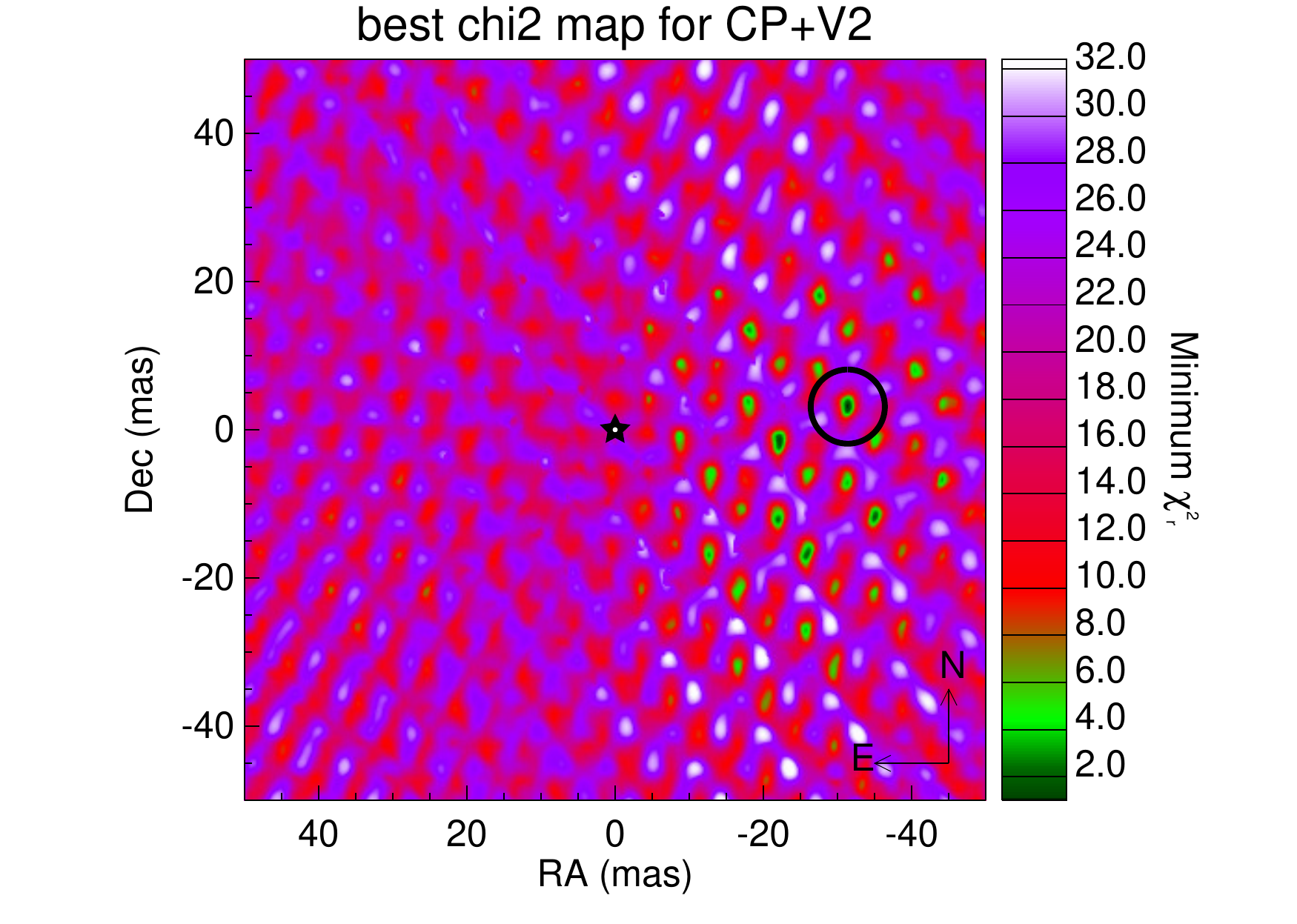}\\
\caption{Normalized $\chi^2$ maps related to the combined CP+$V^2$ analysis for the four stars showing clear signs of an off-axis companion: HD~31203, HD~80133, HD~106906, and HD~175073 (from left to right and top to bottom). The black circles indicate the positions of the minima in the maps.}
 \label{fig:chi2maps}
\end{figure*}

Table~\ref{tab:detcomp} lists the stars that have a significance level higher than $5\sigma$ for the combined $\chi^2$ analysis. HD~31203, HD~80133, HD~106906, and HD~175073 have strong detections, not only in the combined analysis but also in the individual analysis of $V^2$ and CP. They are therefore identified as bona fide binary stars. The $\chi^2$ maps illustrating the detection of a point-like source in these four data sets are illustrated in Fig.~\ref{fig:chi2maps}. HD~36187 and HD~61005 have a low significance for the CP-only analysis, so that the detected excess is identified as being due to extended emission (no evidence for the presence of a point source). For HD~203 and HD~76143, the situation is not as clear, as the detection is at best marginal in the combined and individual analyses. This requires more careful data inspection to decide on the nature of the excess. For HD~203, we note that the detection in the $V^2$ is about twice as significant as in the closure phases. This is a sign that the excess identified in the combined analysis is most probably due to the presence of a disk, which creates a strong signal in the $V^2$ but not in the CP. For HD~76143, looking at the closure phases signal reveals a global offset from 0, which is the sign of a poor calibration. This poor calibration of the CP is suspected to be at the origin of the (marginal) detection of an excess emission in the CP, and we propose that the most likely explanation is rather the presence of a circumstellar disk (although more observations would be needed to firmly confirm this statement). This leaves four stars in our sample with previously unknown companions, which will be removed from our search for hot exozodiacal disks. The four new binary stars are described in more details the following paragraphs. 

\subsection{Notes on newly identified companions}

\paragraph{HD~31203 (iot Pic A).} This F0V-type star is located at 37.1~pc, and known to be a member of a multiple star system \citep{tokovinin15}. The first companion (HD~31204, F4V) is located at $12\farcs3$, and the second one (HD~31261, K2V) at $289\arcsec$. These companions are well outside the PIONIER field-of-view, and even outside the AT field-of-view, so that they do not affect our observations. Besides being a visual multiple system, iot Pic A is known to have variable radial velocities \citep{nordstrom85}, with a variability larger than 30~km\,s$^{-1}$ on a timescale of a few days, based on four measurements. The nature and orbital parameters of the potential close companion are however not constrained, although the amplitude of the radial velocity (RV) variation point toward a solar-type companion.


Based on the measured contrast $c = 0.04$ and the distance, we estimate the companion found by our interferometric observations to have an absolute magnitude $M_H = 5.49$, which corresponds roughly to a K7V spectral type according to \citet{allen00}. Assuming a face-on, circular orbit with a semi-major axis of 2.39~au, and a mass of 1.52 $M_{\odot}$ for iot Pic A \citep{david15}, the orbital period would be around 3~years. Determining whether this companion corresponds (at least partly) to the source of the RV variability found by \citet{nordstrom85} would require more RV and interferometric observations. 

                         
\paragraph{HD~80133.} This K1V-type star is located at 32.8~pc. Based on the measured contrast $c = 0.85$ and the distance, we estimate the companion to have an absolute magnitude $M_H = 4.16$, which corresponds roughly to a G8V spectral type according to \citet{allen00}. In practice, the measured contrast would rather point to a pair of K1-2V stars, or to a slightly evolved K1IV-V primary with a less evolved G8V secondary, owing to their estimated age of about 13~Gyr \citep{takeda07}. Assuming a face-on, circular orbit with a semi-major axis of 0.19~au, and a mass of about $1 M_{\odot}$ for HD~80133 \citep{takeda07}, the period would be around 1~month. Surprisingly, this star has not been identified as a binary star based on RV measurements, while it was included in the California/Carnegie Planet Search programs \citep{valenti05,takeda07}. This might be explained either by a (very) poor time coverage in the RV survey, or by a quasi-perfectly face-on orbit. We note that the warm dust disk detected around HD~80133 by \citet{vican14} is located well outside the estimated semi-major axis of the companion (beyond 1~au), and should therefore be in a stable circumbinary configuration.

\paragraph{HD~106906.} This F5V-type star is located at 92.1~pc, and is identified as a short-period binary in \citet{lagrange21} based on RV measurements. The interferometric observations presented here confirm the binary nature of the star, which turns out to be a quasi-equal flux binary. Our interferometric observations have been included in the analysis of \citet{lagrange21} to better constrain the orbital parameters of the system. We refer to that paper for a full discussion of this system.

\paragraph{HD~175073.} This K1V-type star is located at 24~pc. Based on the measured contrast $c = 0.13$ and the distance, we estimate the companion to have an absolute magnitude $M_H = 6.36$, which corresponds roughly to an M2V spectral type according to \citet{allen00}. Assuming a face-on, circular orbit with a semi-major axis of 0.76~au, and a mass of 0.8 $M_{\odot}$ for HD~175073 \citep{casagrande11}, the period would be around 9~months. Surprisingly, this star has not been identified as a binary star based on RV measurements, while it was included in previous RV planet surveys according to \citet{grether06}. This might be explained either by a (very) poor time coverage in the RV survey, or by a quasi-perfectly face-on orbit. We note that this newly discovered companion cannot be at the origin of the W4 WISE excess detected by \citet{patel14} as their analysis was purely based on mid-infrared colors, which are similar for the host star and its companion. We also note that the warm dust disk detected around HD~175073 by \citet{patel14} is located well outside the estimated semi-major axis of the companion (beyond 2~au), and should therefore be in a stable circumbinary configuration.

\subsection{On the PIONIER sensitivity to faint companions}
\label{subsec:sens}

\begin{figure}[t]
\centering
\includegraphics[scale=0.45]{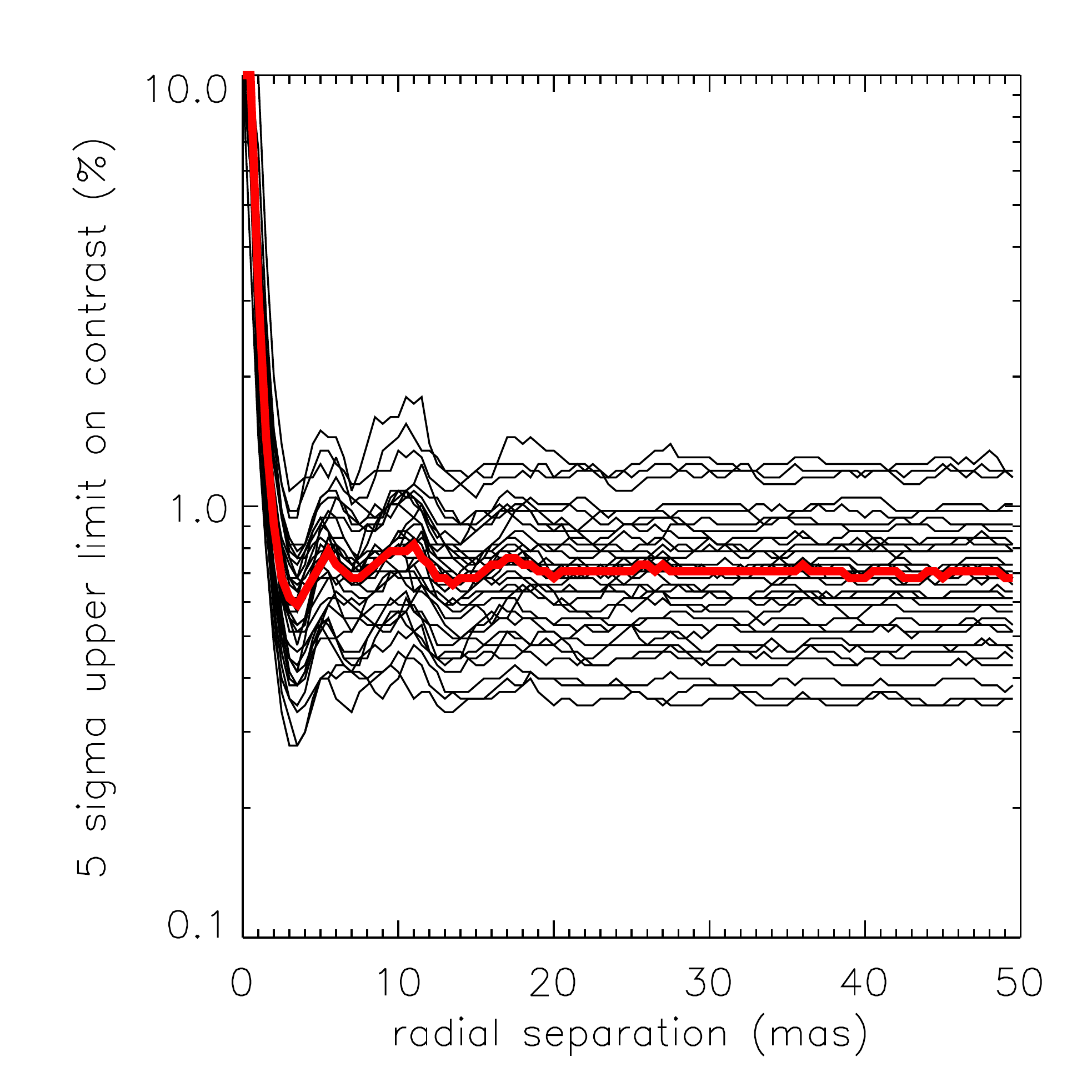}
\caption{Sensitivity of PIONIER to point-like companions as a function of the radial distance to the central star for the 35 stars showing no H-band excess in our observations. The sensitivity is expressed as the azimuthal median of the $5\sigma$ upper limit, based on an analysis of the combined $\chi^2$ for the CP and $V^2$. The red curve is the median sensitivity on the 35 stars.} \label{fig:senscont}
\end{figure}

In the cases where no companion, nor hot exozodiacal disk, is detected around the target stars (35 stars out of the 52 in our sample, see Sect.~\ref{sec:exozodi} for a discussion of the hot exozodi detections), we can compute an upper limit on the contrast of faint companions around the target stars, as a function of the position in the field-of-view. These sensitivity maps are derived from the $\chi^2$ analysis, as explained in \citet{absil11}, with the difference that here we use both the $V^2$ and the CP in our $\chi^2$ analysis. From the sensitivity maps, we can derive the median sensitivity at a given radial distance by computing the median upper limit along an annulus. The result is illustrated in Fig.~\ref{fig:senscont}, where the median sensitivity is plotted as a function of the angular separation for the 35 stars. The typical $5\sigma$ sensitivity of PIONIER in ``survey mode'' (3 OBs per target), illustrated by the red curve in Fig.~\ref{fig:senscont}, is a flux ratio of 0.7\% (i.e., $\Delta H=5.4$) for angular separations larger than 2~mas in the medium-sized AT configuration. This sensitivity corresponds typically to companions with spectral types ranging from M1V to M6V around main sequence stars with spectral types ranging from A0V to K0V.

\section{Search for exozodis}
\label{sec:exozodi}

After removing the four stars identified as binaries in Sect.~\ref{sec:comp}, we are left with a combined sample of 133 stars, as already described in Sect.~\ref{sec:stelsamp}. Of these 133 stars, 48 are new observations from the observing program described in this paper. In this section, we briefly summarize the principle of the search for hot exozodis, and detail the new exozodis found around our 48 new targets.

When it comes to the detection of faint, circumstellar excess emission, the strength of infrared interferometry is the ability to spatially resolve this emission and thus spatially disentangle it from the much brighter stellar emission.  When observing at small baselines of up to a few tens of meters, the host star is nearly unresolved (minimizing the effects of its uncertain diameter on the prediction of the system's $V^2$), while an extended circumstellar emission is ideally fully resolved \citep[see][]{difolco07}. This results in a drop in $V^2$ compared to the purely stellar $V^2$, because it adds incoherent flux. This represents the core of our detection strategy.

\subsection{Fitting strategy}
\label{subsec:strat}

As shown by previous studies \citep{absil09,defrere11}, the $V^2$ drop induced by a circumstellar disk does not depend significantly on the assumed geometry of the disk, provided that the disk is resolved at the considered baselines. As in previous studies, we therefore consider a model consisting of a limb-darkened photosphere surrounded by a uniform circumstellar emission filling the entire field of view of PIONIER ($\sim 400$~mas). The visibility expected from a limb-darkened photosphere is estimated according to \citet{hanbury74} using the linear H-band limb-darkening coefficients of \citet{claret95}. We estimate the visibility for the whole bandwidth of each spectral channel, considering the actual spectrum of the star using tabulated H-band spectra from \citet{pickles98} and the spectral transmission of the PIONIER instrument. The estimated $V^2$ are then compared with the measurements, and the flux ratio for each data set is derived. The computation is performed by a set of IDL routines initially developed for CHARA observations by \citet{absil06}, and later adapted to other interferometers by \citet{defrere11}. To derive the value and uncertainty of the flux ratio for each target, we use a bootstrapping algorithm, where each individual fit to the data is performed using a Levenberg-Marquardt least-squares minimization \citep{Markwardt09}. This means that the individual uncertainties on the data points are not considered directly in the estimate of the uncertainty of the flux ratio, but rather their scatter. In addition, a systematic uncertainty of $5\times 10^{-4}$ due to chromaticism is added to the estimated flux ratio \citep{ertel14}. For the bootstrapping, we consider that simultaneous spectral channels are fully correlated while the six baselines are fully uncorrelated.

\subsection{Results of the search}
\label{subsec:resdisk}

\begin{table}[!t]
\caption{Summary of the results for the 48 stars used for our new hot exozodiacal disk survey (excluding all binaries, and the data sets removed in Sect.~\ref{sec:obsandred}). Stars showing a significant level of excess emission (significance higher than 3$\sigma$) are highlighted in gray. The reduced $\chi^2$ of the star+disk model fit to the data is given in the last column.} \label{tab:all}
\centering
\begin{tabular}{cccc}
\hline\hline
Star & Contrast & Signif. & $\chi^2_r$ \\
 & (\%)  & ($\sigma$) & \\
\hline
\rowcolor[gray]{0.9} \object{HD 203}   & $0.96 \pm 0.23$ 	& 4.25 	& 0.84 \\
\object{HD 2834}  & $2.29 \pm 0.85$ 	& 2.69 	& 5.69 \\
\object{HD 3126}  & $-0.01 \pm 0.24$	& $-0.04$	& 0.63 \\
\rowcolor[gray]{0.9} \object{HD 4113}  & $0.75 \pm 0.20$ 	& 3.82 	& 0.63 \\
\object{HD 9672}  & $0.30 \pm 0.31$ 	& 0.96 	& 1.85 \\
\object{HD 10269} & $-0.16 \pm 0.14$	& $-1.15$	& 0.31 \\
\object{HD 10939} & $0.35 \pm 0.33$ 	& 1.05 	& 1.11 \\
\object{HD 15427} & $0.05 \pm 0.18$ 	& 0.28 	& 0.46 \\
\rowcolor[gray]{0.9} \object{HD 17848} & $1.21 \pm 0.14$ 	& 8.69 	& 0.37 \\
\object{HD 23484} & $0.69 \pm 0.25$ 	& 2.71 	& 0.74 \\
\rowcolor[gray]{0.9} \object{HD 24649} & $1.14 \pm 0.23$ 	& 5.05 	& 0.80 \\
\object{HD 28287} & $0.06 \pm 0.32$ 	& 0.19 	& 1.36 \\
\object{HD 29137} & $0.23 \pm 0.16$ 	& 1.45 	& 0.66 \\
\rowcolor[gray]{0.9} \object{HD 36187$^{a}$} 	& $1.95 \pm 0.13$ 	& 15.00 & 0.40 \\
\object{HD 37306} 	& $0.26 \pm 0.15$ 	& 1.75 	& 0.49 \\
\object{HD 37484} 	& $0.28 \pm 0.21$ 	& 1.36 	& 0.84 \\
\object{HD 38949} 	& $-0.09 \pm 0.18$	& $-0.51$ & 0.71 \\
\object{HD 41278} 	& $0.40 \pm 0.22$ 	& 1.85 	& 0.68 \\
\object{HD 44524} 	& $0.01 \pm 0.20$ 	& 0.05 	& 0.66 \\
\object{HD 60491} 	& $0.44 \pm 0.16$ 	& 2.78 	& 0.61 \\
\rowcolor[gray]{0.9} \object{HD 61005} 	& $0.81 \pm 0.12$ 	& 6.70 	& 0.36 \\
\object{HD 71722} 	& $0.37 \pm 0.17$ 	& 2.21 	& 0.57 \\
\rowcolor[gray]{0.9} \object{HD 76143} 	& $0.60 \pm 0.18$ 	& 3.33 	& 0.47 \\
\rowcolor[gray]{0.9} \object{HD 80883} 	& $1.43 \pm 0.18$ 	& 8.07 	& 0.65 \\
\rowcolor[gray]{0.9} \object{HD 89886$^{a}$} 	& $0.92 \pm 0.26$ 	& 3.47 	& 1.02 \\
\rowcolor[gray]{0.9} \object{HD 90781} 	& $0.63 \pm 0.16$ 	& 3.98 	& 0.41 \\
\object{HD 90874} 	& $0.34 \pm 0.13$ 	& 2.62 	& 0.36 \\
\object{HD 92945} 	& $0.00 \pm 0.18$ 	& 0.00 	& 0.51 \\
\object{HD 105850} 	& $-0.06 \pm 0.18$	& $-0.34$	& 0.56 \\
\object{HD 105912} 	& $0.31 \pm 0.15$ 	& 2.08 	& 0.45 \\
\object{HD 109573} 	& $0.35 \pm 0.15$ 	& 2.35 	& 0.39 \\
\rowcolor[gray]{0.9} \object{HD 109704} 	& $0.85 \pm 0.12$ 	& 7.03 	& 0.34 \\
\object{HD 112603} 	& $0.41 \pm 0.25$ 	& 1.61 	& 0.84 \\
\object{HD 117716} 	& $0.40 \pm 0.18$ 	& 2.26 	& 0.44 \\
\object{HD 118972} 	& $0.16 \pm 0.09$ 	& 1.70 	& 0.16 \\
\rowcolor[gray]{0.9} \object{HD 136544} 	& $1.43 \pm 0.35$ 	& 4.04 	& 1.16 \\
\object{HD 141943} 	& $-0.15 \pm 0.19$	& $-0.76$	& 0.51 \\
\object{HD 161612} 	& $-0.30 \pm 0.12$	& $-2.48$	& 0.30 \\
\object{HD 174474} 	& $-0.12 \pm 0.20$	& $-0.61$	& 0.75 \\
\object{HD 179520} 	& $0.44 \pm 0.24$ 	& 1.79 	& 0.76 \\
\rowcolor[gray]{0.9} \object{HD 181327} 	& $0.48 \pm 0.16$ 	& 3.03 	& 0.45 \\
\object{HD 185615} 	& $-0.61 \pm 0.31$	& $-1.94$	& 0.92 \\
\object{HD 191089} 	& $0.33 \pm 0.48$ 	& 0.68 	& 2.29 \\
\object{HD 192758} 	& $0.12 \pm 0.28$ 	& 0.42 	& 0.89 \\
\object{HD 196141} 	& $0.39 \pm 0.20$ 	& 1.99 	& 0.58 \\
\object{HD 205674} 	& $0.43 \pm 0.50$ 	& 0.86 	& 2.52 \\
\object{HD 220476} 	& $0.04 \pm 0.18$ 	& 0.23 	& 0.49 \\
\object{HD 224228} 	& $0.27 \pm 0.39$ 	& 0.69 	& 0.98 \\
\hline
\end{tabular}
\tablefoot{($^{a}$) not used in the statistical analysis due to insufficient mid- to far-infrared data.}
\end{table}

\begin{figure*}[t]
\begin{center}
\includegraphics[scale=0.45]{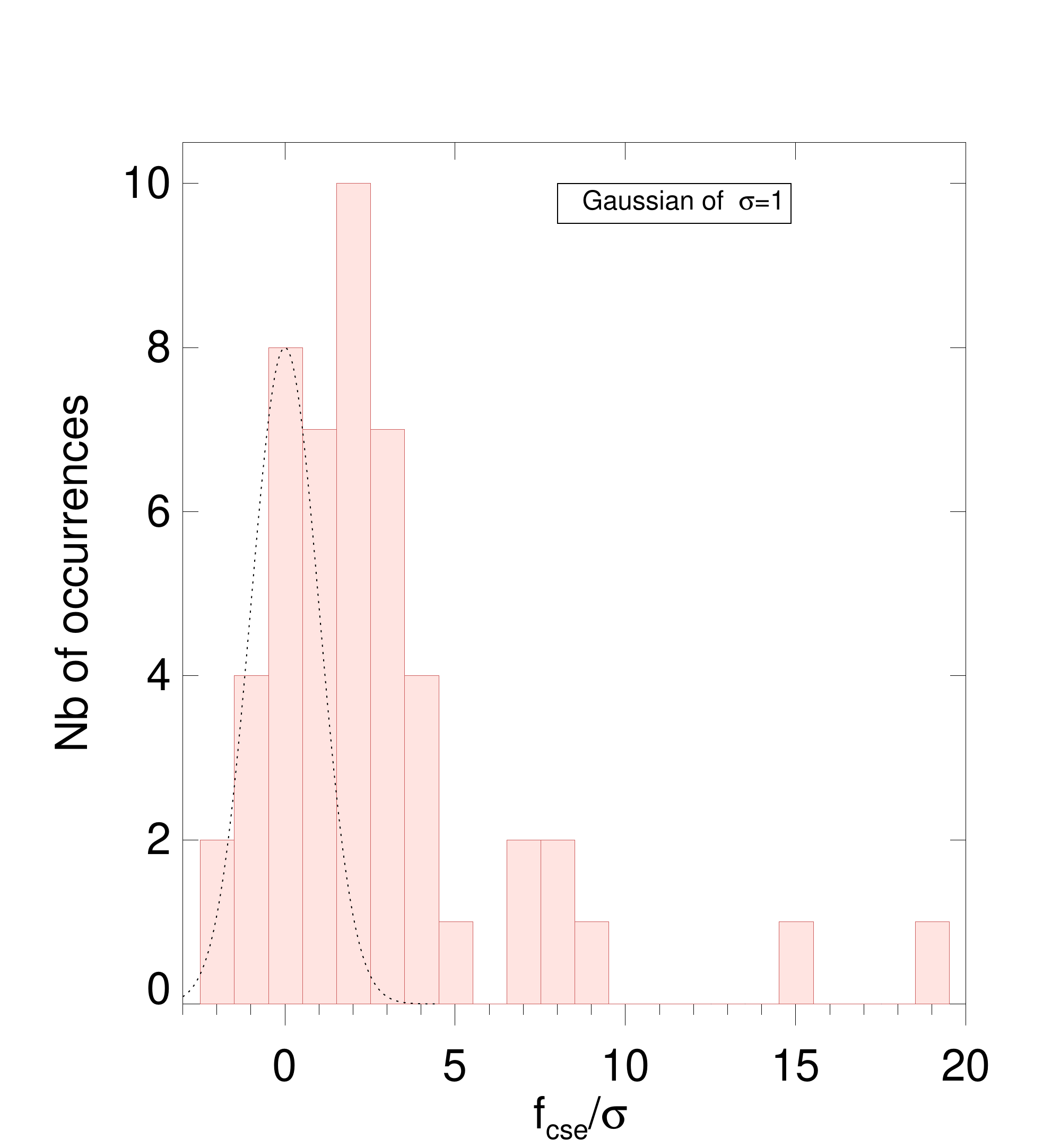}
\includegraphics[scale=0.45]{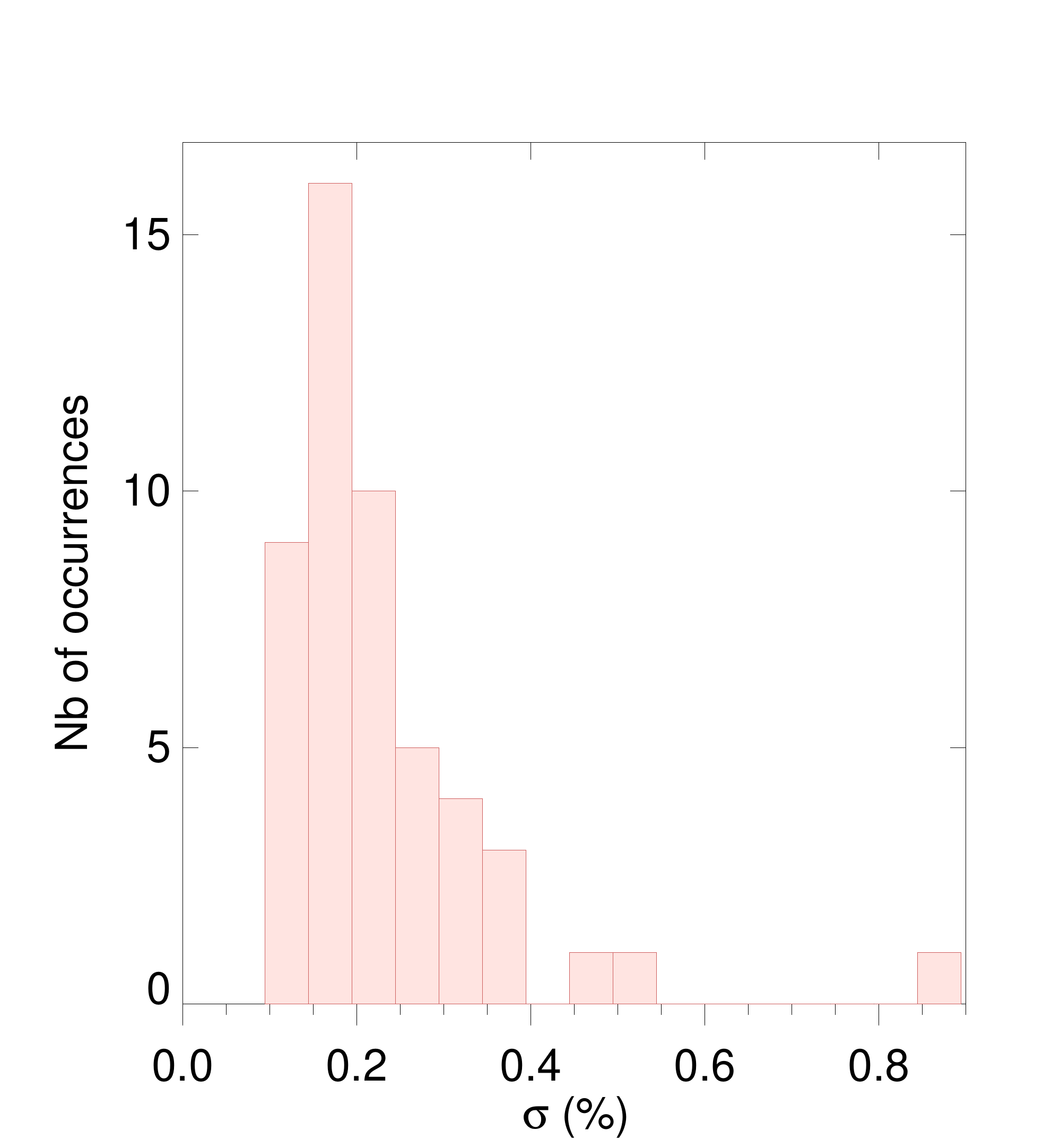}\\
\caption{Distribution of the excess significance level (left) and of the uncertainties on the disk-to-star flux ratio (right) for the observed sample. The Gaussian fit to the negative part of the significance distribution is represented by a dotted line.}
\label{fig:histoexo}
\end{center}
\end{figure*}

Table~\ref{tab:all} presents the results of the fit in terms of disk/star flux ratio, for the 48 new targets observed here. The measured flux ratio is averaged over the three or six spectral channels in our PIONIER observations. To define an appropriate detection threshold, we study the distribution of the significance level $\chi_f$, defined as the ratio between the measured disk/star flux ratio and the uncertainty on this quantity. Figure~\ref{fig:histoexo} shows the histogram of the significance level for our sample of 48 stars. We decided to use a threshold of $3~\sigma$ for the detection as in the study of \citet{ertel14}. Since we used the same methods for observation and data reduction on the same instrument, we can assume that the distribution of the uncertainties will also be comparable. As an additional argument, we study the negative part of the distribution of the significance level. The standard deviation of the negative part, after mirroring it on the positive side is found to be equal to 1.2. In Fig.~\ref{fig:histoexo}, a Gaussian distribution with this standard deviation is over-plotted on the data to guide the eye. The good match in shape, and the width of the distribution, confirm that a 3$\sigma$ criterion is appropriate, which corresponds to a false alarm probability of 0.27\%, and should therefore avoid spurious detections in our sample. 

\begin{figure*}[p]
\centering
\includegraphics[scale=0.80]{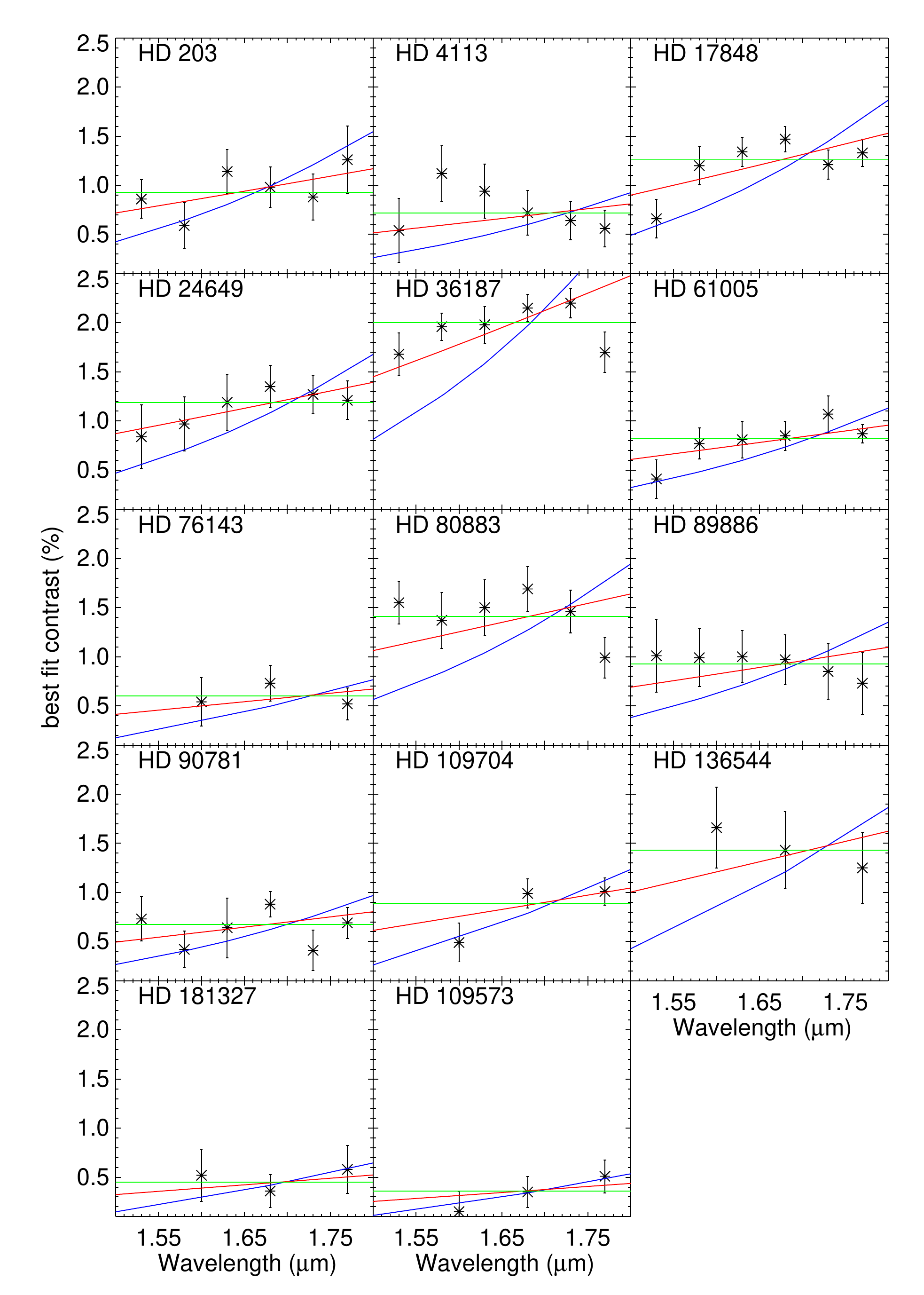}
\caption{Disk/star flux ratio as a function of wavelength for the 13 targets showing a significant H-band excess in our observations, as well as for HD~109573 (HR~4796), which shows a significant excess only in the reddest spectral channel. The blue, red, and green curves show the best fit to these measured flux ratios using blackbodies at 1000~K, 2000~K, and at the star's temperature (constant flux ratio), respectively.}
\label{fig:v2fits1}
\end{figure*}

The 13 stars highlighted in gray in Table~\ref{tab:all} have a significance level above $3\sigma$, and are therefore classified as having a near-infrared circumstellar excess associated with the presence of circumstellar emission. In Fig.~\ref{fig:v2fits1}, we show the wavelength dependence of the measured flux ratio for the 13 stars showing a significant near-infrared excess. The large number of stars with significance levels in the 1$\sigma$-3$\sigma$ range suggests that there may be a population of excesses just below the detection threshold, which remain undetected in our study. Comparing the negative and positive parts of the histogram, we estimate that an additional dozen stars could have an undetected H-band excess in the 1$\sigma$-3$\sigma$ range.

\subsection{Notes on specific hot-exozodi targets}




\paragraph{HD~4113.} This old G5V star is known to have a planetary companion, discovered by RV measurements \citep{Tamuz08}, as well as a directly imaged brown dwarf companion at a projected separation of 22~au \citep{Cheetham18}. The properties of the inner, planetary companion were revisited by \citet{Cheetham18}: $M \sin i=1.602$~M$_{\rm Jup}$, $a=1.298$~au, $e=0.8999$. While this star was originally classified as surrounded by a warm dust disk by \citet{vican14}, this classification was based on a single WISE photometric data point at 22~$\mu$m. Our revised SED analysis does not show the presence of any significant dust population around this star, based on WISE and AKARI photometry. The presence of a hot dust population therefore does not seem to be connected to a massive outer reservoir of larger bodies. It is interesting to assess whether its inner, eccentric giant planet may have a direct influence on the architecture of the hot dust population. We assume that most of the hot dust is located at the sublimation distance of silicate grains (sublimation temperature of 1500~K), i.e., a distance of 0.04~au. Based on the orbital elements of the planets, the periastron is at 0.12~au, while the apastron is at 2.42~au. This orbital configuration suggests that the planet could have a direct influence on the architecture of the dust disk. To our knowledge, this is only the second hot dust system with a well-characterized inner giant planet (the first one being $\beta$~Pictoris). This configuration could be used to constrain the origin of the hot dust. Due to the presence of the planet, Poynting-Robertson (P-R) drag acting on dust grains from a hypothetical (warm or cold) outer dust belt that would remain below the detection threshold is unlikely to efficiently replenish the hot disk, as already suggested in a more general case by \citet{vanLieshout2014} and by \citet{Bonsor18} based on numerical simulations. A scenario where planetesimals belonging to an outer reservoir would be destabilized by the RV planet and sent towards the inner, hot regions where they would sublimate seems like a more plausible hot dust production scenario in this case, in a process somewhat akin to the falling evaporating bodies scenario proposed for the $\beta$~Pictoris system \citep{Beust90} -- although such a massive planet would be better at ejecting planetesimals than scattering them inwards \citep[e.g.,][]{Wyatt17}.


\paragraph{HD~20794.} The near-infrared excess detected around this nearby Solar-type star was already reported in \citet{ertel14}. This star is known to be the host of at least three (maybe four) super-Earth planets, orbiting between 0.1 and 1 au from the star \citep{Feng2017}, and may also host a massive giant planet at a separation between 2 and 10 au based on Gaia proper motion analysis \citep{Kervella19}. Based on our SED analysis, and on the detailed model described in \citet{kennedy2015}, we do not confirm the presence of a warm dust population as suggested by \citet{cotten16}, and classify this target as a cold disk system with a temperature of 80~K. It is another case where a planetary system is located between the hot, inner disk and the outer debris disk, as discussed in detail by \citet{kennedy2015}. Even only considering the RV planets orbiting around HD~20794, which are much less massive than the Jupiter-sized companion of HD~4113, the planetary system is still expected to largely prevent dust from replenishing the inner disk through P-R drag. This is another indication that P-R drag is probably not at the origin of -- or at least not the only contributor to -- the detected near-infrared excess.
 

\paragraph{HD~61005 and HD~181327.} These two stars are known to be surrounded by copious amounts of dust, and show asymmetries in their outer debris disk, which might be due to collision of Pluto-like objects. HD~61005 is also found by our SED analysis to include a warm dust population, at a black-body temperature of $\sim$120~K. Based on near-infrared scattered light observations, \citet{olofsson16} and \citet{esposito16} show that the eastern side of HD~61005 disk is brighter than the western side. \citet{olofsson16} argue that an observed peak of density at the pericenter of the disk may be the signpost of a recent impact, since the material released by the impact would pass again through the initial collision point, creating more collision and thus enhancing the density. HD~181327, a member of the $\beta$~Pic moving group ($\sim$20~Myr), also shows an asymmetry in its outer disk, which may be caused by a recent massive collisional event or by interactions with the interstellar medium \citep{stark14}. The possible collisional activity in the outer part of these two debris disks could be related to a major dynamical instability akin to the Large Heavy Bombardment in our Solar system. In such an event, we would expect planetesimals to be injected in the inner parts of the planetary system, where they may create the hot dust detected in our observations. 

\paragraph{HD~109573 (HR~4796).} This A0-type member of the TW Hya association ($\sim$10~Myr) does not show a significant H-band excess when considering the three PIONIER spectral channels together. However, looking at the spectral channels separately shows a strong slope of the excess emission, increasing with wavelength to a level such that the longest channel has an excess of $0.51\% \pm 0.17\%$, significant at the $3\sigma$ level (see Fig.~\ref{fig:v2fits1}). This may correspond to the onset of thermal emission of a hot exozodiacal disk at a temperature around 1000~K. Although tentative, this possible H-band excess is interesting to put in perspective with the global debris disk architecture. According to \citet{chen14}, the debris disk can be best represented by a two-temperature black body model, with the innermost ring at a temperature of 231~K (i.e., at about 5.7~au from the star). We note however that the presence of warm dust in this system is disputed. Indeed, \citet{wahhaj05} found evidence of warm dust based on mid-infrared imaging, but \citet{kennedy14} proposed that the emission of HR~4796 is compatible with a single black body. A single black body is also suggested by our SED analysis, with a temperature of 97~K, which results in a cold dust classification in our statistical sample. More recently, \citet{Lisse2017} suggested the presence of a tenuous thermal emission component from close-in, $\sim$850~K circumstellar material based on near- to mid-infrared spectroscopy, which might be directly connected to the small H-band excess detected in our PIONIER data. 

Near-infrared high-contrast imaging shows that the outer belt around HR~4796 consists of a sharp, offset ring of dust \citep[e.g.,][]{Milli17,Chen20}, with an angular separation from the star as small as $\sim 200$~mas along the semi-minor axis due to projection effects. The detected H-band excess may therefore also be (partly) due to the contribution of scattered light from the outer debris disk. Based on the surface brightness of the disk extracted by \citet{Milli17}, and considering the off-axis transmission of the PIONIER single-mode fibers, we estimate that the outer disk could contribute up to 0.1\% in terms of integrated H-band excess. The outer disk alone would therefore most probably not explain the measured excess of $\sim0.5\%$. Another piece of evidence for that is the tentative slope in the measured excess, which would not be consistent with scattered light.

The morphology of the HR~4796 outer disk can be best explained through the influence of an eccentric planetary companion that would clear the interior region of the cold dust belt \citep{Lagrange12}. Both \citet{Perrin15} and \citet{Milli17} suggest that the main contribution to scattered light in the outer dust ring comes from rather large, porous grains. This points towards a low dynamical excitation in the outer disk \citep{Lisse2017}, which seems at odds with the main scenarios proposed to explain the presence of a hot exozodiacal disk. Once again, the hot dust population seems disconnected from the cold dust reservoir, but before further investigating the global disk architecture, follow-up observations with near-infrared interferometry will be needed to confirm the tentative H-band excess.

\section{Discussion} \label{sec:discussion}

\begin{table*}[t]
\caption{Hot exozodiacal disk statistics from the combined PIONIER surveys of \citet{ertel14} and this work. Columns ``\#S'' and ``\#E'' represent the number of target stars and of hot exozodi detections, respectively.}
\label{tab:statdata}
\centering
\begin{tabular}{ccccccccccccccc}
\hline
\hline
 & \multicolumn{3}{c}{A-type stars}  & \multicolumn{3}{c}{F-type stars}  & \multicolumn{3}{c}{G/K-type stars}  & \multicolumn{3}{c}{Total}\\
 & \#S & \#E & detect.~rate & \#S & \#E & detect.~rate & \#S & \#E & detect.~rate & \#S & \#E & detect.~rate \\ [+4pt]
\hline 
All$^{a}$ & 40 & 7 & $17.5^{+7.5}_{-4.4} \%$ & 51 & 10 &$19.6^{+6.7}_{-4.4} \%$ & 42 & 5 & $11.9^{+6.8}_{-3.3} \%$ & 133 & 22 &$16.5^{+3.7}_{-2.7} \%$\\ [+4pt]
Warm dust  & 18 & 3 & $16.7^{+12.1}_{-5.3} \%$ & 11 & 2 & $18.2^{+16.3}_{-6.4} \%$ & 6 & 1 & $16.7^{+23.2}_{-6.3} \%$ & 35 & 6 & $17.1^{+8.1}_{-4.6} \%$\\ [+4pt]
Warm dust only & 12 & 2 & $16.7^{+15.5}_{-5.8} \%$ & 7 & 2 & $28.6^{+20.3}_{-10.6} \%$ & 5 & 0 & $0.0^{+26.3}_{-0.0} \%$ & 24 & 4 & $16.7^{+10.1}_{-5.0} \%$\\ [+4pt]
Cold dust only  & 5 & 2 & $40.0^{+21.4}_{-15.6} \%$ & 12 & 2 & $16.6^{+15.5}_{-5.8} \%$ & 14 & 1 & $7.1^{+13.2}_{-2.3} \%$ & 31 & 5 & $16.1^{+8.6}_{-4.5} \%$\\ [+4pt]
No warm dust & 21 & 3 & $14.3^{+10.8}_{-4.6} \%$ & 39 & 7 & $17.9^{+7.7}_{-4.5} \%$ & 36 & 4 & $11.1^{+7.4}_{-3.2} \%$ & 96 & 14 & $14.6^{+4.3}_{-2.8} \%$\\ [+4pt]
No dust  & 16 & 1 & $6.3^{+11.8}_{-2.0} \%$ & 27 & 5 & $18.5^{+9.6}_{-5.2} \%$ & 22 & 3 & $13.6^{+10.4}_{-4.3} \%$ & 65 & 9 & $13.8^{+5.4}_{-3.2} \%$ \\ [+4pt]
\hline
\end{tabular}
\tablefoot{($^{a}$) also includes two dusty stars that have no warm/cold classification: HD~36187 (A0V) and HD~89886 (F7V).}
\end{table*}

\begin{figure}[t]
\centering
\includegraphics[width=0.5\textwidth]{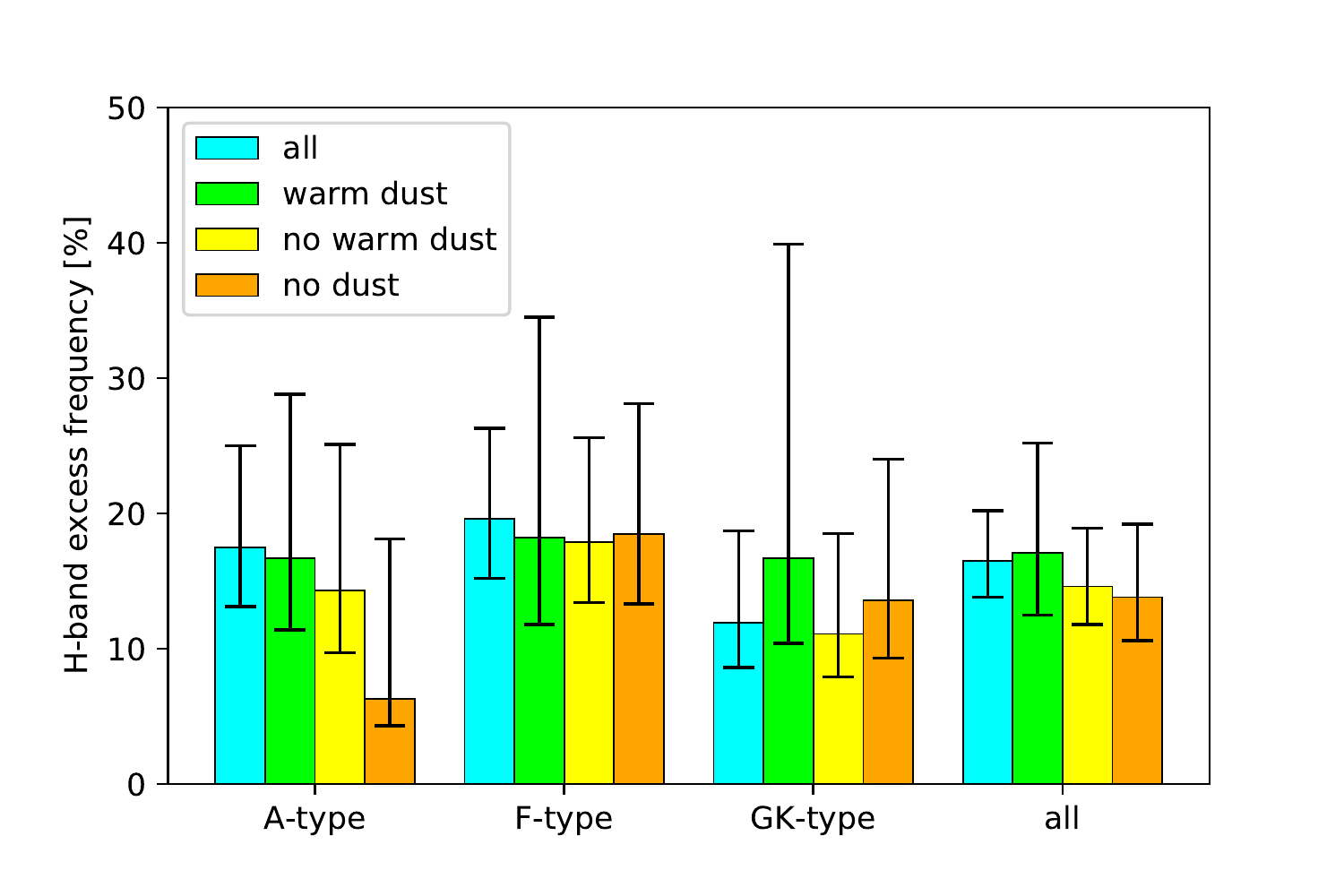}
\caption{Detection rate of hot exozodiacal dust as a function of spectral type, and as a function of the presence of a known warm dust reservoir. No significant difference in detection rate is found between the various populations.}
 \label{fig:histodust}
\end{figure}

In Table~\ref{tab:statdata}, we summarize the results of the PIONIER surveys for hot exozodis presented here and in \citet{ertel14}, in terms of number of detections and detection rates. The results are separated as a function of spectral type, and as a function of the presence of detectable amounts of warm and/or cold dust populations. A graphical representation of the most important information of this table is shown in Fig.~\ref{fig:histodust}, and forms the basis of the discussion in the next paragraphs.

\subsection{Correlation between hot and warm dust} \label{sub:temp_corr}

In our warm dust sample, we measure a detection rate of $17.1^{+8.1}_{-4.6}$\% for H-band excesses, while the control sample with no warm dust shows a detection rate of $14.6^{+4.3}_{-2.8} \%$. These two occurrence rates are well within the error bars of each other, and we note that choosing any temperature threshold in the 100--200~K range to classify warm against cold dust populations would not change this conclusion. In order to confirm that this result is compatible with the two samples being drawn from the same population, we perform a two-sample Anderson-Darling test, which tests the null hypothesis that two samples are drawn from the same population, without having to specify the distribution function of that population. Here, the two samples are defined as the collection of the H-band excess levels in the warm dust and control samples, regardless of their spectral type. The two-sample test returns a p-value of 0.13, which confirms that the null hypothesis cannot be rejected. Performing the Anderson-Darling test on the significance of the H-band excess instead of the excess level itself, to account for the specific sensitivity level reached on each star, does not change the conclusion, with a p-value of 0.22. This is consistent with the study of \citet{mennesson14}, who used the mid-infrared Keck Interferometer Nuller (KIN) to search for warm dust around 11 stars already known to host hot excesses from near-infrared interferometric observations (among a total KIN sample of 40 stars), and did not find a significant correlation between the presence of hot and warm dust. The same conclusion was reached by the recent analysis of the Large Binocular Telescope Interferometer (LBTI) HOSTS survey \citep{ertel18,ertel20}, based on a sample of 38 stars. This lack of correlation is understood as the telltale sign of a disconnection between hot and warm dust populations, which would then not be created by the same parent bodies. Here, with our much larger sample (131 stars), we confirm that the detection rate of hot dust is not significantly enhanced by the presence of a warm asteroid belt. While it cannot be excluded that warm asteroid belts act as prominent suppliers of material to replenish the short-lived hot dust population, the presence of large amounts of warm dust does not seem to be a pre-requisite to the presence of hot exozodiacal dust, and we confirm that hot dust should not be considered as the bright, near-infrared counterpart of warm belts (or at least not in a directly connected way). This conclusion is of course only valid at the sensitivity level of the instruments used in this study (both regarding near-infrared interferometry and mid- to far-infrared spectrophotometry), and may be challenged by future, more sensitive observations.



\begin{figure}[t]
\centering
\includegraphics[width=0.5\textwidth]{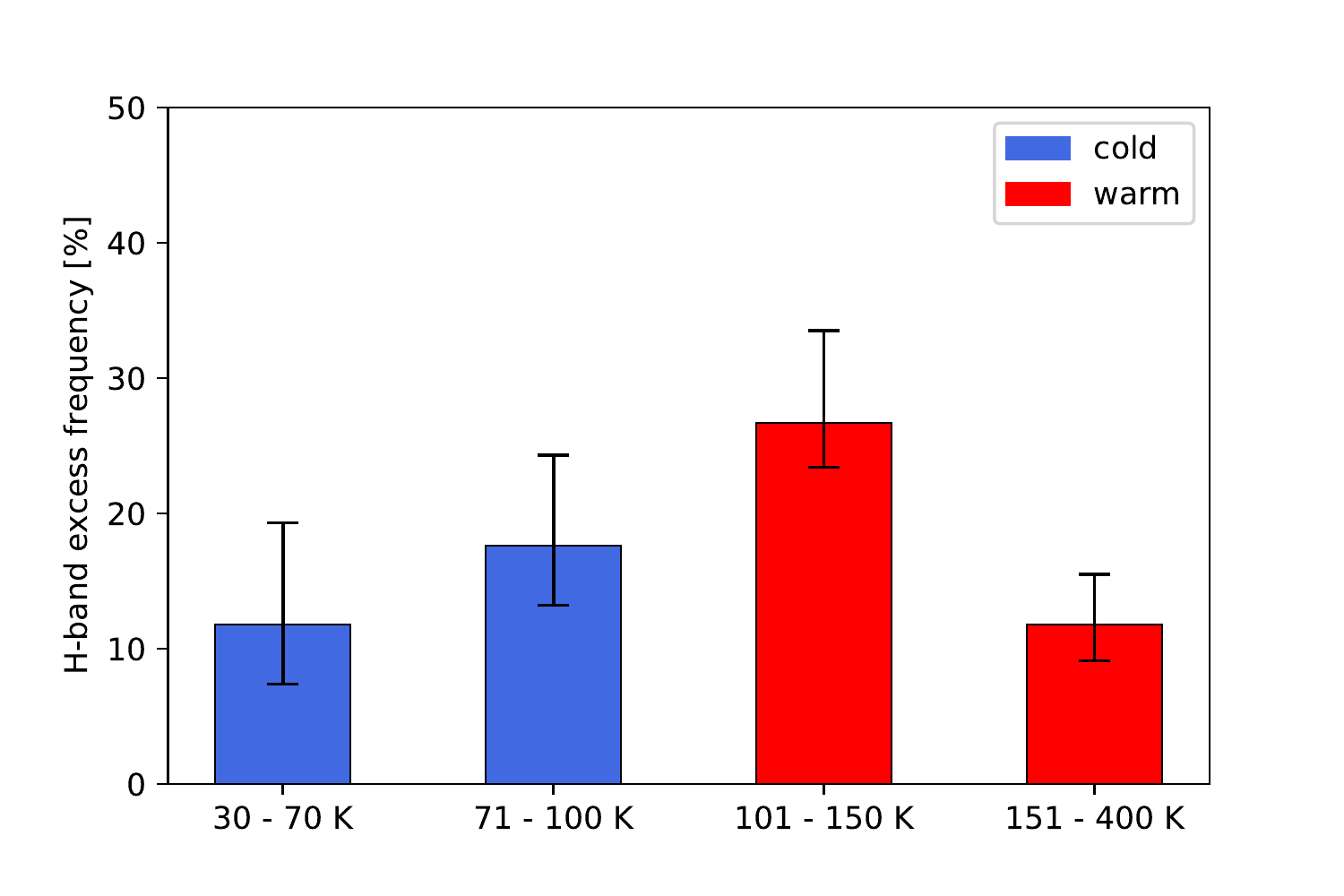}
\caption{Occurrence rate of H-band excesses for the 66 stars hosting a known dust reservoir, as a function of the estimated temperature of the dust.}
 \label{fig:tmpstat}
\end{figure}

To refine our analysis, we also investigate the possible correlation between the temperature of the outer dust reservoirs and the detection of an H-band excess. If there is a direct connection between inner and outer dust disks, we may expect that the warmer the outer disk, the higher the chances will be to detect an H-band excess. However, Fig.~\ref{fig:tmpstat} indicates a lack of correlation between the temperature of the outer dust reservoir and the occurrence rate of H-band excesses -- the apparent drop in occurrence rate for the warmest dust reservoirs being non-significant. A more relevant way of making this analysis may be to use the expected warm belt location rather than its temperature. Since the dust temperature is a good proxy for its location, when taking into account the black-body correction of \citet{pawellek15} as discussed in Sect.~\ref{sub:warmcold}, this does not change the conclusion. A last possible correlation that we investigated is between the luminosity of the warm debris disk (as a proxy for its mass) and the H-band excess. The inward flux of dust due to P-R drag is indeed expected to scale (albeit weakly) with the mass of the warm dust disk \citep{kennedy15}. No correlation was found here either, which seems to concur with the conclusions of \citet{Sezestre19} that P-R drag is unlikely to be at the origin of the hot exozodi phenomenon, although we recognize that fractional luminosities may not be directly proportional to the dust mass. Finally, our new results also make it possible to revisit the conclusion of \citet{ertel14} that the presence of hot dust does not correlate with the presence of cold dust. We confirm this conclusion by comparing the detection rate of H-band excesses around the ``cold dust only'' (no warm dust) and ``no dust'' samples (see Table~\ref{tab:statdata}), and find them to be fully compatible within the statistical uncertainties.


\subsection{Occurrence rate vs.\ stellar parameters}

Previous studies suggested that hot exozodiacal dust is more frequent around early-type stars than solar-type stars, although no firm conclusion could be drawn due to the limited sample \citep{Absil13,ertel14}. Here, this correlation does not appear as obvious any more, with A-type stars showing a similar detection rate ($17.5^{+7.5}_{-4.4} \%$) as FGK-type stars ($16.1^{+4.5}_{-3.1} \%$) in the combined sample. This result may seem to contradict the prediction of the magnetic trapping model, which is shown by \citet{rieke16} to be more efficient around rapidly rotating stars. Although a measurement of $v \sin i$ is not available for all of the stars of our sample, we consider that stars with spectral type earlier than F5 have a much higher chance of showing high rotational velocities, due to the absence of a strong convective layer to brake their initial rotation. A two-sample Anderson-Darling test comparing the H-band excess levels of hot exozodis around stars earlier and later than F5 shows a 2.5\% probability for them to be drawn from the same population, which suggests, at a $2.2\sigma$ level, that the two samples are drawn from different populations. The same Anderson-Darling test, performed on the excess significance instead of the excess levels, shows a 5.3\% probability for them to be drawn from the same population, a marginal evidence at best. 

We found in the previous section that there is a lack of correlation between the presence of inner (hot) dust and outer (warm/cold) dust in our combined sample of 133 stars. It is interesting to investigate whether this lack of correlation holds when looking separately at different spectral types. To do so, we perform the same two-sample Anderson-Darling test as before, considering separately early-type stars and solar-type stars. We choose to set the boundary between early-type and solar-type stars at F5, which corresponds to the spectral type where strong convective envelopes start to appear. The probability of the null hypothesis turns out to above 0.3 in both cases, which suggests once again that the distribution of H-band excesses does not have a different behavior around stars with and without warm dust.


\begin{figure}[t]
\centering
\includegraphics[scale=0.35]{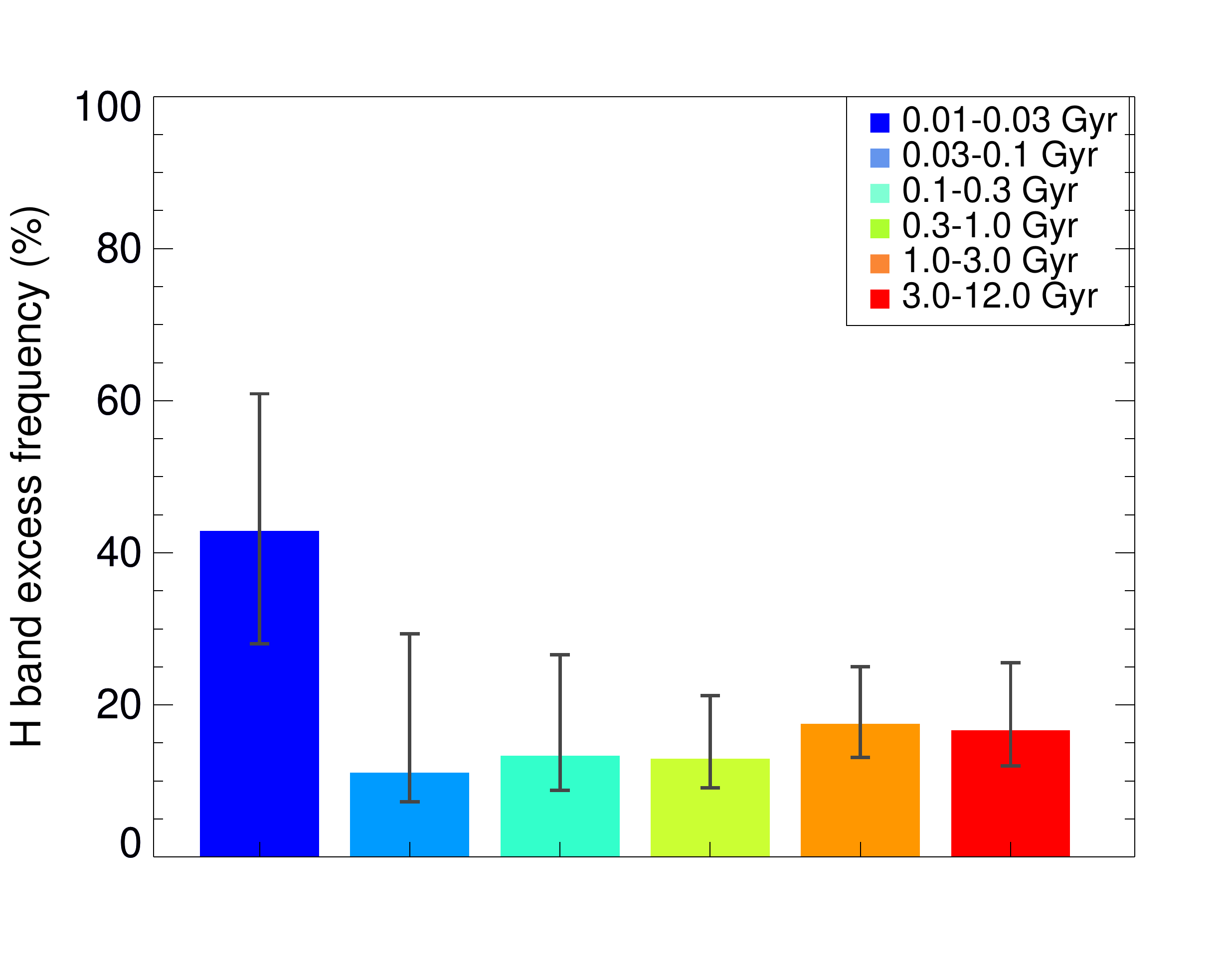}
\caption{Occurrence rate of H-band excesses as a function of stellar age in the combined sample of 133 stars.}
 \label{fig:agestat}
\end{figure}

Based on 85 single stars in their PIONIER survey, \citet{ertel14} investigated the possible relation between stellar age and hot exozodi detection rate, showing the absence of correlation, although a possible trend was observed that FGK stars could have more frequent hot exozodis at old ages. We revisit their analysis including the 48 new single stars observed here (see Fig.~\ref{fig:agestat}). 
Based on this larger sample of young main sequence stars, we note an increase in detection rate at very young ages, with in particular three stars out of five within the $\beta$ Pic moving group  ($\sim$20~Myr) showing a hot exozodi in the combined sample. Actually, among the two stars from the $\beta$ Pic moving group showing no hot exozodi, HD~172555 was identified as a marginal H-band detection by \citet{ertel14} based on the longest wavelength channel, and this detection was later confirmed to be significant through follow-up observations \citep{ertel16}. This leads to an actual detection rate of $80^{+8}_{-25}\%$ for hot exozodis at 20~Myr of age. To determine whether the population of young stars is significantly different from the older stars, we perform an Anderson-Darling test on two samples: the first one composed of young stars (younger than 30~Myr) and the second one composed of the other stars. We find a probability that the two samples are drawn from the same population of 6.8\%, i.e., a marginal evidence for them to be issued from different populations, at a $1.8\sigma$ level. Although this trend is based on a very small sample, it fits well within the picture that young main sequence stars might still be in the process of forming terrestrial planets, which may lead to strong dust production rate even in the innermost parts of the planetary systems. It is somewhat puzzling, though, that the youngest star in our sample (HD~109573 aka HR~4796, part of the TW Hya association), only shows a marginal H-band excess ($0.35 \pm 0.15 \%$).

\subsection{Influence of partly resolved exozodis}
\label{subsec:sublimation}

An important aspect of the hot exozodi detection statistics, which was not explored in previous works, is the influence of the location of the dust on its detectability with infrared interferometry. The most critical case in terms of angular resolution is for the most compact disks, which corresponds to the case where the circumstellar emission comes mostly from a region close to the sublimation distance of the dust grains. This situation actually corresponds to our current picture of hot exozodi detected with near-infrared interferometry, for which the measured excesses are understood to originate from the thermal emission of hot grains at a temperature close to sublimation \citep[e.g.,][]{mennesson11,lebreton13}. The emission could be even more confined by physical mechanisms such as grain pile-up \citep{kobayashi09}, magnetic trapping \citep{rieke16}, or gas drag \citep{pearce20}. In this case, the circumstellar emission might only be partly resolved by the interferometer, which would decrease the strength of the visibility drop, especially at the shortest baselines. Partly resolving the hot exozodiacal disk would therefore lead to a decreased sensitivity, as only part of the disk emission would affect the measured $V^2$. Here, we explore how the uneven sensitivity to compact hot exozodis around our target stars could bias the results of our survey.

So far, our working hypothesis has always been that circumstellar disks are fully resolved, and we have modeled them as a uniform emission filling the whole field of view. To test the impact of this working hypothesis on the measured detection rates, we computed the sublimation radius of the grains for each of the 133 stars in our combined sample, assuming silicates with a sublimation temperature of 1500~K. Our estimation of the sublimation radius is based on a simple black-body assumption, which we validated by running specific simulations with the GRaTeR radiative transfer package \citep{augereau99,lebreton12} to explore the dependency of the sublimation radius as a function of the grain size and composition. The resulting sublimation radii are given in Table~\ref{tab:subdiam}.

\longtab{
\begin{longtable}{ccccc}
\caption{Sublimation radius, sensitivity reduction factor ($\phi$), and effective sensitivity ($\sigma_{\rm eff}$) for the combined sample of 133 stars, computed for a sublimation temperature of 1500~K under black-body assumption. Asterisks denote the newly observed stars on the medium-sized AT configuration, while the other stars were observed on the compact AT configuration by \citet{ertel14}.}
\label{tab:subdiam}\\
\hline \hline
name & subl.\ rad. & subl.\ rad. & $\phi$ & $\sigma_{\rm eff}$  \\
 & (au) & (mas) & &(\%)\\
\hline
\endfirsthead
\caption{continued.}\\
\hline\hline
name & subl.\ rad. & subl.\ rad. & $\phi$ & $\sigma_{\rm eff}$  \\
 & (au) & (mas) & &(\%)\\
\hline
\endhead
\hline
\endfoot
HD~142    	&	0.06	&	2.29	&	0.25	&	1.06 \\
$^{\ast}$HD~203 &   0.08	&	1.94	&	0.90	&	0.25	\\
HD~1581  	&	0.04	&	4.60	&	0.70	&	0.44 \\
HD~2262  	&	0.12	&	5.02	&	0.77	&	0.23 \\
$^{\ast}$HD~2834 &   0.22	&	4.18	&	1.16	&	0.73	\\
$^{\ast}$HD~3126 &   0.06	&	1.43	&	0.59	&	0.40	\\
HD~3302   	&	0.09	&	2.52	&	0.29	&	0.89 \\
HD~3823   	&	0.05	&	2.20	&	0.23	&	0.97 \\
$^{\ast}$HD~4113 &   0.04	&	0.90	&	0.29	&	0.69	\\
HD~7570   	&	0.05	&	3.15	&	0.42	&	0.62 \\
HD~7788  	&	0.07	&	3.42	&	0.48	&	0.36 \\
$^{\ast}$HD~9672 &   0.17	&	2.82	&	1.23	&	0.26	\\
$^{\ast}$HD~10269 &   0.06	&	1.27	&	0.49	&	0.28	\\
HD~10647  	&	0.04	&	2.36	&	0.26	&	1.01 \\
$^{\ast}$HD~10939 &   0.22	&	3.49	&	1.26	&	0.26	\\
HD~11171  	&	0.06	&	2.43	&	0.27	&	1.55 \\
HD~14412  	&	0.02	&	1.83	&	0.16	&	1.30 \\
HD~15008 	&	0.21	&	5.00	&	0.77	&	0.42 \\
$^{\ast}$HD~15427 &   0.14	&	2.95	&	1.25	&	0.14	\\
HD~17051  	&	0.05	&	2.72	&	0.33	&	0.70 \\
$^{\ast}$HD~17848 &   0.16	&	3.24	&	1.26	&	0.11	\\
HD~17925  	&	0.02	&	2.23	&	0.23	&	0.99 \\
HD~19107  	&	0.11	&	2.75	&	0.34	&	0.63 \\
HD~20766  	&	0.03	&	2.49	&	0.28	&	0.92 \\
HD~20794  	&	0.03	&	4.36	&	0.66	&	0.56 \\
HD~20807  	&	0.03	&	2.89	&	0.36	&	1.46 \\
HD~22001 	&	0.08	&	3.64	&	0.52	&	0.39 \\
$^{\ast}$HD~23484 &   0.02	&	1.27	&	0.49	&	0.52	\\
$^{\ast}$HD~24649 &   0.05	&	1.26	&	0.49	&	0.46	\\
HD~25457 	&	0.05	&	2.65	&	0.31	&	0.45 \\
$^{\ast}$HD~28287 &   0.02	&	0.64	&	0.16	&	1.97	\\
HD~28355 	&	0.15	&	3.08	&	0.40	&	0.22 \\
$^{\ast}$HD~29137 &   0.05	&	0.91	&	0.30	&	0.53	\\
HD~30495  	&	0.03	&	2.62	&	0.31	&	0.67 \\
HD~31295 	&	0.14	&	4.05	&	0.60	&	0.25 \\
HD~31925  	&	0.10	&	2.48	&	0.28	&	0.79 \\
HD~33111 	&	0.24	&	8.62	&	0.96	&	0.43 \\
HD~33262 	&	0.04	&	3.79	&	0.55	&	0.38 \\
HD~34721  	&	0.05	&	2.03	&	0.20	&	1.06 \\
$^{\ast}$HD~36187 &   0.25	&	2.89	&	1.24	&	0.11	\\
$^{\ast}$HD~37306 &   0.13	&	1.99	&	0.92	&	0.16	\\
$^{\ast}$HD~37484 &   0.07	&	1.18	&	0.44	&	0.47	\\
HD~38858  	&	0.03	&	2.16	&	0.22	&	1.31 \\
$^{\ast}$HD~38949 &   0.04	&	0.85	&	0.26	&	0.68	\\
HD~39060 	&	0.11	&	5.66	&	0.82	&	0.28 \\
HD~40307  	&	0.02	&	1.48	&	0.11	&	2.18 \\
$^{\ast}$HD~41278 &   0.06	&	1.11	&	0.40	&	0.54	\\
HD~43162  	&	0.03	&	1.76	&	0.15	&	1.40 \\
$^{\ast}$HD~44524 &   0.11	&	1.05	&	0.37	&	0.54	\\
HD~45184  	&	0.04	&	1.75	&	0.15	&	1.00 \\
HD~53705  	&	0.04	&	2.17	&	0.22	&	1.04 \\
HD~56537 	&	0.21	&	6.72	&	0.90	&	0.28 \\
$^{\ast}$HD~60491 &   0.02	&	0.79	&	0.23	&	0.69	\\
$^{\ast}$HD~61005 &   0.03	&	0.83	&	0.25	&	0.48	\\
HD~69830  	&	0.03	&	2.20	&	0.23	&	1.15 \\
HD~71155 	&	0.22	&	5.90	&	0.84	&	0.30 \\
$^{\ast}$HD~71722 &   0.18	&	2.46	&	1.13	&	0.15	\\
HD~72673  	&	0.02	&	1.88	&	0.17	&	1.90 \\
$^{\ast}$HD~76143 &   0.14	&	2.70	&	1.20	&	0.15	\\
HD~76151  	&	0.04	&	2.04	&	0.20	&	1.42 \\
HD~76932  	&	0.05	&	2.19	&	0.23	&	1.85 \\
$^{\ast}$HD~80883 &   0.05	&	0.73	&	0.20	&	0.88	\\
HD~82434  	&	0.11	&	5.73	&	0.83	&	0.70 \\
HD~88955 	&	0.17	&	5.46	&	0.80	&	0.31 \\
$^{\ast}$HD~89886 &   0.18	&	1.10	&	0.39	&	0.68	\\
HD~90132  	&	0.11	&	2.67	&	0.32	&	1.28 \\
HD~90781     &   0.08	&	1.08	&	0.38	&	0.42	\\
$^{\ast}$HD~90874 &   0.15	&	2.19	&	1.01	&	0.13	\\
HD~91324 	&	0.07	&	3.36	&	0.46	&	0.37 \\
$^{\ast}$HD~92945 &   0.02	&	0.91	&	0.30	&	0.60	\\
HD~99211 	&	0.12	&	4.67	&	0.71	&	0.31 \\
HD~102365	&	0.03	&	3.34	&	0.46	&	0.50 \\
HD~104731	&	0.07	&	2.81	&	0.35	&	0.40 \\
$^{\ast}$HD~105850&   0.17	&	2.99	&	1.25	&	0.14	\\
$^{\ast}$HD~105912 &   0.06	&	1.28	&	0.50	&	0.30	\\
HD~108767	&	0.26	&	9.72	&	0.96	&	0.16 \\
$^{\ast}$HD~109573 &   0.17	&	2.53	&	1.15	&	0.13	\\
HD~109704    &   0.14	&	2.07	&	0.96	&	0.13	\\
HD~109787	&	0.20	&	5.05	&	0.77	&	0.26 \\
$^{\ast}$HD~112603 &   0.08	&	1.30	&	0.52	&	0.49	\\
HD~115617	&	0.03	&	4.04	&	0.60	&	0.38 \\
$^{\ast}$HD~117716 &   0.19	&	2.63	&	1.18	&	0.15	\\
$^{\ast}$HD~118972 &   0.02	&	1.26	&	0.49	&	0.19	\\
HD~120136	&	0.06	&	4.06	&	0.60	&	0.37 \\
HD~128898	&	0.10	&	6.32	&	0.87	&	0.25 \\
HD~129502	&	0.09	&	5.17	&	0.78	&	0.18 \\
HD~130109	&	0.27	&	6.56	&	0.89	&	0.48 \\
HD~134083 	&	0.06	&	3.25	&	0.44	&	1.07 \\
HD~135379	&	0.15	&	4.92	&	0.75	&	0.49 \\
HD~136202 	&	0.07	&	2.67	&	0.32	&	2.00 \\
$^{\ast}$HD~136544 &   0.08	&	1.07	&	0.38	&	0.93	\\
HD~139664	&	0.06	&	3.68	&	0.52	&	0.36 \\
HD~141891	&	0.10	&	8.43	&	0.96	&	0.21 \\
$^{\ast}$HD~141943 &   0.05	&	0.90	&	0.29	&	0.69	\\
HD~149661 	&	0.02	&	2.44	&	0.27	&	0.81 \\
HD~152391 	&	0.03	&	1.66	&	0.13	&	1.34 \\
HD~160032	&	0.07	&	3.27	&	0.44	&	0.25 \\
HD~160915 	&	0.05	&	2.89	&	0.36	&	0.77 \\
$^{\ast}$HD~161612 &   0.03	&	1.10	&	0.40	&	0.30	\\
HD~164259	&	0.09	&	3.68	&	0.52	&	0.36 \\
HD~165777	&	0.15	&	5.58	&	0.81	&	0.34 \\
HD~172555 	&	0.10	&	3.44	&	0.48	&	0.53 \\
$^{\ast}$HD~174474 & 	0.18	&	2.18	&	1.01	&	0.19	\\
HD~178253	&	0.18	&	4.67	&	0.71	&	0.50 \\
$^{\ast}$HD~179520 &   0.08	&	1.26	&	0.49	&	0.50	\\
$^{\ast}$HD~181327 &   0.07	&	1.27	&	0.49	&	0.32	\\
HD~182572	&	0.05	&	3.06	&	0.40	&	0.33 \\
$^{\ast}$HD~185615 &   0.03	&	0.80	&	0.23	&	1.37	\\
HD~188228	&	0.23	&	7.27	&	0.94	&	0.29 \\
$^{\ast}$HD~191089 &   0.06	&	1.23	&	0.46	&	1.04	\\
HD~192425 	&	0.16	&	3.49	&	0.49	&	0.51 \\
$^{\ast}$HD~192758 &   0.08	&	1.24	&	0.48	&	0.60	\\
HD~195627 	&	0.10	&	3.43	&	0.48	&	1.09 \\
$^{\ast}$HD~196141 &   0.03	&	0.79	&	0.23	&	0.86	\\
HD~197157	&	0.10	&	4.04	&	0.60	&	0.50 \\
HD~197692	&	0.07	&	4.72	&	0.72	&	0.28 \\
HD~203608	&	0.04	&	4.47	&	0.68	&	0.50 \\
$^{\ast}$HD~205674 &   0.06	&	1.20	&	0.45	&	1.11	\\
HD~206860 	&	0.04	&	2.06	&	0.20	&	1.48 \\
HD~207129 	&	0.04	&	2.51	&	0.28	&	0.63 \\
HD~210049 	&	0.17	&	4.15	&	0.62	&	0.61 \\
HD~210277 	&	0.04	&	1.71	&	0.14	&	2.14 \\
HD~210302 	&	0.06	&	3.24	&	0.44	&	0.57 \\
HD~210418	&	0.18	&	6.35	&	0.87	&	0.33 \\
HD~213845 	&	0.06	&	2.56	&	0.30	&	0.80 \\
HD~214953 	&	0.05	&	2.18	&	0.22	&	1.00 \\
HD~215648	&	0.07	&	4.59	&	0.70	&	0.31 \\
HD~215789	&	0.30	&	7.63	&	0.96	&	0.27 \\
HD~216435 	&	0.06	&	1.89	&	0.17	&	1.56 \\
HD~219482 	&	0.05	&	2.40	&	0.26	&	0.64 \\
HD~219571	&	0.11	&	4.89	&	0.75	&	0.36 \\
$^{\ast}$HD~220476 &   0.03	&	0.97	&	0.33	&	0.54	\\
$^{\ast}$HD~224228 &   0.02	&	0.82	&	0.24	&	1.64	\\
\hline
\end{longtable}
}

\begin{table*}[t]
\caption{Hot exozodiacal disk statistics from the combined PIONIER surveys \citep[][and this work]{ertel14}, after removing all stars showing an effective sensitivity larger than $0.5\%$, taking into account partial resolution effects. Columns ``\#S'' and ``\#E'' represent the number of target stars and of hot exozodi detections, respectively.}
\label{tab:statdatacorr}
\centering
\begin{tabular}{ccccccccccccccc}
\hline
\hline
 & \multicolumn{3}{c}{A-type stars}  & \multicolumn{3}{c}{F-type stars}  & \multicolumn{3}{c}{G/K-type stars}  & \multicolumn{3}{c}{Total}\\
 & \#S & \#E & detect.~rate & \#S & \#E & detect.~rate & \#S & \#E & detect.~rate & \#S & \#E & detect.~rate \\ [+4pt]
\hline 
All$^{a}$ & 34 & 7 & $20.6^{+8.5}_{-5.2} \%$ & 28 & 7 &$25.0^{+9.7}_{-6.3} \%$ & 6 & 1 & $16.7^{+23.2}_{-6.3} \%$ & 68 & 15 &$22.1^{+5.8}_{-4.2} \%$\\ [+4pt]
Warm disk  & 16 & 3 & $18.8^{+13.1}_{-6.1} \%$ & 8 & 2 & $25.0^{+19.3}_{-9.1} \%$ & 1 & 1 & $100^{+0.0}_{-60.0} \%$ & 25 & 6 & $24.0^{+10.3}_{-6.4} \%$\\ [+4pt]
Warm disk only & 11 & 2 & $18.2^{+16.3}_{-6.4} \%$ & 6 & 2 & $33.3^{+21.1}_{-12.7} \%$ & 0 & 0 & N.A. & 17 & 4 & $23.5^{+12.7}_{-7.1} \%$\\ [+4pt]
Cold disk only  & 4 & 2 & $50.0^{+20.2}_{-20.2} \%$ & 6 & 2 & $33.3^{+21.1}_{-12.7} \%$ & 2 & 0 & $0.0^{+45.7}_{-0.0} \%$ & 12 & 4 & $33.3^{+15.1}_{-10.3} \%$\\ [+4pt]
No warm disk & 17 & 3 & $17.6^{+12.6}_{-5.7} \%$ & 20 & 5 & $25.0^{+11.7}_{-7.1} \%$ & 5 & 0 & $0.0^{+26.3}_{-0.0} \%$ & 42 & 8 & $19.0^{+7.4}_{-4.6} \%$\\ [+4pt]
No disk  & 13 & 1 & $7.7^{+14.0}_{-2.6} \%$ & 14 & 3 & $21.4^{+14.2}_{-7.0} \%$ & 3 & 0 & $0.0^{+36.8}_{-0.0} \%$ & 30 & 4 & $13.3^{+8.6}_{-4.0} \%$ \\ [+4pt]
\hline
\end{tabular}
\tablefoot{($^{a}$) also includes one dusty star with no warm/cold classification: HD~36187 (A0V).}
\end{table*}

\begin{figure}[t]
\centering
\includegraphics[scale=0.36]{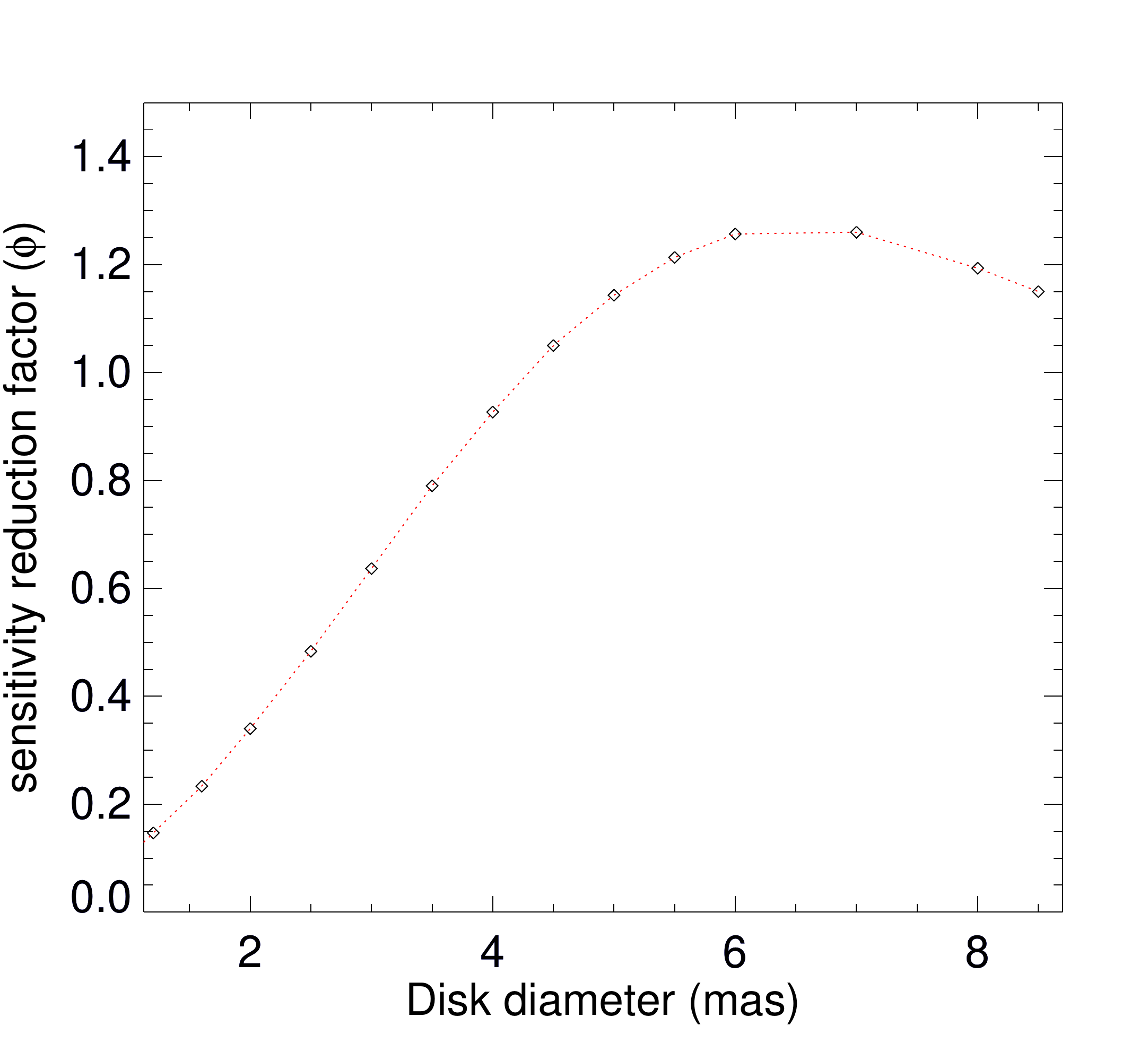}
\caption{Sensitivity reduction factor (i.e., ratio between the measured and injected flux ratio) as a function of the diameter of the circumstellar ring, for the medium-sized AT configuration (D0-H0-G1-I1).}
 \label{fig:contmodel}
\end{figure}

The estimated sublimation radii must then be compared with the angular resolution of the interferometric array to determine whether the disks are fully or only partly resolved. To do so, we considered infinitesimally thin rings with diameters ranging between 1.2 and 8.5~mas, corresponding to twice the minimum and maximum sublimation distance for the newly observed stars on the medium-sized AT configuration (see Table~\ref{tab:subdiam}). We injected these thin rings around a typical star of our survey to produce the expected $V^2$ for the star-disk system, using the medium-sized AT configuration at the VLTI (D0-H0-G1-I1) and a typical observing setup in terms of target elevation and hour angle coverage. This expected $V^2$ was then passed to our exozodi detection routine, which is based on the assumption of a fully resolved circumstellar disk filling the entire field-of-view, and we extracted the measured disk/star flux ratio using our standard fitting method. This measured flux ratio was then compared to the actual flux ratio injected in the model (chosen to be $3\%$ in this case), to produce a ``sensitivity reduction factor'' ($\phi$), defined as the ratio of measured to injected disk/star flux ratio. The result of this exercise is illustrated in Fig.~\ref{fig:contmodel}, where we plot the sensitivity reduction factor for thin annular disks of increasing diameters. As expected, smaller disk diameters lead to a bigger hit in sensitivity, with only about 15\% of the flux detected for the most compact disks. Half of the flux is missed for a disk diameter of about 2.5~mas (i.e., disk radius of 1.25~mas). The same exercise was carried out on the compact AT configuration (A1-B2-C1-D0) for the sample observed by \citet{ertel14}. The resulting sensitivity reduction factor $\phi$ is given for all of the stars in our combined sample in Table~\ref{tab:subdiam}, where asterisks denote the stars observed on the medium-sized AT configuration within the observing program presented in this paper.

Knowing the sensitivity reduction factor for all the stars in the PIONIER surveys, we can compute the effective sensitivity ($\sigma_{\rm eff}$) of our observations under the new working hypothesis that all the disks are confined to the sublimation radius of silicates. The effective sensitivity, defined as the $1\sigma$ error bar on the disk/star flux ratio divided by the sensitivity reduction factor, is given in Table~\ref{tab:subdiam}. Based on these revised sensitivities, we define a homogeneous sample in terms of effective sensitivity, by rejecting all the stars that have an effective sensitivity larger than $0.5\%$ (for which the chances to detect a hot exozodi are much lower, owing to the typical brightness of hot exozodis). This gives us a new sample of 68 stars, among which 25 show the presence of warm dust. The hot exozodi detection rate can then be recomputed on this new, more homogeneous sample, which is however strongly biased towards early-type stars because of the larger star/disk angular separation in those systems. The new detection rates are summarized in Table~\ref{tab:statdatacorr}. They are still compatible with each other within error bars, although the occurrence rate for the ``no-dust'' sample ($13.3^{+8.6}_{-4.0} \%$) seems to be systematically lower than for the rest of the sample (stars hosting warm and/or cold dust), which shows an occurrence rate of $28.9^{+8.2}_{-6.1} \%$ (11 out of 38 stars\footnote{this includes one dusty star for which the dust temperature could not be determined, so that it does not show up in either of the ``warm'' and ``cold'' dust categories in Table~\ref{tab:statdatacorr}}). To further explore this possible correlation, we used a two-sample Anderson-Darling test to compare the H-band excess distribution within the dusty and non-dusty samples, containing respectively 38 and 30 stars. The Anderson-Darling test on the H-band excess levels shows that the null hypothesis that the two samples are drawn from the same population can be rejected at significance level $p=0.0028$, which corresponds to a $3.0\sigma$ level. Using the excess significance instead of the excess levels in the Anderson-Darling test would increase the significance level to $3.4\sigma$ that the two samples are drawn different populations. Since the $1\sigma$ sensitivity threshold of 0.5\% used to define our sensitivity-corrected sample is somewhat arbitrary, we also examine a case where the threshold is set to 0.33\%. This new threshold boosts the significance that a common underlying population can be rejected to a $3.7\sigma$ level, for a sample of 40 stars. A tentative evidence for a correlation between the presence of hot dust and an outer reservoir was already found based on K-band observations at the CHARA array, but only for solar-type stars (FGK types), and based on a smaller sample \citep{Absil13,nunez17}. This tentative correlation was not confirmed at H band on a larger sample of stars in the study of \citet{ertel14}. The analysis presented here seems to finally reconcile the trends observed at H and K bands. It provides the first evidence at a $>3\sigma$ level that the presence of an outer dust reservoir may have a significant influence on the appearance of a near-infrared excess across all spectral types, although we underline the facts that this conclusion is based on the assumption that the dust is arranged in a thin ring close to its sublimation radius, and that the samples are still relatively small. It must also be kept in mind that the absence of observable amounts of warm/cold dust populations does not mean the complete absence of outer dust reservoirs, which could artificially increase or decrease the significance of this tentative correlation.

\begin{figure*}[t]
\centering
\includegraphics[width=0.48\textwidth]{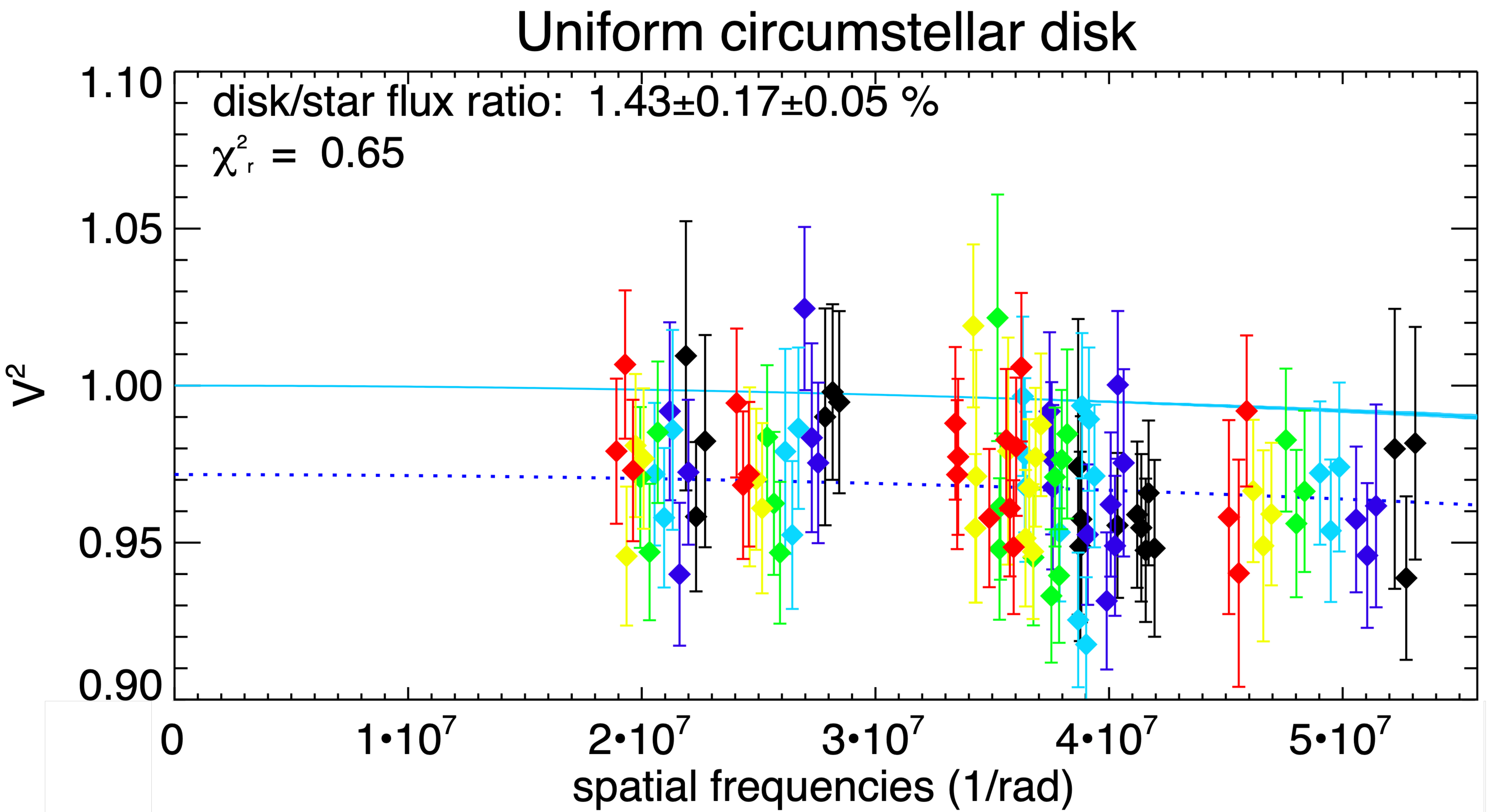} \hspace*{3mm}
\includegraphics[width=0.48\textwidth]{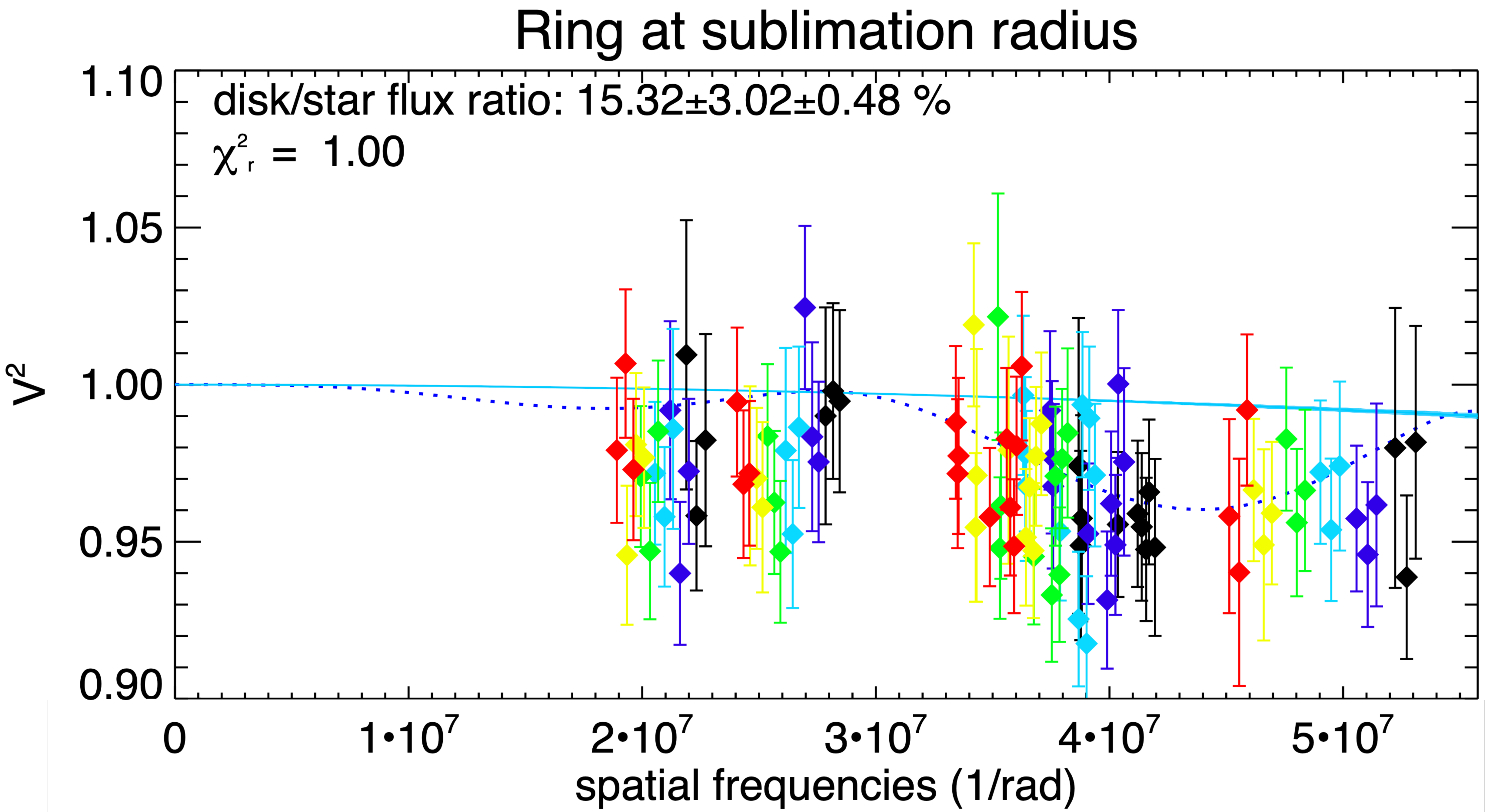}
\caption{Measured and modeled squared visibilities for HD~80883 using a limb-darkened star surrounded by uniform circumstellar emission (left) or by a ring of dust at the sublimation radius (right). The different colors of the data points represent the different spectral channels (one color per channel). The solid blue line shows the expected visibility for the stellar photosphere alone, and the dotted blue line the best fit for the star+disk model. Both disk models provide a reasonable fit to the measured visibilities, with $\chi_r^2 \leq 1$. }
 \label{fig:comparslope}
\end{figure*}

	\subsection{Location and temperature of the hot dust}

In order to further investigate the robustness of the tentative conclusion from the previous section, an interesting question is whether we could discriminate between a fully resolved (uniform) circumstellar emission, and a thin annulus model. This type of morphological study has already been attempted on Fomalhaut by \citet{absil09} using VLTI/VINCI, and on $\beta$~Pictoris by \citet{defrere12} using VLTI/PIONIER. In both cases, a very large number of observations were available, but no constraint could be derived on the disk morphology. We do not expect this situation to change in the present case, where we only collected three OBs on each of our targets. Nevertheless, we search for possible signs of partly resolved disks in our whole sample of detected hot exozodis, by looking for a slope in the $V^2$ drop as a function of baseline. Indeed, partly resolved disks should lead to smaller $V^2$ drop at shorter baselines, as they become less and less resolved. This exercise is illustrated in Fig.~\ref{fig:comparslope} for the case of HD~80883, for which the expected sublimation radius is particularly small. The data set shows no significant slope, suggesting that the excess is more probably caused by an extended disk than by a thin ring at the sublimation radius, although both models are consistent with the available data set. 

Another possible way to constrain the location of the dust grains would be to infer their temperature. This can potentially be done by exploring the wavelength dependence of the measured disk/star flux ratio. To test this, we have fitted blackbodies of various temperatures, including the host star temperature (flat contrast) to simulate the effect of scattered light, to the measured disk/star flux ratio as a function of wavelength (Fig.~\ref{fig:v2fits1}). Blackbody models with temperatures between 1500~K and the host star temperature fit the data almost equally well, with a reduced $\chi^2$ around 1. Using a blackbody temperature of 1000~K increases the median reduced $\chi^2$ to about 2 for our sample of 13 hot exozodis. The accuracy that can currently be reached with precision infrared interferometers such as VLTI/PIONIER is therefore not high enough to conclude on the dust temperature. At best, we could reject the hypothesis that the detected excess are due to the thermal emission of grains at temperature below 1000~K, which is not expected as such grains would not produce a significant H-band emission anyway. A possible way to circumvent this limitation would be to follow up our detections at other wavelengths, for instance using the second generation interferometric instruments of the VLTI \citep[GRAVITY in the near-infrared and MATISSE in the mid-infrared, see e.g.,][]{kirchschlager20}.

    \subsection{Origin of the hot dust}

The lack of a strong correlation between the hot exozodi phenomenon and the amount of warm and/or cold dust in outer reservoirs remains puzzling. Our understanding is that the hot dust is likely supplied from an outer reservoir for most stars, because parent bodies cannot survive on Myr timescales close to the dust sublimation radius due to collisional activity \citep[e.g.,][]{absil06}. However, the determining factor of whether detectable amounts of hot dust are present seems not to be the mass or location of this reservoir, but rather a different condition triggering the phenomenon. Trapping mechanisms have been proposed to sustain the observed, high dust masses that would otherwise require extreme replenishment rates due to the efficient removal of the hot dust \citep[e.g.,][]{pearce20}. These mechanisms include the pile-up of sublimating dust \citep{kobayashi09}, the effect of the stellar magnetic field \citep{czechowksi10,rieke16}, or the effect of gas \citep{lebreton13,pearce20}, possibly originating in the sublimation of the dust grains themselves. Trapping mechanisms could explain why even faint, undetected outer belts can supply sufficient material to produce detectable hot dust. Alternatively, or in addition, a specific configuration of the dust reservoir and a planetary system could be required to supply sufficient material to the inner regions \citep{Bonsor12a,Bonsor12b,Bonsor14,Faramaz17}. Such a mechanism could also be sufficiently efficient to act indistinguishably by us on systems with detectable and undetectable cold reservoirs.


Irrespective of the delivery mechanism and of the presence or absence of a trapping mechanism, a larger reservoir of cold material provides more material to be delivered to the inner region. Thus, naively one would still expect a correlation between the presence of massive debris disks and near-infrared excesses, at least statistically if not for individual targets. There is only a weak evidence for this to be the case in our sample, with a tentative correlation appearing only after correcting the sample for sensitivity under the assumption that all the hot dust is located close to the sublimation distance of silicates (Sect.~\ref{subsec:sublimation}). The lack of a more prominent correlation between hot dust and outer reservoirs might imply that there is an upper limit on the amount of hot dust that may be present or be supplied, and that this is reached in systems where the conditions for delivery and trapping are met, even for relatively small outer reservoirs. In the case of P-R drag, the amount that reaches the innermost regions is limited by collisions between the migrating dust grains \citep{Wyatt05}: while massive, collision-dominated outer belts produce more dust that can be dragged inward, most of this dust is destroyed before it reaches the inner regions. In contrast, most of the dust created in more tenuous, transport-dominated outer belts may reach the inner regions. This naturally decouples to some degree the amount of dust supplied to the hot exozodi region from the mass of the outer reservoir. For other scenarios, like comet delivery of the hot dust, an upper limit on the amount of hot dust could be set by the collisional evolution of the dust, which happens at a shorter time scale for more massive disks so that an equilibrium between dust influx and removal could typically be reached around similar dust levels. It is also possible that the amount of material transported inwards is dominated by the efficiency of the transport process (e.g., scattering) rather than the supply of material, which would also result in a (partial) decoupling of the hot dust quantity from the brightness of the cold reservoir. Finally, some trapping mechanisms may have intrinsic upper limits on the amount of dust they can trap. An alternative explanation for the similar flux ratios observed around all stars with near-infrared excess would be that the hot dust is optically thick. In that case, the observed flux would be driven by the surface area of the disk, and would be independent of the dust mass to the first order. This scenario would then require the surface area of the hot disk to increase for earlier spectral types, which could at least partly be explained by the larger dust sublimation radius around earlier spectral types.

Among the potential origins for the hot dust population, the possibility that hot dust is primordial (i.e., a remnant of the initial protoplanetary disk) cannot be completely ruled out without a detailed analysis of the possible trapping mechanisms. This would however require the trapping mechanism at play to be efficient enough on Gyr timescales, and to be on-going since the late stages of the primordial disk dispersal, which seems rather unlikely. This scenario would also be expected to show a more prominent age dependence in the hot exozodi phenomenon.

The apparent lack of correlation between hot and warm dust populations could be both an asset and a drawback in the preparation of space-based missions dedicated to Earth-like planet imaging. On one hand, the lack of correlation means that stars hosting hot dust populations should not necessarily be removed from the potential target lists of such missions, because they are not necessarily associated with large amounts of warm dust in the habitable zone. The influence on the mission performance of hot dust populations located at much smaller angular separation from the star than the habitable zone is however still to be investigated -- a task that we defer to a future, dedicated work. On the other hand, the lack of correlation also means that the detection of significant near-infrared emission with precision interferometry is not a prime criterion to build the target lists, which means that more mid-infrared interferometric observations with existing \citep[LBTI/NOMIC,][]{ertel20} or upcoming \citep[VLTI/Hi-5,][]{defrere18} instruments will be needed.


\section{Conclusions} \label{sec:concl}

In this paper, we used the VLTI/PIONIER interferometric instrument to search for resolved near-infrared circumstellar emission around a sample of main sequence stars known to harbor a warm dust disk from previous mid-infrared spectrophotometric observations, in an attempt to identify a possible connection between warm and hot dust populations. For that, we built a target list of 62 stars that showed signs of warm dust in the literature. Among the 52 stars for which we obtained data of sufficient quality, we identified 17 new H-band excesses, among which four are shown to be due to the presence of a previously unknown close stellar companion. The remaining 13 excess are thought to originate from hot dust populations, adding to the nine hot exozodi systems already detected with PIONIER by \citet{ertel14}. Combining these two samples, resulting in a total of 133 stars, we find an overall detection rate of $16.5^{+3.7}_{-2.7}\%$ for H-band excesses around nearby main sequence stars, with a possible hint for a larger underlying population of excesses below our sensitivity limit. Taking into account the fact that some of the hot exozodiacal disks may only be partly resolved by our interferometric baseline lengths, we estimate that the true occurrence rate could actually be as high as $22.1^{+5.8}_{-4.2}\%$, if we only include stars that have a corrected $1\sigma$ sensitivity of 0.5\% or better on the disk/star flux ratio. Our data sets do however not allow us to discriminate between a fully resolved disk and a thin annulus at the sublimation radius as the most appropriate model to reproduce our observations, so that the true occurrence rate at our sensitivity level could be anywhere between 16.5\% and 22.1\%.

We then searched for a possible correlation between the presence of a known warm dust population around the target stars and the detection of a near-infrared excess in our interferometric observations. For that, we re-evaluated the presence of warm and/or cold dust around all of the 133 stars in our combined sample through SED modeling, and defined two samples containing respectively the stars showing warm dust emission or not. We found that the distribution of near-infrared excesses around the warm dust sample is fully compatible with that of the control sample, suggesting the absence of direct connection between warm and hot dust populations. This conclusion does not depend on the considered spectral type. No correlation was found either between the detection rate of near-infrared excess and the stellar age, although there is a marginal trend for young stars ($\leq 30$~Myr) to have more frequent H-band excesses. After correcting the sensitivity of our observations for the fact that the hot dust could be arranged in a thin ring around its sublimation radius, and subsequently limiting our sample to the stars for which the corrected $1\sigma$ error bar is smaller than 0.5\%, we find tentative evidence at the $3\sigma$ level that the distribution of near-infrared excesses around stars showing any kind of outer dust reservoir (warm or cold) is statistically different from the distribution of near-infrared excesses around stars showing no outer dust reservoir, with larger near-infrared excesses around the dusty stars. This conclusion pertains mostly to early-type (A and F) stars, which make up the most of our sensitivity-corrected sample, and only holds if the dust is arranged in a thin annulus close to its sublimation radius, a hypothesis that we cannot confirm nor infirm based on our PIONIER data. A possible caveat to these conclusions is that some of the near-infrared excesses might be variable, as suggested by \citet{ertel16} and \citet{nunez17}, so that a non-detection does not necessarily mean that there is never any detectable excess around a particular star. It must also be kept in mind that we can only probe correlations down to the sensitivity level of both the near-infrared interferometric observations and the mid- to far-infrared photometry used in this study, and that underlying correlations may exist at lower sensitivity levels.


Although the present work puts in light a tentative, previously unknown correlation between hot and warm/cold dust populations, it does not settle the question of the origin of hot exozodiacal dust. Our current understanding is that at least one transport mechanism is at play to inject material in the innermost parts of planetary systems, and that the material additionally needs to be confined close to its sublimation radius by a trapping mechanism. The nature of these transport and trapping mechanisms is however still unclear, and will probably require new diagnostic tools to be properly constrained, although it is worth noting that some specific hot-dust systems in our sample look incompatible with P-R drag dust production. High-contrast interferometric observations in the thermal infrared (L, M, and N bands) would be a powerful way to derive useful new constraints on these dust populations.

\begin{acknowledgements}
The authors thank the French National Research Agency (ANR, contract ANR-2010 BLAN-0505-01, EXOZODI) for financial support. L.\,M.\ acknowledges the F.R.S.-FNRS for financial support through a FRIA PhD fellowship. G.\,M.\,K.\ is supported by the Royal Society as a Royal Society University Research Fellow. J.\,O. acknowledges support by ANID, -- Millennium Science Initiative Program -- NCN19\_171, from the Universidad de Valpara\'iso, and from Fondecyt (grant 1180395). We thank the Belgian GTO on VISA for the generous allocation of observing time. This work made use of the Smithsonian/NASA Astrophysics Data System (ADS) and of the Centre de Donn\'ees astronomiques de Strasbourg (CDS).
\end{acknowledgements}

\bibliographystyle{aa} 
\bibliography{PIONIER_exozodi20} 

\begin{appendix}

\section{Spectral energy distributions} \label{appendix}

Figures~\ref{seds0} to \ref{seds-1} show observations and model spectral energy distribution for each target in our combined sample of 133 stars.

\newcommand{\sedtext}{Observations and models for the targets in our survey. Each sub-panel shows photometry as dots, upper limits as triangles, and IRS spectra as black lines. Models are shown as solid lines, with blue for the star, green for the disc, and orange for the total model. Where two are fitted, individual disc components are shown as dotted lines.}

\begin{figure*}
\centering
\includegraphics[width=1\textwidth]{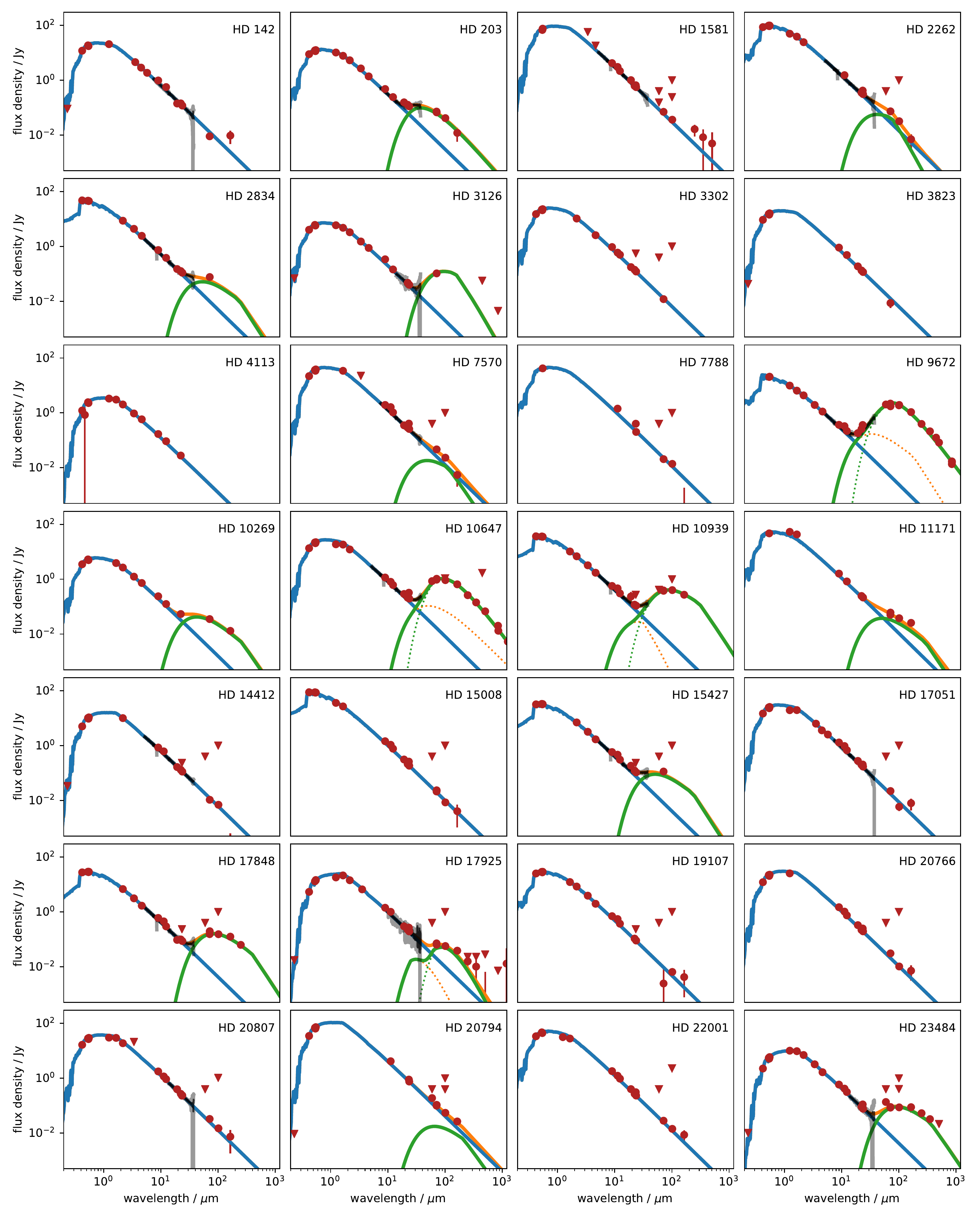}
\caption{\sedtext}
\label{seds0}
\end{figure*}

\begin{figure*}
\centering
\includegraphics[width=1\textwidth]{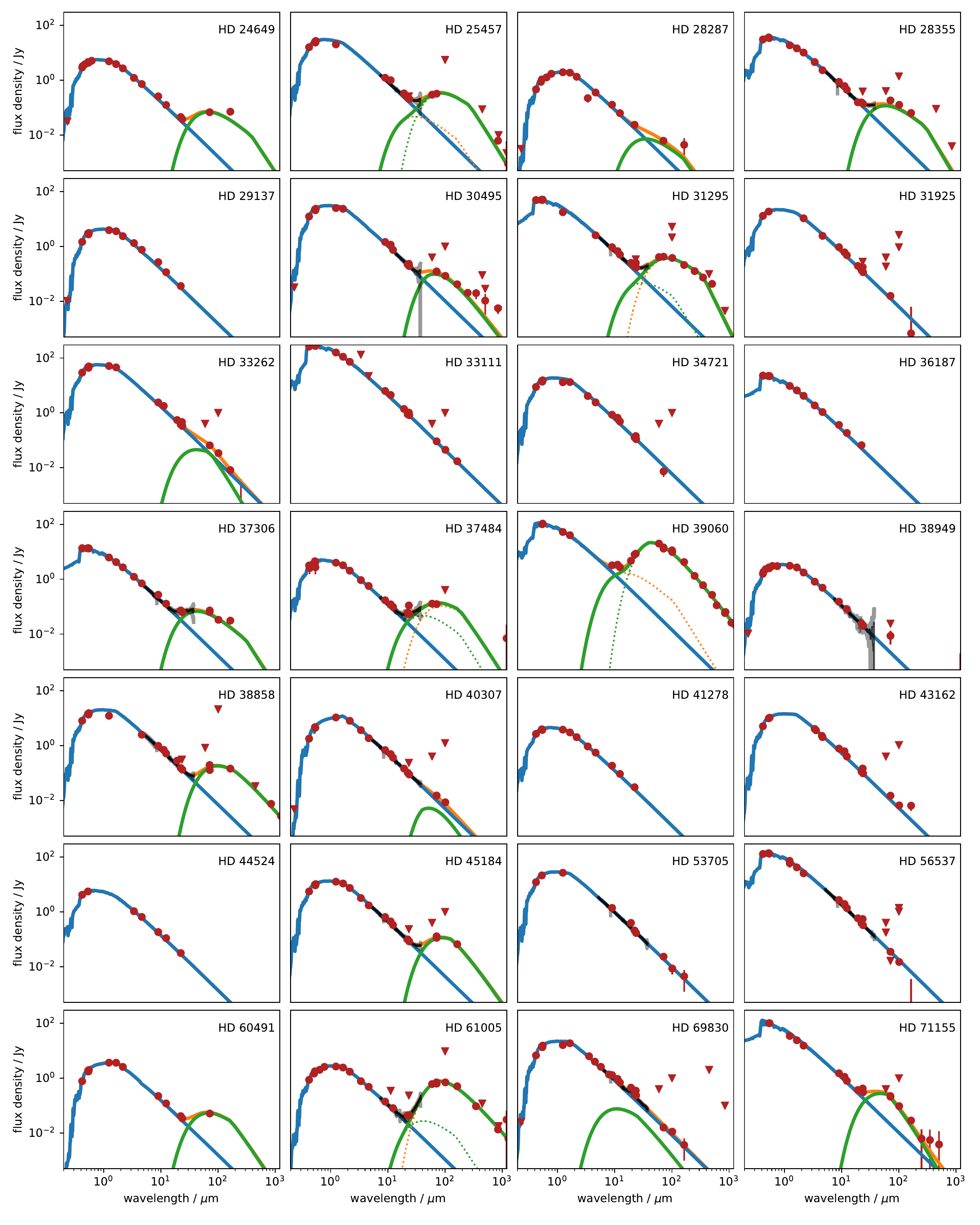}
\caption{\sedtext}
\end{figure*}

\begin{figure*}
\centering
\includegraphics[width=1\textwidth]{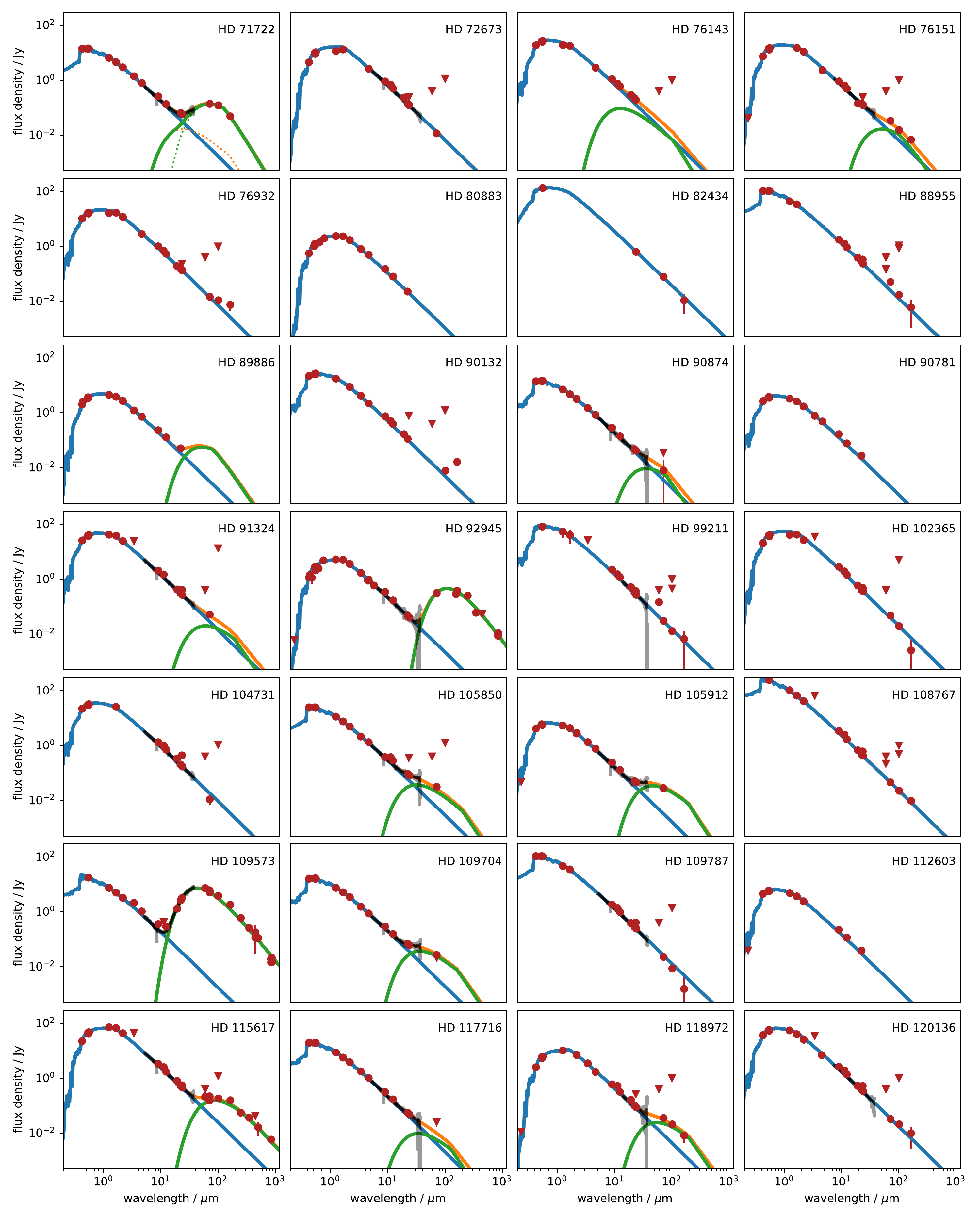}
\caption{\sedtext}
\end{figure*}

\begin{figure*}
\centering
\includegraphics[width=1\textwidth]{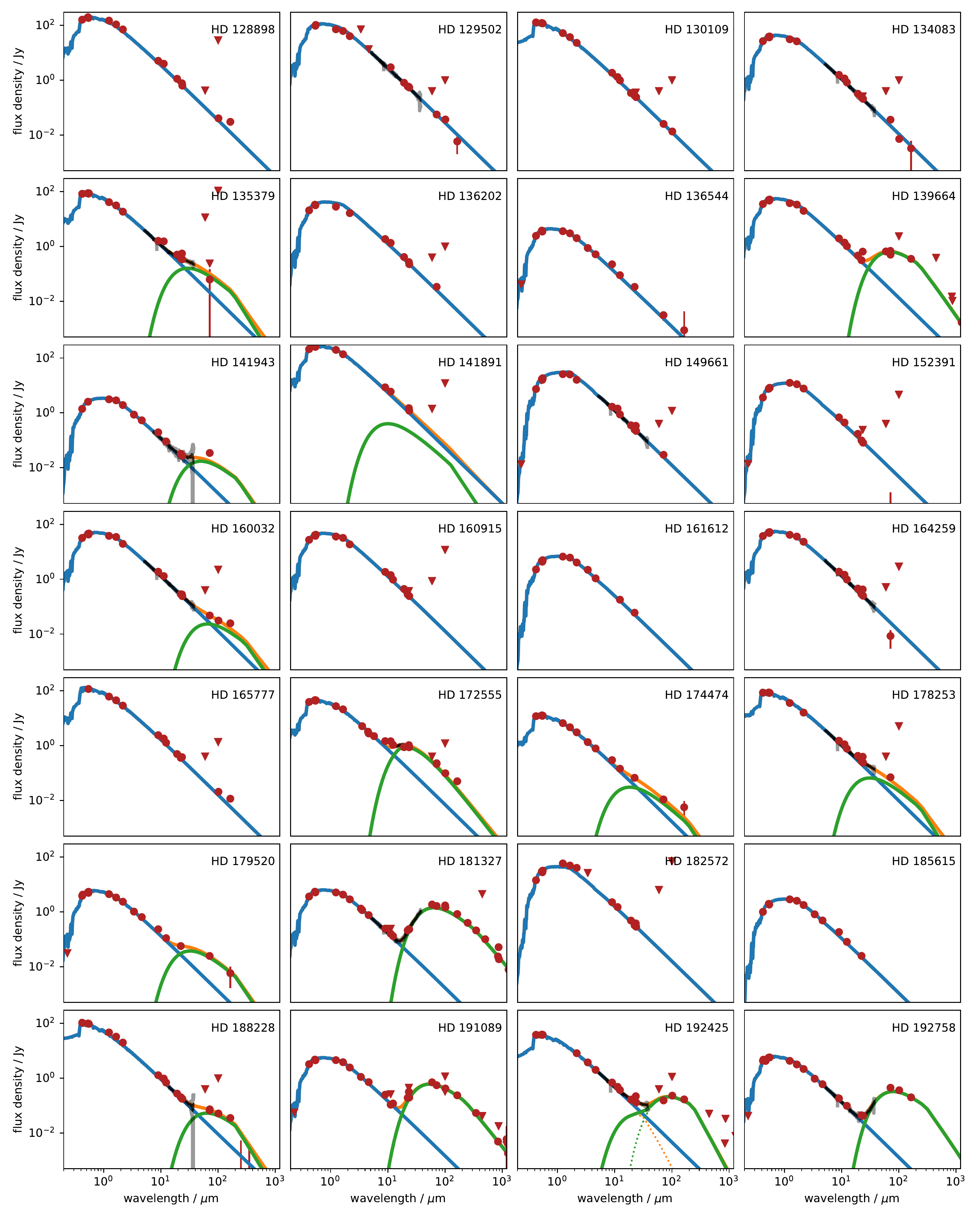}
\caption{\sedtext}
\end{figure*}

\begin{figure*}
\centering
\includegraphics[width=1\textwidth]{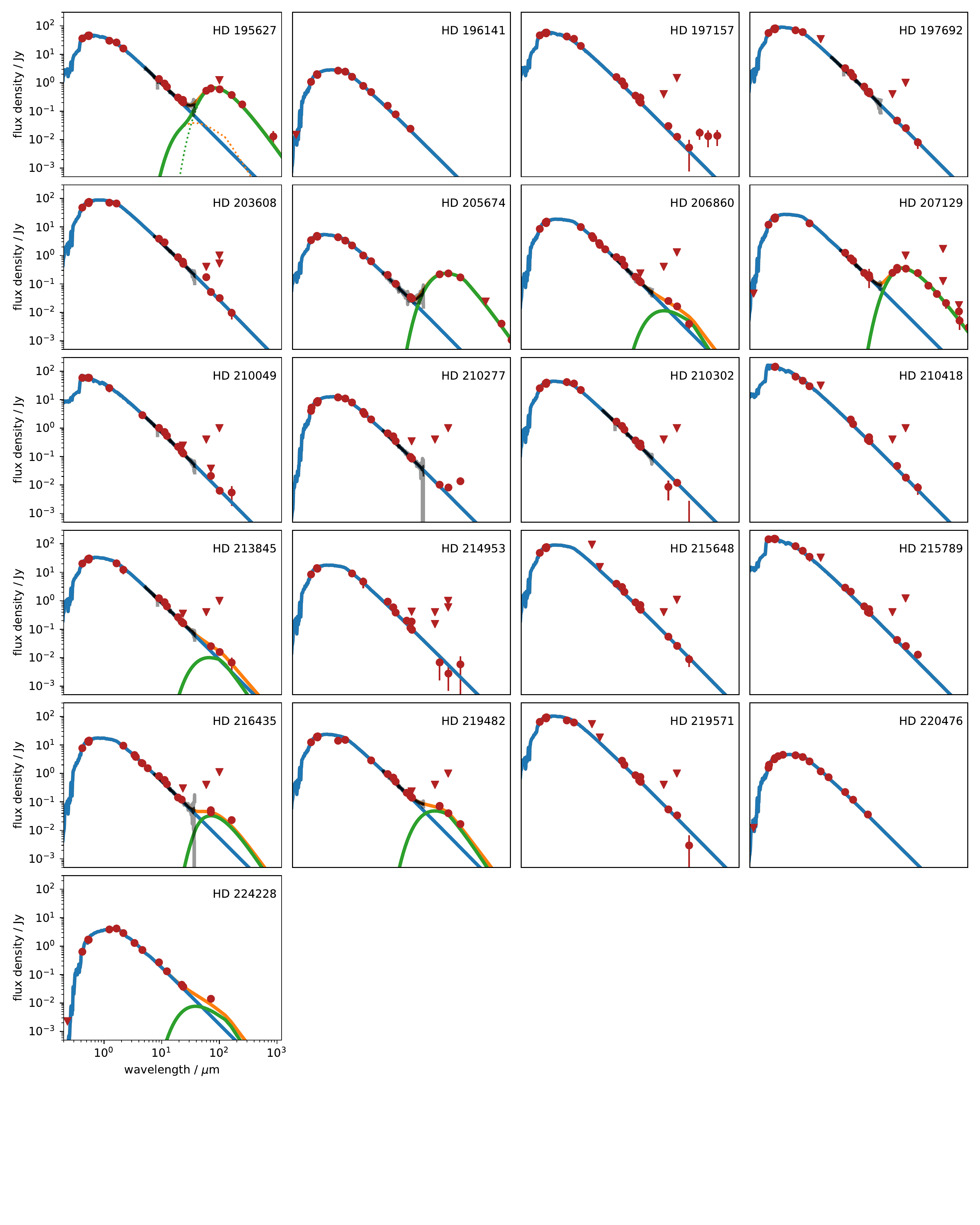}
\caption{\sedtext}
\label{seds-1}
\end{figure*}



\end{appendix}

\end{document}